\documentclass[11pt]{article}
\usepackage{amsmath}
\usepackage{amssymb}
\usepackage{graphicx}
\usepackage{verbatim}
\usepackage{psfrag}

\newcommand{\bibtitle}[1]{{\em #1}}
\newcommand{\hepth}[1]{{\tt hep-th/#1}}
\newcommand{\hepph}[1]{{\tt hep-ph/#1}}
\newcommand{\mathph}[1]{{\tt math-ph/#1}}

\newcommand{\gemstone}{{\tt gemstone}}

\newcommand{\spin}[1]{{\mbox{\it spin}}{#1}}
\newcommand{\Spin}[1]{{\mbox{\it Spin}}{#1}}

\DeclareMathOperator{\tr}{tr}                               
\newcommand{\be}{\begin{equation}}
\newcommand{\ee}{\end{equation}}
\newcommand{\R}{\mathbb{R}}                                 
\newcommand{\N}{\mathbb{N}}                                 
\renewcommand{\S}{\mathbb{S}}                               
\newcommand{\ti}[1]{{\scriptscriptstyle\mathrm{#1}}}        
\newcommand{\Ncal}{\mathcal{N}}                             
\newcommand{\unit}{\mathbf{1}}                              
\newcommand{\modulus}[1]{{| #1 |}}                          
\newcommand{\eps}{\varepsilon}                              
\newcommand{\order}[1]{{\cal O}\left(#1\right)}          
\newcommand{\comm}[2]{[#1,#2]}                              
\newcommand{\acomm}[2]{\{#1,#2\}}                           
\newcommand{\U}{\mathrm{U}}                                 
\newcommand{\SU}{\mathrm{SU}}                               
\newcommand{\SO}{\mathrm{SO}}                               
\newcommand{\su}{\mathfrak{su}}                             
\newcommand{\slf}{\mathfrak{sl}}                             
\newcommand{\so}{\mathfrak{so}}                             
\newcommand{\ket}[1]{\bigl|#1\bigr>}                        
\newcommand{\Op}[1]{\bigl|{{\cal O}}_{#1}\bigr>}        

\newcommand{\Z}{Z}

\newcommand{\Zd}{{Z^\dag}}                                  
\newcommand{\Wd}{{W^\dag}}                                  
\newcommand{\Yd}{{Y^\dag}}                                  
\newcommand{\bZ}{{\bar Z}}                                  
\newcommand{\bZd}{{\bar Z^\dag}}                            
\newcommand{\bWd}{{\bar W^\dag}}                            
\newcommand{\bYd}{{\bar Y^\dag}}                            

\newcommand{\atopfrac}[2]{\genfrac{}{}{0pt}{}{#1}{#2}}

\makeatletter \@addtoreset{equation}{section} \makeatother

\begin{document}

\thispagestyle{empty}
\begin{flushright}
{\sc\footnotesize hep-th/0412331} \\
{\sc\footnotesize AEI-2004-098} \\
{\sc\footnotesize ULB-TH-05-03}
\end{flushright}
\vspace{.8cm}
\setcounter{footnote}{0}
\begin{center}
{\Large{\bf Planar plane-wave matrix theory at the four loop order: \par}}
\vspace{2mm}
{\Large{\bf Integrability without BMN scaling\par}}
\vspace{10mm}
{\sc Thomas Fischbacher$^{(a,b)}$, Thomas Klose$^{(a)}$ and Jan Plefka$^{(a)}$} \\[8mm]
$a$)
{\it Max-Planck-Institut f\"ur Gravitationsphysik \\
     Albert-Einstein-Institut \\
     Am M\"uhlenberg 1, D-14476 Potsdam, Germany} \\ [5mm]

$b$)
{\it Physique Th\'eorique et Math\'ematique and\\
     International Solvay Institutes, Universit\'e Libre de Bruxelles,\\
     Campus Plaine C.P. 231, B--1050 Bruxelles, Belgium\\[5mm]
}

{\tt tf,thklose,plefka@aei.mpg.de} \\[10mm]
\end{center}

\begin{abstract}
\noindent  We study $\SU(N)$ plane-wave matrix theory up to fourth
perturbative order in its large $N$ planar limit. The effective
Hamiltonian in the closed $\su(2)$ subsector of the model is
explicitly computed through a specially tailored computer program
to perform large scale distributed symbolic algebra and generation
of planar graphs. The number of graphs here was in the deep billions.

\noindent The outcome of our computation establishes the four-loop
integrability of the planar plane-wave matrix model. To elucidate
the integrable structure  we apply the recent technology of the
perturbative asymptotic Bethe ansatz to our model. The resulting
S-matrix turns out to be structurally similar but nevertheless
distinct to the so far considered long-range spin-chain S-matrices
of Inozemtsev, Beisert-Dippel-Staudacher and
Arutyunov-Frolov-Staudacher in the AdS/CFT context.

\noindent In particular our result displays a breakdown of BMN
scaling at the four-loop order. That is, while there exists an
appropriate identification of the matrix theory mass parameter
with the coupling constant of the $\Ncal = 4$ superconformal
Yang-Mills theory which yields an eighth order lattice derivative
for well separated impurities (naively implying BMN scaling) the
detailed impurity contact interactions ruin this scaling property
at the four-loop order.

\noindent Moreover we study the issue of ``wrapping'' interactions,
which show up for the first time at this loop-order through a
Konishi descendant length four operator.
\end{abstract}

\newpage\mbox{}\vspace{10mm}
\tableofcontents
\newpage


\section{Introduction}

\noindent
In this work we report on a high order perturbative study of
plane-wave matrix theory \cite{bmn} in its large $N$ limit, which
is a supersymmetric, mass deformed $\SU(N)$ gauge quantum
mechanical model intricately related to maximally supersymmetric
four dimensional Yang-Mills theory through a consistent reduction on
$\R\times\S^3$ \cite{pwmt-from-sym}. It contains quite involved
technical parts, in particular on the side of realizing this
computation through a specially tailored computer program designed
by one of the authors (TF) to perform large scale distributed
symbolic algebra for the computation of planar diagrams. Readers
interested mostly in the computer science aspects of this
computation are advised to directly proceed to
App.~\ref{sec:gemstone}, which is largely self contained.

The motivation for performing this computation grew out of the
mounting evidence for integrability of planar $\Ncal=4$ super
Yang-Mills theory (SYM), which should be of prime importance in
the AdS/CFT correspondence. In a nutshell it states that the
dilatation operator of the planar gauge theory, which yields the
anomalous dimensions of local composite operators in the conformal
quantum field theory, is given by a long-range integrable spin
chain Hamiltonian
\cite{minahan-integrability,bks,beisert-complete-one-loop,beisert-su32,ss-ino,bds-long-range}\footnote{For
a recent review see \cite{BeisertPHD}.}. This statement is by now
firm at the one-loop level for the full theory in form of a
(non-compact) $\su(2,2|4)$ super spin-chain
\cite{beisert-complete-one-loop} and has been extended to the
three-loop level in a closed supersymmetric $\su(2|3)$ subsector
\cite{beisert-su32}. In this picture the loop order of the
considered dilatation operator is linked to the spread of the
local spin interactions of the spin-chain Hamiltonian. In the
smallest compact closed subsector of the $\Ncal=4$ gauge theory,
the $\su(2)$ sector, the $l$-loop order contribution corresponds
to a $s=1/2$ spin-chain Hamiltonian of Heisenberg type involving
the interactions of $l+1$ neighboring spins \cite{bks}. Next to
integrability a key property of the established form of the
dilatation operator is its perturbative ``BMN scaling'', which is
also predicted from the dual plane-wave superstring \cite{bmn}.
This scaling property concerns the form of the $l$th loop
contribution to the energy eigenvalue (or anomalous scaling
dimension) in the thermodynamic limit of large spin-chains:
its leading contribution scales as $1/L^{2\, l}$, where $L$
is the spin-chain length.

In \cite{ss-ino} one of the few known long
range integrable spin-chain models of statistical mechanics due to
Inozemtsev \cite{inozemtsev} was identified as being able to
provide the established form of the super Yang-Mills dilatation
operator up to the three-loop level. This is of course very
exciting, as this model could thus provides us with exact
non-perturbative information on the $\Ncal=4$ gauge theory.
However, as was explicitly demonstrated in \cite{ss-ino}, the
Inozemtsev chain predicts a breakdown of the BMN scaling property
at four-loop order. We wish to stress that such a breakdown of
{\sl perturbative} BMN scaling is not necessarily at clash with
the AdS/CFT correspondence, despite the BMN scaling property of
the dual plane-wave superstring spectrum. A similar breakdown is
by now firmly established in the ``near'' plane-wave superstring
\cite{nearPW} and in the matching of energies of rotating
solitonic $AdS_5\times S^5$ string solutions
\cite{spinningstrings} at three-loop order in perturbative
gauge theory. This discrepancy has been attributed to an order of
limits problem, stating that a {\sl perturbative} agreement should
not be expected \cite{bds-long-range}. Hence, based on these
results there presently is no reason to expect perturbative BMN
scaling in $\Ncal=4$ super Yang-Mills! \footnote{In the literature
arguments for the mechanism of an all loop BMN scaling property of
$\Ncal=4$ SYM have been put forward in \cite{gross}. In particular
our finding of the ``naive'' BMN scaling property is in line with
these arguments, the breakdown of BMN scaling arises from subtle
boundary effects when two impurities come close to each other to
be discussed in section 4.}

Nevertheless the {\sl assumption} of perturbative BMN scaling is a
very fruitful one. Indeed Beisert, Dippel and Staudacher
\cite{bds-long-range} have argued that the $\su(2)$ dilatation
operator is completely determined up to the five loop level once
one assumes integrability, BMN scaling and structural constraints
from the underlying Feynman diagrammatic structure. These authors
went on to conjecture an intriguingly simple all loop (asymptotic)
Bethe ansatz, capable of reproducing the eigenvalues of the
explicitly deduced dilatation operator up to five loops. Inspired
by this structure a similar but distinct asymptotic Bethe ansatz
for the full quantum $AdS_5\times S^5$ string in the $\su(2)$
subsector was conjectured by Arutyunov, Frolov and Staudacher in
\cite{frolov-gleb-matthias}. This ansatz accounts for the above
mentioned three-loop discrepancies and remarkably reproduces all
presently known data of the string side of the correspondence: In
its thermodynamic limit it reduces to the Bethe equations of the
classical spinning string sigma model of \cite{Kazakov}, it
reproduces the near plane-wave corrections to the quantum string
spectrum deduced in \cite{nearPW} and leads to the expected
$\lambda^{1/4}$ behaviour of the spectrum at strong coupling. The
associated spin-chain Hamiltonian to the first few orders in a
small $\lambda$ expansion was established in
\cite{spinchainforqstrings}.

Given these results it would clearly be of central importance to
determine the exact form of the $\su(2)$ dilatation operator at
the four-loop level and determine whether it obeys BMN scaling and
is  indeed integrable. This, however, appears to be beyond present
capability in the perturbative quantum field theory, where the
state of the art is at three-loop order \cite{eden-BMN,moch,qcdpeople}.

Our work adds a further piece to this puzzle by considering a reduced
(toy) model of the full quantum conformal field theory. The scaling dimensions
of the gauge theory on $\R^4$ are equivalent to the spectrum
of energies in a radial quantization of the model on $\R\times\S^3$.
It turns out that one may consistently truncate the Kaluza-Klein
spectrum of states of $\Ncal=4$ SYM on the three-sphere to obtain a
reduced gauge quantum mechanical model \cite{pwmt-from-sym}.
The resulting reduced model is the plane-wave matrix theory (PWMT), which has also
been proposed as a microscopic definition of M-theory in a maximally
supersymmetric eleven-dimensional plane-wave background\footnote{It also
arises as a regularization of the light-cone supermembrane in this
background, or alternatively from the dynamics of D0-branes in type
IIA string theory \cite{dasgupta-mt1}}. Moreover, as was shown in
\cite{pwmt-integrability}, it not only inherits the planar integrability of the
$\Ncal=4$ gauge theory, but leads to precisely the same
dilatation operator in the closed $\su(2)$ subsector up to the three
loop level, where the SYM information is firm. In order to perform
this matching one has to establish a relation between the coupling
constants of the full and the reduced model, which is corrected order
by order in perturbation theory. In fact, this relation is uniquely
determined by the requirement of BMN scaling behavior.

\medskip

In this paper we push the perturbation theory of PWMT to fourth
order. We explicitly compute the planar effective Hamiltonian in
the $\su(2)$ subsector without relying on any assumptions. The
model remains integrable at this loop order but {\sl ceases} to
obey BMN scaling: the leading behavior of the energy eigenvalues
scales as $1/L^7$ -- after one performs the above mentioned
renormalization of coupling constant. Indeed the obtained
four-loop contribution is \emph{neither} of the Inozemtsev \emph{nor} of
the Beisert, Dippel, Staudacher type, while it clearly agrees with
these spin-chains up to the three-loop level.

We elucidate the underlying integrable structure of the system by
constructing the associated ``perturbative asymptotic Bethe ansatz'', a
novel concept very recently developed in \cite{s-matrix}. The
extracted S-matrix turns out to qualitatively agree with the
quantum string S-matrix of Arutyunov, Frolov and Staudacher
\cite{frolov-gleb-matthias} but quantitatively disagrees by an
additional power of the coupling constant in the exponential term
involving the higher charges -- which is responsible for the
breakdown of BMN scaling. This finding lends support to the
suspicion that the generic form of a long-range integrable spin
chain S-matrix will involve an extra exponential factor with a
characteristic dependence on the higher charges of the spin-chain,
first proposed in \cite{frolov-gleb-matthias} and further
developed in \cite{s-matrix}.

Furthermore we can shed light on the structure of the ``wrapping''
interactions, which are inaccessible in the asymptotic forms of
the presently conjectured all loop Bethe ans\"atze. At four-loop
order (including five neighboring spin interactions) we have
for the first time a non-protected operator, which is shorter than
the spread of the dilatation operator: a Konishi descendant of
length four. It turns out that no natural prescription of how to
act with the five spin terms on this shorter state (wrapping or
dropping) reproduces the explicitly computed scaling dimension.

On the computational side, the four-loop study was achieved with a
newly developed computer program which constructs all planar
diagrams of the perturbation expansion and performs the
corresponding algebra. A change of the platform and the use of
heavily improved algorithms led to a performance increase by a
factor of more than 100 with respect to the previous version when
comparing the three-loop computation times. Only this enhancement brought
fourth order to reach, but it still required some 88.000 CPU hours
on current computers.


\section{Effective Hamiltonian of plane-wave matrix theory}

\subsection{The model}

\noindent
The Hamiltonian of plane-wave matrix theory can be written as $H = H_0 + V_1 + V_2$ with
\begin{align}
  H_0 & = \tr \left[ \tfrac{1}{2} P_I P_I 
                   + \tfrac{1}{2} \left( \tfrac{M}{2} \right)^2 X_a X_a
                   + \tfrac{1}{2} M^2 X_i X_i
                   - \tfrac{3M}{4} i \Theta \gamma_{123} \Theta \right] \; , \nonumber \\
  V_1 & = - M i\eps_{ijk} \tr X_i X_j X_k - \tr \Theta \gamma_I \comm{X_I}{\Theta} \; , \label{eqn:matrix-theory} \\
  V_2 & = - \frac{1}{4} \tr \comm{X_I}{X_J} \comm{X_I}{X_J} \; . \nonumber
\end{align}
The degrees of freedom are:
\begin{center}
\begin{tabular}{|l|cl|c|l|c|} \hline
  field           & \multicolumn{2}{l|}{notation} & mass & modes & symmetry \\ \hline
  light scalars   & $X_a$           & $(a=1,\ldots,6)$ & $M/2$        & $a_a$,      $a^\dag_a$      & $\SO(6)$ \\
  heavy scalars   & $X_i$           & $(i=1,\ldots,3)$ & $M$          & $b_i$,      $b^\dag_i$      & $\SO(3)$ \\
  Majorana spinor & $\Theta_\alpha$ & $(\alpha=1,\ldots,16)$ & $3M/4$ & $c_\alpha$, $c^\dag_\alpha$ & $\SO(9)$ \\ \hline
\end{tabular}
\end{center}
\noindent The ``mass parameter'' $M$ is in fact dimensionless\footnote{In terms of M-theoretic quantities it is given by $M = \frac{\mu l_P^2}{6R}$, where $\mu$ is 
the parameter of the plane-wave background, $l_P$ is the eleven-dimensional Planck length and $R$ is the radius of the compactified eleventh dimension.} and will 
serve as (inverse) coupling constant in perturbation theory below.

In \eqref{eqn:matrix-theory} we used the index $I=(a,i)$ to refer to all scalars at once. All fields $[X_a(t)]_{rs}$ etc are function of time and take values in the 
Lie algebra of the gauge group $\SU(N)$, $r=1,\ldots,N^2-1$. In the fermionic sector we work in a representation with charge conjugation matrix equal to unity and 
with real and symmetric Euclidean Dirac matrices $(\gamma_I)_{\alpha\beta}$.

For the quantization, we introduce the following creation and annihilation operators
\begin{align}
  P_a           & = \sqrt{\tfrac{M}{4}}  \left( a_a + a_a^\dag \right) &
  P_i           & = \sqrt{\tfrac{M}{2}}  \left( b_i + b_i^\dag \right) &
                & \Theta_\alpha = c_\alpha + c^\dag_\alpha \nonumber  \\
  X_a           & = \tfrac{i}{\sqrt{M}}  \left( a_a - a_a^\dag \right) &
  X_i           & = \tfrac{i}{\sqrt{2M}} \left( b_i - b_i^\dag \right) &
                & \mbox{with } {c = \Pi^- \Theta \atop c^\dag = \Pi^+ \Theta }
\end{align}
where $\Pi^\pm := \tfrac{1}{2}(\unit \pm i\gamma_{123})$. They satisfy the canonical (anti-)commutation relations
\begin{align} \label{eqn:commutation-relations}
  \comm{(a_a)_{rs}}{(a^\dag_b)_{tu}}           & = \delta_{ab} \left( \delta_{st} \delta_{ru} - \tfrac{1}{N} \delta_{rs} \delta_{tu} \right) \; , \nonumber\\
  \comm{(b_i)_{rs}}{(b^\dag_j)_{tu}}           & = \delta_{ij} \left( \delta_{st} \delta_{ru} - \tfrac{1}{N} \delta_{rs} \delta_{tu} \right) \; , \nonumber\\
  \acomm{(c_\alpha)_{rs}}{(c^\dag_\beta)_{tu}} & = \tfrac{1}{2} \Pi^-_{\alpha\beta} \left( \delta_{st} \delta_{ru} - \tfrac{1}{N} \delta_{rs} \delta_{tu} \right) \; .
\end{align}
The free Hamiltonian now reads
\be
  H_0 = \tr \left[ \frac{M}{2} a_a^\dag a_a + M b_i^\dag b_i + \frac{3M}{2} c_\alpha^\dag c_\alpha \right] \; .
\ee
Physical states are constrained to be gauge invariant and are given by traces over words in the creation operators
\be
 \ket{\psi} = \tr ( a^\dag_{a_1} b^\dag_{i_2} c^\dag_{\alpha_3} \ldots ) \tr (\ldots) \ldots \ket{0} \; .
\ee
In the planar limit we concentrate on single trace states. The number of oscillators in the trace is called the length of a state. It is common to visualize a state 
of length $L$ as a ``spin-chain'' with $L$ sites, where every spin orientation corresponds to a particular oscillator. In the case of the full PWMT, the spins are 
elements of the 17-dimensional module $\mathrm{span}\{a^\dag_a,b^\dag_i,c^\dag_\alpha\}$ of $\su(4|2)$. In the following, however, we will be interested only in 
states that are built from fields of a certain $\su(2)$ subsector. In order to define this subsector it is convenient to relabel the $\so(6)$ fields according to
\be \label{eqn:so6-redefined}
  \Zd  = \tfrac{1}{\sqrt{2}} (a_1^\dag + i a_2^\dag) \; , \;
  \bZd = \tfrac{1}{\sqrt{2}} (a_1^\dag - i a_2^\dag) \; , \;
  \Z   = \tfrac{1}{\sqrt{2}} (a_1      - i a_2     ) \; , \;
  \bZ  = \tfrac{1}{\sqrt{2}} (a_1      + i a_2     ) \; .
\ee
and analogously for $\Wd = \frac{1}{\sqrt{2}} (a_3^\dag + i a_4^\dag)$ and $\Yd = \frac{1}{\sqrt{2}} (a_5^\dag + i a_6^\dag)$. The only non-vanishing commutators are
\be
  \comm{(\Z)_{rs}}{(\Zd)_{tu}} = \comm{(\bZ)_{rs}}{(\bZd)_{tu}} = \delta_{st} \delta_{ru} - \tfrac{1}{N} \delta_{rs} \delta_{tu}
\ee
and identical formulas for $\Wd$ and $\Yd$. The mentioned $\su(2)$ subsector is spanned by $\Zd$ and $\Wd$. A typical state is given by
\be
  \tr \bigl( \Zd\Zd\Wd\Zd\Zd\Zd\Wd\Zd \ldots \bigr) \ket{0} \; \widehat{=} \; | \uparrow\uparrow\downarrow\uparrow\uparrow\uparrow\downarrow\uparrow\ldots \rangle \; 
. \ee
It is customary to call $\Wd$ an ``impurity'' within the ``background fields'' $\Zd$. In the spin-chain language $\Wd$
refers to the excitation of a ``magnon''. Both fields can be distinguished by a $\U(1)_R$ charge $J$ associated to rotations in the 1-2-plane of the $\so(6)$ field 
space:
\begin{center}
\begin{tabular}{|c|c|c|c|c|c|} \hline
\vphantom{\rule[3mm]{1mm}{1mm}}
field & $\Zd$  & $\bZd$ & $\Z$   & $\bZ$  & others \\ \hline
$J$   & $+1$   & $-1$   & $-1$   & $+1$   & $0$ \\ \hline
\end{tabular}
\end{center}
The R-charge is additive and hence $J$ counts the number of $\Zd$'s in an $\su(2)$ state.

Later on we will need the parity conjugate of (single-trace) states, which is defined as the state with reversed order of fields multiplied by 
$(-1)^{\mathrm{Length}}$.

\subsection{Perturbation theory to fourth order}

The aim is to construct the effective Hamiltonian $T$ (sometimes referred to as energy operator) for the matrix model. The effective Hamiltonian is defined as an 
operator similar to the full Hamiltonian, $T = U^{-1} H U$, which does not mix states with different free energies. It is the matrix theory operator that corresponds 
to the dilatation operator of SYM.

The computation of the effective Hamiltonian $T$ has to be done in perturbation theory for large values of $M$. Note that the interaction terms $V_1$ and $V_2$ in 
\eqref{eqn:matrix-theory} are suppressed by powers of $1/M^2$ and $1/M^4$, respectively, as can be seen by rescaling the fields. In \cite{pwmt-integrability} $T$ was 
derived up to three-loop order. For convenience we repeat the formulas of the perturbation expansion of $T$, now including fourth order, in 
App.~\ref{sec:perturbation-theory}. We have intentionally refrained from choosing the transformation operator $U$ to be unitary. Though this yields a non-hermitian 
$T$, it has the essential advantage to produce far less terms that need to be computed. This is important because fourth order is very close to the borderline of the 
technically infeasible. To perform the actual computation we have developed the highly optimized computer program \gemstone{} which is described in detail in 
App.~\ref{sec:gemstone}.

While at third order it was still possible to find the \emph{non}-planar effective Hamiltonian \cite{pwmt-integrability}, at fourth order this is impossible with 
todays technology. And actually our interest just concerns the planar part of $T$. This is because the integrability is only a feature of the planar limit. Also the 
connection between PWMT and SYM at the non-planar level has already ceased to exist at three-loop.

Note, however, that the formulas for $T$ as given in \eqref{eqn:eff-ham} will produce any graph, in particular non-planar ones. It would be extremely nice to have a 
formalism that just builds planar graphs automatically. In cases where there are no pure creation and pure annihilation vertices, this can indeed be achieved by 
replacing the matrix oscillators by scalar oscillators and their algebra \eqref{eqn:commutation-relations} by the Cuntz algebra \cite{Halpern}. But the interaction 
terms of PWMT unfortunately do not meet this requirement.

Therefore the implementation of the perturbation expansion on the computer has to be supplemented by a special algorithm which guarantees the construction of and only 
of the planar graphs. Simple extensions of the Cuntz algebra approach suffer from an over counting or an omission of some planar graphs. Our algorithm (which 
introduces structures like ``graphs'', ``components of connection'', ``domains'' etc) is explained in App. \ref{sec:planar-graph-generation}.

There is another general problem of taking the planar limit. When just looking at an operator, as opposed to an operator applied to a state, it is in general not 
possible to extract its planar part. This is because a superficially non-planar looking operator can in fact produce a planar graph when it is applied to a 
sufficiently short state. This issue is known as wrapping problem
\cite{BeisertPHD}.

The resolution of this problem is to split the computation in pieces: a general computation under the assumption that the states are long enough to avoid wrapping, 
and then as many special cases as there are lengths which are not covered by the general part. At four-loop order wrapping can only occur for states of length less 
than or equal to four. These cases will be considered in Sec. \ref{sec:wrapping} below. For the time being let us assume that the effective Hamiltonian acts only onto 
states of length greater than four.

A convenient notation for the planar operators was introduced in \cite{bks}: Let $P_{k_1,k_2}$ exchange the oscillators at sites $k_1$ (mod $L$) and $k_2$ (mod $L$) 
of a state of length $L$. Then any planar operator that preserves the number of individual flavours -- such as the effective Hamiltonian in the $\su(2)$ subsector -- 
can be written as a linear combination of objects of the following kind:
\begin{align}
  \{n_1,n_2,\ldots\} & := \sum_{k=1}^L P_{k+n_1,k+n_1+1} P_{k+n_2,k+n_2+1} \cdots \nonumber \; , \\
  \{\}               & := L \; . \label{eqn:permutations}
\end{align}
There are many trivial relations for these operators such as
\begin{align}
  \{\ldots,n,n,\ldots\} & = \{\ldots,\ldots\}      \nonumber \\
  \{\ldots,n,m,\ldots\} & = \{\ldots,m,n,\ldots\}  \qquad\qquad \mbox{for $\modulus{n-m} \ge 2$} \\
  \{n_1,n_2,\ldots\}    & = \{n_1+m,n_2+m,\ldots\} \nonumber
\end{align}
as well as the identity
\be
\begin{split}
  & \{\ldots,\ldots\} + \{\ldots,n\pm1,n,\ldots\} + \{\ldots,n,n\pm1,\ldots\} \\
  & - \{\ldots,n,\ldots\} - \{\ldots,n\pm1,\ldots\} - \{\ldots,n,n\pm1,n,\ldots\} = 0
\end{split}
\ee
which only holds in the $\su(2)$ subsector and expresses the fact that two spins cannot be placed completely antisymmetric onto three sites.

Now, our strategy to determine the planar effective Hamiltonian is the following. We start from an ansatz for $T_8$ in terms of the permutation operators 
\eqref{eqn:permutations} with free coefficients and apply this ansatz to some particular states. The coefficients are then determined by applying the whole 
perturbation expansion \eqref{eqn:eff-ham-t8} to the same states using \gemstone{} and matching the results.

Specifically we use the following ansatz for the planar four-loop effective Hamiltonian
\be \label{eqn:ansatz4loop}
\begin{split}
T_8 & = c_{1} \{\}
      + c_{2} \{0\}
      + c_{3} \{0,1\} + c_{4} \{1,0\} \\
    & + c_{5} \{0,2\}
      + c_{6} \{0,1,2\} + c_{7} \{2,1,0\}
      + c_{8} \{0,2,1\} + c_{9} \{1,0,2\}
      + c_{10} \{1,0,2,1\} \\
    & + c_{11} \{0,3\}
      + c_{12} \{0,1,3\} + c_{13} \{0,3,2\}
      + c_{14} \{1,0,3\} + c_{15} \{0,2,3\} \\
    & + c_{16} \{0,1,2,3\} + c_{17} \{3,2,1,0\}
      + c_{18} \{0,2,1,3\} + c_{19} \{1,0,3,2\} \\
    & + c_{20} \{0,1,3,2\} + c_{21} \{0,3,2,1\}
      + c_{22} \{1,0,2,3\} + c_{23} \{2,1,0,3\} \\
    & + c_{24} \{0,2,1,3,2\} + c_{25} \{2,1,0,3,2\}
      + c_{26} \{1,0,2,1,3\} + c_{27} \{1,0,3,2,1\} \\
    & + c_{28} \{1,0,2,1,3,2\} + c_{29} \{2,1,0,3,2,1\} \; .
\end{split}
\ee
This ansatz is capable of describing any planar operator which acts locally as permutation on five adjacent fields. In fact it is even too general for the effective 
Hamiltonian which possesses some properties that could actually be used to reduce the number of free coefficients. However, we are not going to make use of these 
properties in order to have a consistency check for our computation. The price for this is just a marginally increased computation time.

Firstly we could discard all terms which consist of more than 4 elementary permutations. This is because any transposition corresponds to a four-point vertex in the 
language of Feynman diagrams, and at four-loop order there are at
most four of these vertices\footnote{This is the Feynman
diagrammatic structural constraint mentioned in the introduction,
which entered the analysis of \cite{bds-long-range}.}. Hence, one
could set \be \label{eqn:constraints-feynman}
  c_{24} = \ldots = c_{29} = 0
\ee
right from the beginning leaving just 23 coefficients. The second property is the conservation of parity, which implies that the effective Hamiltonian has to be 
invariant under the transformation
\be
  \{n_1,\ldots,n_r\} \rightarrow \{-n_1,\ldots,-n_r\} \; .
\ee
This would yield the following further relations
\be \label{eqn:constraints-parity}
 c_{3} = c_{4} \quad
 c_{6} = c_{7} \quad
 c_{8} = c_{9} \quad
 c_{12} = c_{13} \quad
 c_{14} = c_{15} \quad
 c_{18} = c_{19} \quad
 c_{20} = c_{21} \quad
 c_{22} = c_{23}
\ee
such that there are actually only 15 independent coefficients. On the other hand we cannot assume hermiticity of \eqref{eqn:ansatz4loop} since we explicitly destroy 
the hermiticity by using a non-unitary similarity transformation. But we just note that hermitian conjugation acts as
\be \label{eqn:hermitean-conjugation}
  \{n_1,\ldots,n_r\} \rightarrow \{n_r,\ldots,n_1\} \; .
\ee

Now we need to find an appropriate state, where the effective Hamiltonian should be applied to. The largest number $m$ of coefficients that are fixed by some state of 
length $L$ is given in the following table, together with a sample chain:
\begin{center}
\begin{tabular}{|c|c|l|}
\hline
$L$ & $m$ &  possible initial chain \\ \hline
 6  &   4 &  (ZZWZWW) \\
 7  &   5 &  (ZZWZZWW) \\
 8  &  10 &  (ZZZWZWWW) \\
 9  &  14 &  (ZZZZWZWWW) \\
10  &  24 &  (ZZZZWWZWWW) \\
11  &  29 &  (ZZZZZWWZWWW) \\ \hline
\end{tabular}
\end{center}
Hence a single initial state with eleven (or more) elementary fields is indeed sufficient to fix all 29 parameters of the ansatz \eqref{eqn:ansatz4loop} at once! A 
similar analysis for the constrained ansatz with 15 independent coefficients shows that one would have to use a state of length ten.

We chose the initial state
\be
\ket{\mathrm{in}} = \tr \bigl( \Zd\Zd\Zd\Zd\Zd\Wd\Wd\Zd\Wd\Wd\Wd \bigr) \ket{0}
\ee
for our computation and apply both the ansatz \eqref{eqn:ansatz4loop} and the whole sequence of vertices \eqref{eqn:eff-ham-t8}. The result of the latter calculation 
is given in \eqref{eqn:gemstone11} and fixes all coefficients uniquely to
\be
\begin{split} \label{eqn:eff-ham-result-t8}
  T_8 & = \frac{N^4}{M^{11}} \bigl[ \;
                    - 22719 \{\}
                    + 33143 \{0\}
                    - 5628 ( \{0,1\} + \{1,0\} )
                    - 1044 \{0,2\} \\[-1.5mm]
& \qquad\quad\;\;\; + 984 ( \{0,1,2\} + \{2,1,0\} )
                    + 384 \{0,2,1\}
                    - 416 \{1,0,2\}
                    - 28 \{1,0,2,1\} \\
& \qquad\quad\;\;\; - 32 \{0,3\}
                    + 60 ( \{0,1,3\} + \{0,3,2\} )
                    +  4 ( \{1,0,3\} + \{0,2,3\} ) \\
& \qquad\quad\;\;\; - 80 ( \{0,1,2,3\} + \{3,2,1,0\} )
                    + 24 ( \{0,2,1,3\} + \{1,0,3,2\} ) \\
& \qquad\quad\;\;\; - 32 ( \{0,1,3,2\} + \{0,3,2,1\} )
                    + 24 ( \{1,0,2,3\} + \{2,1,0,3\} ) \; \bigr] \; .
\end{split}
\ee
Note that the requirements \eqref{eqn:constraints-feynman} and \eqref{eqn:constraints-parity} are satisfied. Also note that the sum of all coefficients is zero. This 
implies that all states which are composed entirely of $\Zd$'s or entirely of $\Wd$'s or which contain only a single oscillator of the other kind have vanishing 
energy shift. This accounts for the fact that these states are
perturbatively protected \cite{kimplefka}.

We will now switch to SYM-language: By a simple rescaling
\be
  D := \frac{2}{M} T
\ee
the effective Hamiltonian turns into the dilatation operator $D$ and the eigenvalues are now called conformal dimensions $\Delta$ instead of energies. The dictionary 
is given in the following table:
\begin{center}
\begin{tabular}{|p{50mm}|p{24mm}p{26mm}|} \hline
 \multicolumn{1}{|c|}{PWMT}  & \multicolumn{2}{c|}{SYM}                  \\ \hline
 effective Hamiltonian $T$   & \multicolumn{2}{l|}{dilatation operator $D$}             \\
 energy $E$                  & \multicolumn{2}{l|}{conformal dimension $\Delta$}        \\
 free energy $E_0$           & \multicolumn{2}{l|}{bare conformal dimension $\Delta_0$} \\
 energy shift $\delta E$     & \multicolumn{2}{l|}{anomalous dimension $\delta\Delta$}  \\
 $E_0(a^\dag)=M/2$           & $\Delta_0(\phi) = 1$   & (scalars)        \\
 $E_0(b^\dag)=M$             & $\Delta_0(F)    = 2$   & (field strength) \\
 $E_0(c^\dag)=3M/4$          & $\Delta_0(\psi) = 3/2$ & (fermions)       \\ \hline
\end{tabular}
\end{center}
In addition to this we introduce a new notation for the effective coupling. The perturbation expansion of $D$ is organized in powers of
\be
  \frac{1}{M} T_{2k} \propto \left( \frac{N}{M^3} \right)^k
\ee
which we now denote by
\be
 \frac{\Lambda}{2} := \frac{N}{M^3} \; .
\ee
The connection to SYM at the classical level is then given by \cite{pwmt-from-sym}
\be \label{eqn:PWMT-SYM-relation}
 \Lambda = G^2 N \quad , \quad G = \frac{g_\ti{YM}}{4\pi} \; .
\ee

With these definitions we find the planar four-loop dilatation operator of PWMT including lower orders from \cite{pwmt-integrability} to be
\be
  D(\Lambda) = \sum_{k=0}^\infty \Lambda^k D_{2k}
\ee
\begin{align}
  D_0 & = \{\} \; , \nonumber \\
  D_2 & = 2\{\} - 2\{0\} \; , \nonumber \\
  D_4 & = -15\{\} + 19\{0\} - 2(\{0,1\} + \{1,0\}) \; , \nonumber \\
  D_6 & = 187\{\} - 259\{0\} + 38(\{0,1\} + \{1,0\}) \nonumber \\
      & \quad + 4\{0,2\} - 4(\{0,1,2\} + \{2,1,0\}) - 2(\{0,2,1\} - \{1,0,2\}) \; , \label{eqn:dilop} \\
  D_8 & = -\tfrac{22719}{8} \{\}
        + \tfrac{33143}{8} \{0\}
        - \tfrac{1407}{2} ( \{0,1\} + \{1,0\} ) - \tfrac{261}{2} \{0,2\} \nonumber \\
& \quad + 123 ( \{0,1,2\} + \{2,1,0\} )
        + 48 \{0,2,1\} - 52 \{1,0,2\}
        - \tfrac{7}{2} \{1,0,2,1\} \nonumber \\
& \quad - 4 \{0,3\}
        + \tfrac{15}{2} ( \{0,1,3\} + \{0,3,2\} )
        + \tfrac{1}{2} ( \{1,0,3\} + \{0,2,3\} )  \nonumber \\
& \quad - 10 ( \{0,1,2,3\} + \{3,2,1,0\} )
        +  3 ( \{0,2,1,3\} + \{1,0,3,2\} ) \nonumber \\
& \quad -  4 ( \{0,1,3,2\} + \{0,3,2,1\} )
        +  3 ( \{1,0,2,3\} + \{2,1,0,3\} ) \; . \nonumber
\end{align}

\subsection{Hermitization}

In App.~\ref{sec:perturbation-theory} we have set up the perturbation theory in such a way that the effective Hamiltonian is computed from the least number of terms. 
This, however, has led to a non-hermitian effective Hamiltonian. Therefore we apply to the result \eqref{eqn:dilop} another similarity transformation of the form
\be
  D'(\Lambda) = e^{-\Lambda^3 A_3} e^{-\Lambda^2 A_2} e^{-\Lambda A_1} D(\Lambda) e^{\Lambda A_1} e^{\Lambda^2 A_2} e^{\Lambda^3 A_3} \; .
\ee
We demand that $D'$ is hermitian, parity conserving, of maximal range five and made of at most four elementary permutations. This allows the following operators
\begin{align}
A_1 & = \tfrac{1}{2}(x_L + x_R + 1) \{0\} \nonumber \\
A_2 & = \bigl[ -\tfrac{31}{8} + \tfrac{19}{4}(x_L+x_R) + \tfrac{1}{2}(y_L+y_R) - \tfrac{1}{2}(z_L+z_R) \bigr] \{0\}
        + x_L \{0,1\} + x_R \{1,0\} \nonumber \\
A_3 & = w \{0\}
        + y_L \{0, 1 \} + y_R \{1, 0\}
        + \bigl[  \tfrac{3}{4} - x_L - x_R \bigr] \{0, 2\} \\
& \quad + \bigl[ -\tfrac{3}{4} + 2 x_L - \tfrac{1}{2}(z_L+z_R) \bigr] \{0, 1, 2\}
        + \bigl[ -\tfrac{3}{4} + 2 x_R - \tfrac{1}{2}(z_L+z_R) \bigr] \{2, 1, 0\} \nonumber \\
& \quad + z_L \{0,2,1\} + z_R \{1,0,2\} \nonumber
\end{align}
where $w,x_L,x_R,y_L,y_R,z_L,z_R \in \R$ are arbitrary. The result of this transformation is
\begin{align}
  D'_0 & = \{\} \; , \nonumber \\
  D'_2 & = 2\{\} - 2\{0\} \; , \nonumber \\
  D'_4 & = -15\{\} + 19\{0\} - 2(\{0,1\} + \{1,0\}) \; , \nonumber \\
  D'_6 & = 187\{\} - 259\{0\} + 38(\{0,1\} + \{1,0\}) + 4\{0,2\} - 4(\{0,1,2\} + \{2,1,0\}) \; , \label{eqn:dilop-hermitean} \\
  D'_8 & = -\tfrac{22719}{8} \{\}
         + \tfrac{33127+32\alpha}{8} \{0\}
         - \tfrac{1403+8\alpha}{2} ( \{0,1\} + \{1,0\} )
         - \tfrac{259+4\alpha}{2} \{0,2\} \nonumber \\
 & \quad + (122+2\alpha) ( \{0,1,2\} + \{2,1,0\} )
         - (3-2\alpha) ( \{0,2,1\} + \{1,0,2\} )
         - \tfrac{5+4\alpha}{2} \{1,0,2,1\} \nonumber \\
 & \quad - 4 \{0,3\}
         + 4 ( \{0,1,3\} + \{0,3,2\} + \{1,0,3\} + \{0,2,3\} ) \nonumber \\
 & \quad - 10 ( \{0,1,2,3\} + \{3,2,1,0\} )
         - (2+2\alpha) ( \{0,2,1,3\} + \{1,0,3,2\} ) \nonumber \\
 & \quad - \alpha ( \{0,1,3,2\} + \{0,3,2,1\} + \{1,0,2,3\} + \{2,1,0,3\} ) \nonumber
\end{align}
where the parameter $\alpha$ depends only on the following particular combination
\be
  \alpha = \tfrac{1}{2} \left(x_L + x_R\right)^2 + z_L - z_R \; .
\ee
It does not influence the conformal dimensions and may therefore be set to any value. This ambiguity accounts for the freedom of making a unitary change of basis in 
the space of states.

\subsection{Short states -- wrapping issues} \label{sec:wrapping}

We still need to consider the special cases which have been excluded in the previous computation. This concerns all states of length less than or equal to four. Out 
of the total number of 13 states of this kind only the following two may acquire an anomalous dimension, cf. discussion below Eq.~\eqref{eqn:eff-ham-result-t8}:
\begin{align}
  \ket{ZWZW} & = \tr \Zd \Wd \Zd \Wd \ket{0} \; , \\
  \ket{ZZWW} & = \tr \Zd \Zd \Wd \Wd \ket{0} \; .
\end{align}
We apply the whole perturbation expansion \eqref{eqn:eff-ham-t8} onto these states. The raw results are given in \eqref{eqn:gemstoneZZWW}, \eqref{eqn:gemstoneZZWW}. 
Converted to the language of the dilatation operator we find
\begin{align} \label{eqn:dilop-onto-Konishi}
  D_8 \ket{ZWZW} & = - 11028 \ket{ZWZW} + 11028 \ket{ZZWW} \; , \\
  D_8 \ket{ZZWW} & = +  5514 \ket{ZWZW} -  5514 \ket{ZZWW} \; .
\end{align}
Let us now see whether this result could follow from the general
expression for the effective Hamiltonian $T_8$ by any natural
description of how to interpret the action of five spin
interaction terms on a state of length four: One could either
declare the interactions to wrap around the state, i.e. identify site
5 with site 1. This yields the mixing matrix
\be
\frac{1}{4} \left(
\begin{matrix}
 -43486 & +43486 \\
 +21743 & -21743
\end{matrix}
\right)
\ee
in disagreement with
Eq.~\eqref{eqn:dilop-onto-Konishi}. Alternatively one could simply
drop the five spin interaction terms, leading to
\be \frac{1}{4}
\left(
\begin{matrix}
 -43646 & +43710 \\
 +21855 & -21791
\end{matrix}
\right)
\ee
again in disagreement with
Eq.~\eqref{eqn:dilop-onto-Konishi}. We hence conclude that there
appears to be no ``natural'' extension of the long-range form of
$D_8$ to short states.

This completes the computation of the PWMT-dilatation operator and
we will investigate its properties in the next section.


\section{Properties of the PWMT-dilatation operator}

\noindent
Having obtained the PWMT-dilatation operator, an immediate question is whether it is equivalent to the four-loop proposal of the SYM-dilatation operator given by 
Beisert, Dippel and Staudacher \cite{bds-long-range} or alternatively to the spin-chain Hamiltonian of Inozemtsev \cite{inozemtsev,ss-ino}. In order to make contact 
to these models we consider a renormalization of the coupling constant $\Lambda$ in order to (naively) reach a thermodynamic scaling property known to exist in the 
dual string theory (``BMN scaling''). Rewritten in terms of commuting permutations we see, however, that the PWMT-dilatation operator is inequivalent to both of these 
models. Since the Beisert-Dippel-Staudacher Hamiltonian is the unique $\SU(2)$ Hamiltonian which is integrable and obeys BMN scaling up to four loops 
\cite{bds-long-range} (the Inozemtsev Hamiltonian violates BMN scaling \cite{ss-ino}), the PWMT-dilatation operator must either break integrability or BMN scaling or 
both. As an indication of integrability we display the
degeneracies in the four-loop spectrum and present higher
commuting charges. A proof of the breakdown of BMN scaling and
deeper insights into the integrable structure of the model will be
given in the next section by means of the underlying perturbative
asymptotic Bethe ansatz.

\subsection{Renormalization and naive BMN scaling}

Let us consider states carrying two excitations $\Wd$ among $J$ excitations $\Zd$, i.e.\ a spin-chain of length $L=J+2$ with two magnons:
\be
\label{eqn:2magnonstates}
 \Op{l} := \tr \Wd (\Zd)^{l-1} \Wd (\Zd)^{L-1-l} \ket{0} = \Op{L-l} \qquad \mbox{for $1 \le l \le L-1$} \, .
\ee
The plane-wave string/gauge theory correspondence predicts a scaling dimension of these states in the full four-dimensional ${\cal N}=4$ super Yang-Mills model of the 
all-loop form (with the $\U(1)_R$ charge $J=L-2$)
\be \label{eqn:BMN-formula}
 \Delta^{\mathrm{BMN}}(n,J) = J + 2 \sqrt{1 + \Lambda \frac{16 \pi^2 n^2}{J^2}} \qquad \mbox{with $n \in \N$}
\ee
in the limit $J \to \infty$. This expression displays BMN scaling behavior, i.e.~in the limit $J,N\to \infty$, with $N/J^2$ held fixed, the effective loop counting 
parameter $\Lambda':=\Lambda/J^2$ arises. So let us check whether this scaling behavior is realized in PWMT by simply applying the dilatation operator 
\eqref{eqn:dilop-hermitean} to the set of the above two-magnon states. For the time being we consider only those states where the magnons are further apart than the 
range of the dilatation operator, i.e. the dilatation operator does not act on both magnons at the same time. Technically speaking we restrict the action of $D_{2k}$ 
to states $\Op{l}$ with $k < l < L-k$ and find
\begin{align}
D'_2 \Op{l} & = + 8 \Op{l} - 4 \Op{l\pm1} \; , \nonumber \\
D'_4 \Op{l} & = - 52 \Op{l} + 30 \Op{l\pm1} - 4 \Op{l\pm2} \; , \label{eqn:gendiffeqs} \\
D'_6 \Op{l} & = + 612 \Op{l} - 374 \Op{l\pm1} + 76 \Op{l\pm2} - 8 \Op{l\pm3} \; , \nonumber \\
D'_8 \Op{l} & = - \tfrac{17923}{2} \Op{l} + \tfrac{22615}{4} \Op{l\pm1} - 1397 \Op{l\pm2} + 244 \Op{l\pm3} - 20 \Op{l\pm4} \; , \nonumber
\end{align}
where we used the shorthand $\Op{l\pm m} := \Op{l-m} + \Op{l+m}$. The observation now is that this may be written as a sum of even powers of a discretized second 
derivative $\partial_l^2$ which acts as $\partial_l^2 \Op{l} = -  2 \Op{l} + \Op{l\pm1}$. One finds
\begin{align}
 D'_2 & = -  \mathbf{4} \partial_l^2 \; , \nonumber \\
 D'_4 & = -  \mathbf{4} \partial_l^4 + 14 \partial_l^2 \; , \\
 D'_6 & = -  \mathbf{8} \partial_l^6 + 28 \partial_l^4 - 142 \partial_l^2 \; , \nonumber \\
 D'_8 & = - \mathbf{20} \partial_l^8 + 84 \partial_l^6 - 333 \partial_l^4 + \tfrac{7767}{4} \partial_l^2 \; . \nonumber
\end{align}
Now it is easy to determine the spectrum (neglecting the boundary effects of the action of $D'$ involving two magnons at the same time). On the Fourier transformed 
basis
\be
 \ket{n} := \sum_{l=1}^{L-1} e^{2\pi i n (l-1)/J} \Op{l} \qquad (J=L-2)
\ee
the dilatation operator becomes diagonal since
\be\label{eqn:latplace}
 \partial_l^2 \ket{n} = \left( - 2 + e^{2\pi in/J} + e^{-2\pi in/J} \right) \ket{n}
                      = -4 \sin^2 \left( \frac{\pi n}{J} \right) \ket{n}
                      \xrightarrow{J\to\infty}  - \frac{4 \pi^2 n^2}{J^2} \ket{n} \; .
\ee
On the other hand, if one expands the square root formula \eqref{eqn:BMN-formula} in a series in $\Lambda$
\be \label{BMN-formula-expanded}
\begin{split}
 \Delta^{\mathrm{BMN}}(n,J) = J+2 & -  \mathbf{4} \Lambda   \left(\frac{-4 \pi^2 n^2}{J^2}\right)
                                    -  \mathbf{4} \Lambda^2 \left(\frac{-4 \pi^2 n^2}{J^2}\right)^2 \\
                                  & -  \mathbf{8} \Lambda^3 \left(\frac{-4 \pi^2 n^2}{J^2}\right)^3
                                    - \mathbf{20} \Lambda^4 \left(\frac{-4 \pi^2 n^2}{J^2}\right)^4
                                    + \ldots
\end{split}
\ee
one observes, that these values are exactly produced by the highest derivatives in $D'$. The lower order derivatives, however, spoil the BMN scaling behavior 
\eqref{eqn:BMN-formula}. But this can be restored by a redefinition of the coupling constant. If we set
\be \label{eqn:renormalization-coupling}
  \Lambda = \Lambda_r + \frac{7}{2} \Lambda_r^2 - 11 \Lambda_r^3 + \frac{1257}{16} \Lambda_r^4 + \order{\Lambda_r^5}
\ee
then the dilatation operator expanded in $\Lambda_r$ is simply given by clean lattice laplacians of ascending order
\be\label{eqn:D'}
 D'_{r,2} = -  4 \partial_l^2 \quad,\quad
 D'_{r,4} = -  4 \partial_l^4 \quad,\quad
 D'_{r,6} = -  8 \partial_l^6 \quad,\quad
 D'_{r,8} = - 20 \partial_l^8
\ee
and hence in accordance with BMN scaling.

We wish to stress, that the precise series of coefficients $4,4,8,20$ of the plane-wave string formula \eqref{eqn:BMN-formula},\eqref{BMN-formula-expanded} is an 
inherent property of PWMT and cannot be altered by any perturbative redefinition of $\Lambda$. Only the derivatives of non-highest degree could be eliminated, and 
demanding their absence as a necessary prerequisite for BMN scaling uniquely fixes the renormalization \eqref{eqn:renormalization-coupling}. Any differently 
renormalized coupling constant would prevent BMN scaling immediately.

Hence, from now on we will work with the dilatation operator obtained from \eqref{eqn:dilop-hermitean} by redefining $\Lambda$ according to the naive BMN scaling 
prescription \eqref{eqn:renormalization-coupling}:
\begin{align}
  D'_{r,0} & = \{\} \; , \nonumber \\
  D'_{r,2} & = 2\{\} - 2\{0\} \; , \nonumber \\
  D'_{r,4} & = -8\{\} + 12\{0\} - 2(\{0,1\} + \{1,0\}) \; , \nonumber \\
  D'_{r,6} & = 60\{\} - 104\{0\} + 24(\{0,1\} + \{1,0\}) + 4\{0,2\} - 4(\{0,1,2\} + \{2,1,0\}) \; , \label{eqn:dilop-renormalized} \\
  D'_{r,8} & = - 573 \{\}
               + (1079+4\alpha) \{0\}
               - (283+4\alpha) ( \{0,1\} + \{1,0\} )
               - \tfrac{175+4\alpha}{2} \{0,2\} \nonumber \\
 & \quad       + (80+2\alpha) ( \{0,1,2\} + \{2,1,0\} )
               - (3-2\alpha) ( \{0,2,1\} + \{1,0,2\} )
               - \tfrac{5+4\alpha}{2} \{1,0,2,1\} \nonumber \\
 & \quad       - 4 \{0,3\}
               + 4 ( \{0,1,3\} + \{0,3,2\} + \{1,0,3\} + \{0,2,3\} ) \nonumber \\
 & \quad       - 10 ( \{0,1,2,3\} + \{3,2,1,0\} )
               + (2+2\alpha) ( \{0,2,1,3\} + \{1,0,3,2\} ) \nonumber \\
 & \quad       - \alpha ( \{0,1,3,2\} + \{0,3,2,1\} + \{1,0,2,3\} + \{2,1,0,3\} ) \nonumber
\end{align}

One has to keep in mind, that these considerations only naively confirm the BMN scaling behavior of the renormalized dilatation operator, as the neglected boundary 
terms could (and actually do) ruin this scaling. Clearly the contact interactions of two magnons will be suppressed by a factor of $1/J$ with respect to the generic 
terms, however, such a suppression does not suffice to implement BMN scaling at higher orders in perturbation theory. This will be seen in 
Sec.~\ref{sec:BMNviolation}.

\subsection{Distinction to Inozemtsev and Beisert-Dippel-Staudacher spin-chains}

The comparison is most easily done in the language of commuting
permutations, which was used in \cite{ss-ino}. Our PWMT-dilatation
operator then reads \begin{align} \label{eq:HamP}
D'_{r,2} =\ &  2 L - 2 \sum_{i} P_{i,i+1} \; , \nonumber\\
D'_{r,4} =\ & -6 L + 8 \sum_{i} P_{i,i+1} -2 \sum_{i} P_{i,i+2} \; , \nonumber\\
D'_{r,6} =\ & 40 L -56 \sum_{i} P_{i,i+1} +16\sum_{i} P_{i,i+2} + 4\sum_i(P_{i,i+2} P_{i+1,i+3} - P_{i,i+3} P_{i+1,i+2}) \; , \nonumber\\
D'_{r,8} =\ & -363 L + 522 \sum_i P_{i,i+1} - \frac{415}{3} \sum_i P_{i,i+2} -\frac{41}{3}\sum_i P_{i,i+3} + \mathbf{6}\sum_i P_{i,i+4} \nonumber\\
            & - \frac{71}{6}\sum_i P_{i,i+1} P_{i+2,i+3} +\frac{217}{3} \sum_i P_{i,i+3} P_{i+1,i+2} - \frac{147}{2} \sum_i P_{i,i+2} P_{i+1,i+3} \nonumber\\
            & + \mathbf{\frac{16}{3}} \sum_i ( P_{i,i+3} P_{i+2,i+4} + P_{i,i+2} P_{i+1,i+4} + P_{i,i+3} P_{i+1,i+4} \nonumber\\
            & \qquad \quad -P_{i,i+4} P_{i+2,i+3} - P_{i,i+4} P_{i+1,i+2} - P_{i,i+4} P_{i+1,i+3}) \; .
\end{align}
For the four-loop contribution we have committed ourselves to the choice of $\alpha=-2/3$ for the free parameter of unitary transformations in order to obtain the 
same coefficient for all terms in the last two lines of \eqref{eq:HamP}. Comparing to the results obtained in \cite{ss-ino} it is clear that we are not dealing with 
the Inozemtsev spin-chain here, as that would predict the five spin interaction terms with different coefficients, namely
\be
\begin{split}
D_{8,\ti{Inozemtsev}} =\ & \mathbf{4} \sum_i P_{i,i+4}
                           + \mathbf{4} \sum_i ( P_{i,i+3} P_{i+2,i+4} + P_{i,i+2} P_{i+1,i+4} + P_{i,i+3} P_{i+1,i+4} \\
                         & - P_{i,i+4} P_{i+2,i+3} - P_{i,i+4} P_{i+1,i+2} -P_{i,i+4} P_{i+1,i+3}) + \ldots \; .
\end{split}
\ee
Curiously our five spin interaction terms do agree with the proposed four-loop dilation operator of $\Ncal=4$ super Yang-Mills theory \cite{bds-long-range}. However, 
they differ in the two, three and four spin interaction terms. The precise discrepancy to the Beisert-Dippel-Staudacher dilatation operator $D^{\rm {BDS}}_8$ is
\be
\begin{split}
  \label{eqn:D8diff}
  D'_{r,8}-D^{\mathrm{BDS}}_8 =\ & 585 \bigl[ -L + 2\sum_i P_{i,i+1} - \sum_i P_{i,i+2} + \sum_i P_{i,i+3} \\
                              & \quad -\frac 1 2 \sum_i P_{i,i+1} P_{i+2,i+3} - \sum_i P_{i,i+3} P_{i+1,i+2} + \frac{1}{2} \sum_i P_{i,i+2} P_{i+1,i+3} \bigr] \; .
\end{split}
\ee
Thus it is clear that we are facing a model, which is \emph{neither} of the Inozemtsev \emph{nor} of the ``novel'' Beisert-Dippel-Staudacher type. Nevertheless, the 
PWMT-dilatation operator describes a perturbatively integrable long range spin-chain as we will indicate in the next subsection and confirm afterwards by deriving 
appropriate Bethe equations.\footnote{Looking through the literature, we find an old proposal for the SYM-dilatation operator in Ref.~\cite{bks} which \emph{is} 
equivalent to the PWMT one. Formula (F.3) of \cite{bks} with the replacement $\alpha \to -1/4$ and $\beta \to \alpha + 1$ matches exactly our above finding 
\eqref{eqn:dilop-renormalized}. In \cite{bks} integrability was implemented but the implications of BMN scaling were not fully exploited yet. From this result, one 
could principally already infer the integrability of PWMT and the failure of BMN scaling.}

\subsection{Degenerate spectrum and higher charges}

In Tab.~\ref{tab:spectrum} we list the (renormalized) four-loop anomalous dimensions $\delta\Delta_r$ of all $\su(2)$-multiplets with states of length $L\le9$. Apart 
form the length which is equal to the bare dimension $\Delta_0$, multiplets are labeled by the magnon number $M$. Alternatively one could use the $\su(2)$ Dynkin 
label $a=L-2M$. Every multiplet, which is specified by the pair $(L,M)$, is realized with a certain multiplicity $m$. All of these multiplets can mix among each 
other. Since the dilatation operator commutes with the parity operator, it is possible to disentangle the multiplets such that all states within one multiplet have 
equal parity $p$. This has been done in Tab.~\ref{tab:spectrum} and is indicated by $m^p$.
\newcommand{\confdims}[4]{#1 #2\Lambda_r #3 \Lambda_r^2 #4 \Lambda_r^3}
\begin{table}
\begin{center}
\begin{tabular}{|c|c|c|l|} \hline
$L $ & $M$ & $m^p$ & $\delta \Delta_r / \Lambda_r$ \\ \hline
$2$  & $0$ & $1^+$    & $0$ \\ \hline
$3$  & $0$ & $1^-$    & $0$ \\ \hline
$4$  & $0$ & $1^+$    & $0$ \\ \cline{2-4}
     & $2$ & $1^+$    & $\confdims{12}{-48}{+336}{-12771/4}$ \\ \hline
$5$  & $0$ & $1^-$    & $0$ \\ \cline{2-4}
     & $2$ & $1^-$    & $\confdims{8}{-24}{+136}{-1024}$ \\ \hline
$6$  & $0$ & $1^+$    & $0$ \\ \cline{2-4}
     & $2$ & $2^+$    & $\confdims{(10-\sqrt{5})}{-(34-\sqrt{5})}{+\tfrac{1}{5}(1170-414\sqrt{5})}{-\tfrac{1}{5}(10695-4134\sqrt{5})}$ \\
     &     &          & $\confdims{(10+\sqrt{5})}{-(34+\sqrt{5})}{+\tfrac{1}{5}(1170+414\sqrt{5})}{-\tfrac{1}{5}(10695+4134\sqrt{5})}$ \\ \cline{2-4}
     & $3$ & $1^-$    & $\confdims{12}{-36}{+252}{-2484}$ \\ \hline
$7$  & $0$ & $1^-$    & $0$ \\ \cline{2-4}
     & $2$ & $2^-$    & $\confdims{4}{-6}{+37/2}{-335/4}$ \\
     &     &          & $\confdims{12}{-42}{+555/2}{-9465/4}$ \\ \cline{2-4}
     & $3$ & $1^+$    & ${\bf \confdims{10}{-30}{+200}{-3565/2}}$ \\ \cline{3-4}
     &     & $1^-$    & ${\bf \confdims{10}{-30}{+200}{-3565/2}}$ \\ \hline
$8$  & $0$ & $1^+$    & $0$ \\ \cline{2-4}
     & $2$ & $3^+$    & $\confdims{3.01}{-3.32}{+7.66}{-27.30}$ \\
     &     &          & $\confdims{9.78}{-29.22}{+167.64}{-1256.15}$ \\
     &     &          & $\confdims{15.21}{-59.45}{+456.70}{-4430.55}$ \\ \cline{2-4}
     & $3$ & $1^+$    & ${\bf \confdims{8}{-20}{+112}{-842}}$ \\ \cline{3-4}
     &     & $2^-$    & ${\bf \confdims{8}{-20}{+112}{-842}}$ \\
     &     &          & $\confdims{12}{-36}{+264}{-2592}$ \\ \cline{2-4}
     & $4$ & $3^+$    & $\confdims{6.49}{-7.56}{+10.22}{+10.25}$ \\
     &     &          & $\confdims{10.90}{-31.76}{+249.76}{-2538.27}$ \\
     &     &          & $\confdims{22.60}{-88.68}{+636.03}{-5933.98}$ \\ \hline
$9$  & $0$ & $1^-$    & $0$ \\ \cline{2-4}
     & $2$ & $3^-$    & $\confdims{(8-4\sqrt{2})}{-(26-17\sqrt{2})}{+\tfrac{1}{8}(179-993\sqrt{2})}{-\tfrac{1}{32}(51376-36103\sqrt{2})}$ \\
     &     &          & $\confdims{8}{-20}{+98}{-639}$ \\
     &     &          & $\confdims{(8+4\sqrt{2})}{-(26+17\sqrt{2})}{+\tfrac{1}{8}(179+993\sqrt{2})}{-\tfrac{1}{32}(51376+36103\sqrt{2})}$ \\ \cline{2-4}
     & $3$ & $3^+$    & ${\bf \confdims{6.45}{-13.18}{+60.73}{-378.34}}$ \\
     &     &          & ${\bf \confdims{11.04}{-33.50}{+231.93}{-2149.91}}$ \\
     &     &          & ${\bf \confdims{16.51}{-55.32}{+383.33}{-3494.75}}$ \\ \cline{3-4}
     &     & $3^-$    & ${\bf \confdims{6.45}{-13.18}{+60.73}{-378.34}}$ \\
     &     &          & ${\bf \confdims{11.04}{-33.50}{+231.93}{-2149.91}}$ \\
     &     &          & ${\bf \confdims{16.51}{-55.32}{+383.33}{-3494.75}}$ \\ \cline{2-4}
     & $4$ & $1^+$    & ${\bf \confdims{10}{-30}{+220}{-4185/2}}$ \\ \cline{3-4}
     &     & $3^-$    & $\confdims{(12-4\sqrt{3})}{-18(2-\sqrt{3})}{+\tfrac{1}{2}(456-255\sqrt{3})}{-\tfrac{1}{4}(7908-4541\sqrt{3})}$ \\
     &     &          & ${\bf \confdims{10}{-30}{+220}{-4185/2}}$ \\
     &     &          & $\confdims{(12+4\sqrt{3})}{-18(2+\sqrt{3})}{+\tfrac{1}{2}(456+255\sqrt{3})}{-\tfrac{1}{4}(7908+4541\sqrt{3})}$ \\ \hline
\end{tabular}
\caption{Particle spectrum in $\su(2)$ subsector}
\label{tab:spectrum}
\end{center}
\end{table}

The crucial observation is that whenever there are multiplets of positive and negative parity with equal $(L,M)$-labels, then they form pairs of degenerate conformal 
dimension, which have been highlighted in the table through a bold font. This degeneracy is ascribed to a conserved charge and was the original indication of 
integrability \cite{bks}. In fact integrability implies the existence of a set $\{Q_i\}$ of as many conserved charges as there are degrees of freedom:
\be
  \comm{D'_r}{Q_i} = 0 \quad,\quad \comm{Q_i}{Q_j} = 0 \qquad  \forall i,j\in\{3,4,5,\ldots\} \; .
\ee
Here we have adopted the usual enumeration of the so-called higher charges starting from $i=3$, cf.~\ref{sec:spin-chain-language}. All charges are Hermite an. Charges 
with even (odd) index (anti-)commute with the parity operator. Therefore the odd-indexed charges are responsible for the degeneracy of the parity pairs.

The charges have a perturbative expansion
\be \label{eqn:higher-charges}
  Q_i(\Lambda_r) = \sum_{k=1}^\infty \Lambda_r^{k-1} Q_{i,2k}
\ee
and can be written in terms of permutations \eqref{eqn:permutations}. The higher the charge the wider its range: $Q_{i,2k}$ acts simultaneous on $i+k-1$ adjacent 
spins. In App.~\ref{sec:higher-charges} we give $Q_3$ and $Q_4$ up to $k=4$. \\

Let us go back to Tab.~\ref{tab:spectrum} and look at the state with labels $(L,M)=(4,2)$. It is given by $\tr \comm{\Zd}{\Wd}^2 \ket{0}$ and is in fact a 
$\su(2,2|4)$-descendant of the Konishi operator $\tr a^\dag_a a^\dag_a \ket{0} = \tr[\Zd\bZd + \Wd\bWd + \Yd\bYd] \ket{0}$. The anomalous dimension of Konishi is 
known up to three-loop \cite{bks,eden-BMN,moch,qcdpeople}.  As we have done the explicit computation \eqref{eqn:dilop-onto-Konishi}, we dare to predict the four-loop 
value under the assumption of a valid PWMT-SYM-relationship
\be
\begin{split}
 \delta\Delta_r & = 12 \Lambda_r
                  - 48 \Lambda_r^2
                  + 336 \Lambda_r^3
                  - \tfrac{12771}{4} \Lambda_r^4 \\
                & = \frac{3 \lambda_\ti{YM}}{4 \pi^2}
                  - \frac{3 \lambda_\ti{YM}^2}{16 \pi^4}
                  + \frac{21 \lambda_\ti{YM}^3}{256 \pi^6}
                  - \frac{12771 \lambda_\ti{YM}^4}{262144 \pi^8} \; ,
\end{split}
\ee where we used \eqref{eqn:PWMT-SYM-relation} for the
renormalized coupling. It would be extremely nice to test this
conjecture through an explicit four-loop computation in the full
superconformal $\Ncal = 4$ gauge field theory.


\section{Bethe ansatz for plane-wave matrix theory}

\noindent
In this section we shall deduce the long-range Bethe ansatz allowing for an exact diagonalization of the obtained PWMT-Hamiltonian in the planar limit. We use a 
technique which was very recently developed in \cite{s-matrix} and called ``perturbative asymptotic Bethe ansatz''. At first we derive the Bethe equations in the 
two-magnon sector. These are straightforwardly generalized to the full $N$-magnon sector and verified for a number of three and four-magnon states
through numeric diagonalization. We then use the Bethe ansatz to solve the complete two-magnon problem, which explicitly reveals terms that violate BMN scaling.

\subsection{Spin-chain language} \label{sec:spin-chain-language}

For the following discussion we switch the interpretation of the effective PWMT-Hamiltonian $T$, which intermediately became the PWMT-dilatation operator $D'_r$, a 
second time. Now, the anomalous piece of the dilatation operator $\delta D'_r$ is regarded as the Hamiltonian $Q_2$ of an $\su(2)$ spin-chain
\be \label{eqn:spin-chain-hamiltonian}
  Q_2 := \Lambda_r^{-1} \delta D_r = \Lambda_r^{-1} ( D_r - D_{r,0} ) \; .
\ee
The spin-chain is in accordance with our previous manner of speaking still given by the single trace $\su(2)$ states of PWMT. With notation 
\eqref{eqn:spin-chain-hamiltonian} the Hamiltonian joins nicely the set of higher charges $\{Q_{i\ge3}\}$. The label $Q_1$ is reserved for the spin-chain momentum 
operator, which in our case is identically to zero due to the cyclicity of the trace. We denote the eigenvalues of $Q_i$ by corresponding small letters $q_i$. The 
following table lists equivalent quantities:
\begin{center}
\begin{tabular}{|l|l|l|} \hline
\multicolumn{1}{|c|}{PWMT}       & \multicolumn{1}{c|}{SYM}                     & \multicolumn{1}{c|}{Spin-chain} \\ \hline
energy shift operator $\delta T$ & anomalous dilatation operator $\delta D'_r$  & Hamiltonian $Q_2$ \\
energy shift $\delta E$          & anomalous dimension $\delta\Delta_r$         & energy $q_2$ \\
free energy $E_0$                & bare conformal dimension $\Delta_0$          & spin-chain length $L$ \\ \hline
\end{tabular}
\end{center}

\subsection{Two-magnon scattering}

The application of the dilatation operator to a spin-chain causes motion and interaction of the magnons. Integrability implies that only two magnons interact at a 
time. Therefore it is possible to deduce the general case from studying the two-magnon situation. However, one must not use the cyclic two-magnon states 
\eqref{eqn:2magnonstates} \cite{s-matrix} where only the distance of the magnons matters (or alternative their relative momentum). Once we place more magnons onto the 
spin-chain the position (or momentum) relative to the new insertions becomes important. Thus we need to investigate \emph{open} two-magnon states, the building blocks 
for a general state:
\be \label{eqn:2magnonblock}
 \Op{l_1,l_2} := (\Zd)^{l_1-1} \Wd (\Zd)^{l_2-l_1-1} \Wd (\Zd)^{L-l_2} \ket{0}   \qquad \mbox{for $1 \le l_1 < l_2 \le L-1$} \; .
\ee
The dilatation operator mixes all of these $L(L-1)/2$ states. We superpose them to eigenstates parametrized by two yet undetermined complex variables $p_1$ and $p_2$
\be \label{eqn:asymptBA}
\ket{p_1,p_2} := \sum_{\textstyle \atopfrac{l_1,l_2=1}{l_1<l_2}}^L a(l_1,l_2,p_1,p_2) \Op{l_1,l_2} \; .
\ee
This may be regarded as a kind of Fourier transformation, the magnons are now labeled by their (quasi-)momenta $p_1$ and $p_2$. The Bethe ansatz for 
$a(l_1,l_2,p_1,p_2)$ resembles an in-coming and a scattered out-going wave\footnote{Notice the reversed order of the arguments in $S$. It will turn out that 
$S(p_1,p_2)=[S(p_2,p_1)]^{-1}$.
Also note that the detailed form of our ansatz
differs slightly from the one considered in \cite{s-matrix}.}
\be \label{eqn:fouriercoeffs}
  a(l_1,l_2,p_1,p_2) = e^{i(p_1 l_1 + p_2 l_2)} f(l_2-l_1,p_1,p_2) + S(p_2,p_1) e^{i(p_1 l_2 + p_2 l_1)} f(L-l_2+l_1,p_1,p_2) \; .
\ee
This ansatz will exactly diagonalize the spin-chain Hamiltonian
\be \label{eqn:eigenvalue-equation}
  Q_2 \ket{p_1,p_2} = q_2(p_1,p_2) \ket{p_1,p_2} \; .
\ee
The key ingredient of the ansatz is the S-matrix $S(p_1,p_2)$, which describes the scattering of two magnons and solely determines the eigenvalue of $\ket{p_1,p_2}$. 
In the following we will determine the S-matrix perturbatively up to order $\Lambda_r^3$
\be \label{eqn:Smatrix}
  S(p_1,p_2) = S_0(p_1,p_2) + \Lambda_r S_1(p_1,p_2) + \Lambda_r^2 S_2(p_1,p_2) + \Lambda_r^3 S_3(p_1,p_2) + \ldots \; .
\ee
The function $f(l,p_1,p_2)$ takes the special perturbative
form
\be \label{eqn:fdef}
  f(l,p_1,p_2) = 1 + \Lambda_r^{l} f_0(l,p_1,p_2) + \Lambda_r^{l+1} f_1(l,p_1,p_2) + \Lambda_r^{l+2} f_2(l,p_1,p_2) + \ldots
\ee
and has been introduced to account for boundary effects where the magnons come close to each other. The fact that the range of the spin-chain Hamiltonian depends on 
the loop order, explains why the order of the leading correction in \eqref{eqn:fdef} depends on the magnon separation $l=l_2-l_1$. It turns out that the precise form 
of the function $f(l,p_1,p_2)$ is irrelevant for the energy eigenvalue $q_2(p_1,p_2)$.

There is another ``boundary effect'' where the magnons approach either end of the open spin-chain \eqref{eqn:2magnonblock}, i.e.\ where $l_1 \approx 1$ or $l_2 
\approx L$. Since the spin-chain Hamiltonian \eqref{eqn:spin-chain-hamiltonian},\eqref{eqn:dilop-renormalized} is not well-defined for open chains we will neglect 
these cases and assume $l_1$ to be large and $l_2$ to be small enough. Moreover we may replace the function $f(L-l_2+l_1,p_1,p_2)$ in the scattered piece of the wave 
function by $1$, if we also assume $L$ to be large enough, because the perturbative corrections will only contribute at order $\Lambda^{L-l_2+l_1}$. The reason for 
introducing $f$ at this point lies in the derivation of the Bethe equations from the periodicity conditions, cf.~\eqref{eqn:periodicity-condition}.

It will turn out that despite these assumptions the wave-function and the eigenvalues are exactly (and correctly) determined and hence the S-matrix, the Bethe 
equations, and the energy formula as well. Since this is obviously not a rigorous derivation we shall check the result for a number of cases by explicit numerical 
diagonalization for low values of $L$.

Now, in order to determine the unknown functions of the ansatz \eqref{eqn:fouriercoeffs} from demanding the eigenvalue equation \eqref{eqn:eigenvalue-equation}, we 
apply the spin-chain Hamiltonian to the states \eqref{eqn:asymptBA}. We do this order by order in $\Lambda_r$. The action of the Hamiltonian on $\Op{l_1,l_2}$ is given 
by\footnote{We use similar shorthands as before, e.g.\ $\Op{l_1\pm m,l_2} := \Op{l_1-m,l_2} + \Op{l_1+m,l_2}$.}
\begin{align}
  Q_{2,2} \Op{l_1,l_2} & =
  \begin{cases}
    4 \Op{l_1,l_2}
   -2 \Op{l_1-1,l_2}
   -2 \Op{l_1,l_2+1}
  & \mbox{for $l_2 = l_1 + 1$} \\[3mm]
    8 \Op{l_1,l_2}
   -2 \Op{l_1\pm1,l_2}
   -2 \Op{l_1,l_2\pm1}
  & \mbox{for $l_2 \ge l_1 + 2$}
  \end{cases} \\[5mm]
  Q_{2,4} \Op{l_1,l_2} & =
  \begin{cases}
    -8 \Op{l_1,l_2}
    +8 \Op{l_1-1,l_2}
    +8 \Op{l_1,l_2+1} \\[1mm]
    -2 \Op{l_1-1,l_2-1}
    -2 \Op{l_1+1,l_2+1}
    -2 \Op{l_1-2,l_2}
    -2 \Op{l_1,l_2+2}
  & \mbox{for $l_2 = l_1 + 1$} \\[4mm]
    -28 \Op{l_1,l_2}
    +8 \Op{l_1\pm1,l_2}
    +8 \Op{l_1,l_2\pm1} \\[1mm]
    -2 \Op{l_1-2,l_2}
    -2 \Op{l_1,l_2+2}
  & \mbox{for $l_2 = l_1 + 2$} \\[4mm]
    -24 \Op{l_1,l_2}
    +8 \Op{l_1\pm1,l_2}
    +8 \Op{l_1,l_2\pm1} \\[1mm]
    -2 \Op{l_1\pm2,l_2}
    -2 \Op{l_1,l_2\pm2}
  & \mbox{for $l_2 \ge l_1 + 3$}
  \end{cases}
\end{align}
and similar more involved expressions for $Q_{2,6}$ and $Q_{2,8}$ which we refrain from stating explicitly here. Due to the extended range of the interaction, we need 
to consider $k$ special cases for $Q_{2,2k}$; for $l_2 \ge l_1+k+1$ the formula becomes generic. Clearly the action of the dilatation operator is not diagonal in this 
basis. In the superposition \eqref{eqn:asymptBA}, however, we can adjust the coefficients $a(l_1,l_2,p_1,p_2)$ such that the right hand side becomes again 
proportional to $\ket{p_1,p_2}$.

Observe that the action of the dilatation operator in the generic case is symmetric in $l_1$ and $l_2$. Hence, both the in-coming and the out-going waves acquire the 
same factor upon action with the dilatation operator originating from the exponentials. This factor is independent of the function $f$ since in the generic case the 
separation of the magnons is so large that only the leading $1$ of the function $f$ of \eqref{eqn:fdef} contributes. It is also independent of the S-matrix which does 
not depend on the magnon positions at all. Therefore the generic equation determines the eigenvalue $q_2(p_1,p_2)$ alone. The eigenvalue splits
into a sum \be \label{eqn:two-magnon-energy}
  q_2(p_1,p_2) = q_2(p_1) + q_2(p_2)
\ee
with
\be \label{eqn:energyrel}
  q_2(p) =    8             \sin^2(\tfrac{p}{2})
         -   32 \Lambda_r   \sin^4(\tfrac{p}{2})
         +  256 \Lambda_r^2 \sin^6(\tfrac{p}{2})
         - 2560 \Lambda_r^3 \sin^8(\tfrac{p}{2}) + \ldots \; .
\ee
This function is called ``dispersion relation'' or one-magnon energy. The boundary equations determine particular perturbative orders of $S$ and particular values of 
the $f_i$'s. We summarize this in the following table:
\newcommand{\lift}[1]{\raisebox{3mm}[0mm]{#1}}
\newcommand{\oL}[1]{|_{\order{\Lambda_r^{#1}}}}
\begin{center}
\begin{tabular}{|l|c|c|c|c|} \cline{2-5}
\multicolumn{1}{l|}{} & $D_2$            & $D_4$             & $D_6$                 & $D_8$            \\ \hline
$l_2\ge l_1 + 5$ &                       &                   &                       & $q_2(p)\oL{3}$    \\ \cline{1-1} \cline{5-5}
$l_2 =  l_1 + 4$ &                       & $q_2(p)\oL{1}$    & \lift{$q_2(p)\oL{2}$} & $f_0(3,p_1,p_2)$ \\ \cline{1-1} \cline{4-5}
$l_2 =  l_1 + 3$ & \lift{$q_2(p)\oL{0}$} &                   & $f_0(2,p_1,p_2)$      & $f_1(2,p_1,p_2)$ \\ \cline{1-1} \cline{3-5}
$l_2 =  l_1 + 2$ &                       & $f_0(1,p_1,p_2)$  & $f_1(1,p_1,p_2)$      & $f_2(1,p_1,p_2)$ \\ \hline
$l_2 =  l_1 + 1$ & $S_0(p_1,p_2)$        & $S_1(p_1,p_2)$    & $S_2(p_1,p_2)$        & $S_3(p_1,p_2)$   \\ \hline
\end{tabular}
\end{center}
The form of $f$ is not very enlightening as it depends on the basis (expressed by the parameter $\alpha$ of $D_8$). The S-matrix is of course independent of $\alpha$ 
and we find
\be \label{eqn:Smatrix-result}
  S(p_1,p_2) = \frac{\varphi(p_1)-\varphi(p_2)+i}{\varphi(p_1)-\varphi(p_2)-i}  e^{i \psi(p_1,p_2)}
\ee
with the phase function
\be \label{eqn:phaserel}
  \varphi(p) = \tfrac{1}{2}\cot(\tfrac{p}{2}) \bigl[ 1
  + 8 \Lambda_r   \sin^2(\tfrac{p}{2})
 - 32 \Lambda_r^2 \sin^4(\tfrac{p}{2})
+ 256 \Lambda_r^3 \sin^6(\tfrac{p}{2})  + \ldots \bigr]
\ee
and the exponent
\be \label{eqn:exponent}
  \psi(p_1,p_2) = 104 \Lambda_r^3 \left[ \sin^2 (\tfrac{p_1}{2}) \sin p_2 \sin^2 (\tfrac{p_2}{2}) - p_1 \leftrightarrow p_2 \right] \; .
\ee
These formulas provide an eigenstate $\ket{p_1,p_2}$ with eigenvalue $q_2(p_1,p_2)$ for any value of $p_1$ and $p_2$. The discrete physical spectrum is specified not 
before imposing boundary conditions which lead to the Bethe equations and pick out a discrete set of Bethe momenta.

In the two-magnon case we impose the periodicity condition
\be \label{eqn:periodicity-condition}
  a(l_1,l_2,p_1,p_2) = a(l_2,l_1+L,p_1,p_2) \; .
\ee
Now it pays off that we introduced the same function $f$ for in-coming and out-going wave, since this leads independent of $f$ to
\be \label{eqn:2magnonBE}
  \exp(iLp_1) = S(p_1,p_2) = \exp(-iLp_2) \; .
\ee
These are the two-magnon Bethe equations. Due to the originating trace structure the states of the spin-chain need to be invariant under cyclic permutations. This 
leads to the additional zero total momentum condition
\be \label{eqn:2zero-momentum-condition}
  p_1 + p_2 = 0 \; .
\ee
The equations \eqref{eqn:2magnonBE} and \eqref{eqn:2zero-momentum-condition} completely determine the two-magnon spectrum, cf.\ Sec.~\ref{sec:BMNviolation}.

Finally let us remark that the energy shift of the four-loop
``wrapping'' state
of Sec.~2.4 may also be brought into the apparently universal form
\eqref{eqn:Smatrix-result} with the identical phase function \eqref{eqn:phaserel}
and the  exponential factor 
$\psi_{\rm wrap}(p,-p)=\frac{16034}{3}\, \Lambda_r^3\,\sin^4(\frac p
2)\,\sin p$ which only differs from \eqref{eqn:exponent} in the
overall constant up 
front\footnote{We thank S. Frolov for a discussion on this point.}.

\subsection{Bethe equations}

The case of a spin-chain with $M$ magnons and total length $L$ is a straightforward generalization of \eqref{eqn:2magnonBE}. According to the notion of 
\cite{s-matrix}, in an integrable system a general scattering process is always factorized into two-body scattering sub-processes. Therefore one just needs to replace 
the right hand side of \eqref{eqn:2magnonBE} by a product of S-matrices. Any magnon, represented by $p_k$, interacts once with every other magnon $p_{j \not= k}$ on 
its way around the spin-chain:
\be \label{eqn:MmagnonBE}
  \exp(iLp_k)
  = \prod_{\textstyle \atopfrac{j=1}{j\not=k}}^M S(p_k,p_j)
  = \prod_{\textstyle \atopfrac{j=1}{j\not=k}}^M
  \frac{\varphi(p_k)-\varphi(p_j)+i}{\varphi(p_k)-\varphi(p_j)-i}  e^{i \psi(p_k,p_j)}
\ee
for $k=1,\ldots,M$. This is supplemented by the zero total momentum condition
\be
  \sum_{i=1}^M p_i = 0 \; .
\ee
As the derivation of \eqref{eqn:MmagnonBE} contained some subtle points we have checked the Bethe equation for a number of states. The results are documented in 
App.~\ref{sec:spectrum}.

Let us now compare this form to the other known perturbative long-range Bethe ans\"atze in the literature. We note that the dispersion relation \eqref{eqn:energyrel} 
and the phase function \eqref{eqn:phaserel} are exactly the same as for the Beisert-Dippel-Staudacher spin-chain \cite{bds-long-range}. The difference lies in the 
presence of the exponential factor in the Bethe ansatz for the PWMT. This exponential is reminiscent of the Bethe ansatz for quantum strings proposed by Arutyunov, 
Frolov and Staudacher \cite{frolov-gleb-matthias}. There the exponent is a series of antisymmetric products of all conserved charges. This is exactly the same as we 
observe here, since our exponent is in fact
\be
  \psi(p_1,p_2) = \frac{13 \Lambda_r^3}{8} \left[ q_2(p_1) q_3(p_2) - q_3(p_1) q_2(p_2) \right]
\ee
where
\be
  q_2(p) = 8 \sin^2(\tfrac{p}{2}) \quad , \quad q_3(p) = 8 \sin(p) \sin^2(\tfrac{p}{2})
\ee
are the lowest order contributions to the eigenvalues of the higher charges $Q_2$ and $Q_3$, cf.~\eqref{eqn:energyrel} and \eqref{eqn:eigenvalue-formulas}. A more 
detailed comparison to the quantum string ansatz reveals that our exponent has one additional power of the coupling constant $\Lambda_r$. This explains why the 
exponential yields a breakdown of BMN scaling in our case, while it results in a sub-leading discrepancy in the near-BMN limit with respect to SYM theory in the 
quantum string case.

The existence of an exponential factor in the Bethe ansatz seems to be the generic form for integrable long-range spin-chains. It would be very interesting to 
determine whether the Bethe ansatz for the Inozemtsev spin-chain may be brought into this general form as well. The idea would be to trade contributions in
the phase function for an exponential factor.


\subsection{Two-magnon spectrum and violation of BMN scaling} \label{sec:BMNviolation}

Using the above Bethe ansatz one may solve the two-magnon problem explicitly. Here, due to the zero total momentum condition $p_1 = -p_2 =:p$, one finds just one 
Bethe equation
\be \label{eqn:two-magnon-bethe}
  \exp(iLp)=\frac{\varphi(p)+i/2}{\varphi(p)-i/2} \, e^{i\psi(p,-p)} \; .
\ee
This equation is solved via the perturbative ansatz
\be
p = p_0 + \Lambda_r p_1 + \Lambda_r^2 p_2 + \Lambda_r^3 p_3 + \ldots\; ,
\ee
plugging this into \eqref{eqn:two-magnon-bethe} and making use of \eqref{eqn:phaserel} one finds the explicit result:
\begin{align}
p_0 & =  \frac{2n\pi}{L-1} \nonumber \\
p_1 & = -\frac{16 \cos(\tfrac{n\pi}{L-1}) \sin^3(\tfrac{n\pi}{L-1}) }{L-1} \nonumber \\
p_2 & =  \frac{64 \cos(\tfrac{n\pi}{L-1}) \sin^5(\tfrac{n\pi}{L-1})}{(L-1)^2}  \Bigl[ 2L + (L + 3) \cos(\tfrac{2n\pi}{L-1}) \Bigr] \\
p_3 & =  \frac{32 \cos(\tfrac{n\pi}{L-1}) \sin^5(\tfrac{n\pi}{L-1})}{3(L-1)^3} \Bigl[ (-79L^2+122L-79)
                                                                                     +4(5L^2-31L+20) \cos(\tfrac{2n\pi}{L-1}) \nonumber \\
    &    \qquad\qquad\qquad\qquad                                                    +4(4L^2+13L-20) \cos(\tfrac{4n\pi}{L-1})
                                                                                     +4( L^2- 7L+10) \cos(\tfrac{6n\pi}{L-1}) \Bigr] \nonumber
\end{align}
Inserting these expressions into the \eqref{eqn:two-magnon-energy} yields, upon expansion in $\Lambda_r$, the closed expression for the two-magnon spectrum 
$\delta\Delta_r = \Lambda_r q_2$ parametrized by the integers $n$ and $L$, where $L$ denotes the length of the spin-chain and $n = 
0,\ldots,\left[\tfrac{L-2}{2}\right]$:
\be
\begin{split}
 \delta\Delta_r(n,L) =\ & 16 \Lambda_r \sin^2 (\tfrac{n\pi}{L-1}) \\
                        & - \frac{64\Lambda_r^2}{L-1} \sin^4(\tfrac{n\pi}{L-1}) \Bigl[ (L+1) + 2 \cos(\tfrac{2n\pi}{L-1}) \Bigr] \\
                        & + \frac{256\Lambda_r^3}{(L-1)^2} \sin^6(\tfrac{n\pi}{L-1}) \Bigl[ (2L^2+5L-2)
                                                                                          + 2(5L+2) \cos(\tfrac{2n\pi}{L-1})
                                                                                          + (L+4) \cos(\tfrac{4n\pi}{L-1})  \Bigr] \\
                        & + \frac{256\Lambda_r^4}{3(L-1)^3} \sin^6(\tfrac{n\pi}{L-1}) \Bigl[ (-30L^3-39L^2+51L-39) \\
                        & \qquad         + 3(10L^3-49L^2+26L-13) \cos(\tfrac{2n\pi}{L-1})
                                         + 8(11L^2-10L) \cos(\tfrac{4n\pi}{L-1}) \\
                        & \qquad         + (18L^2+90L-30) \cos(\tfrac{6n\pi}{L-1})
                                         + (2L^2+17L+30) \cos(\tfrac{8n\pi}{L-1})  \Bigr]
\end{split}
\ee
Expanding this in the BMN limit, i.e.~$\Lambda_r, L\to\infty$ with $\Lambda_r/L^2$ fixed, one finds
\be \label{eqn:BMNlimit}
\begin{split}
  \lim_{\textstyle \atopfrac{L\to\infty}{\Lambda_r\propto L^2}}
  \delta\Delta_r(n,L) =\ & 4 \left ( 4 \pi^2 n^2
  \frac{\Lambda_r}{L^2}\right )
                          - 4\left ( 4 \pi^2 n^2
  \frac{\Lambda_r}{L^2}\right )^2
                          + 8 \left ( 4 \pi^2 n^2
  \frac{\Lambda_r}{L^2}\right )^3 \\[-2mm]
                        & - 6656 \pi^6 n^6 \frac{\Lambda_r^4}{L^7}\left( 1 - \frac{1}{L} \right)
                          - 20 \left ( 4 \pi^2 n^2
  \frac{\Lambda_r}{L^2}\right )^4 + \ldots \; .
\end{split}
\ee
This is in fact divergent due to the four-loop term $\sim \Lambda_r^4/L^7$: the proclaimed breakdown of BMN scaling. This formula is to be compared with the proper 
scaling behavior \eqref{BMN-formula-expanded} with $(L\approx J)$. The additional (violating) piece originates from the microscopic contact interactions of two 
magnons.


\section{Conclusion and discussion}

\noindent
In this paper we have performed a fourth order perturbative analysis of the $\SU(N)$ plane-wave matrix theory in its planar large $N$ limit in the closed minimal 
$\su(2)$ subsector. The outcome of our analysis yielded an effective planar Hamiltonian, which is integrable to this rather high loop order. Additionally we have 
explicitly constructed the first two higher commuting charges of the integrable system to the corresponding loop order, which is summarized in 
App.~\ref{sec:higher-charges}. We take these results as a very strong indication of the complete integrability of the planar matrix theory system and similarly so for 
the closely related ``mother'' theory of superconformal $\Ncal = 4$ Yang-Mills.

Moreover we determined the underlying perturbative asymptotic Bethe ansatz \cite{s-matrix} of the system for states of length larger than four. While the form of the 
integrable spin-chain Hamiltonian (or planar dilatation operator or effective matrix model Hamiltonian) becomes more and more involved at higher loops, the ``core'' 
of the model -- the two-magnon S-matrix $S(p_1,p_2)$ -- is of the intriguingly compact form
\be \label{eq:genform} S(p_1,p_2) =
\frac{\varphi(p_1)-\varphi(p_2)+i}{\varphi(p_1)-\varphi(p_2)-i} 
\times \exp\left (i f(\Lambda_r) \sum_{i=1}^\infty \Lambda_r^i q_{i+1}(p_1) q_{i+2}(p_2) - p_1 \leftrightarrow p_2
\right ) \; ,
\ee
which was first established for the quantum $AdS_5\times S^5$
string Bethe ansatz in \cite{frolov-gleb-matthias}. Here $q_i(p)$
denotes the higher charges of the spin-chain,
$f(\Lambda_r)$ a function of the coupling constant and
$\varphi(p)$ the phase relation of the model. This form of the
S-matrix appears to be {\it generic} for integrable long-range
spin-chains: Except for the Inozemtsev chain all known long-range
spin-chains or Bethe ans\"atze (the Beisert-Dippel-Staudacher chain
\cite{bds-long-range}, the Arutyunov-Frolov-Staudacher quantum
string chain \cite{frolov-gleb-matthias}, our plane-wave matrix
theory effective Hamiltonian and the recently established
S-matrices of the $\slf(2)$ and $\su(1|1)$ $\Ncal =4 $ dilatation
operators \cite{s-matrix}) fall into the class \eqref{eq:genform}
with the {\it same} functions $\varphi(p)$ and ${q}_i(p)$!
\footnote{Strictly speaking this is not quite correct as the
fermionic $\su(1|1)$ chain has a trivial phase relation
$\varphi(p)=0$ \cite{s-matrix}.} It would be very interesting to
establish whether the Inozemtsev chain S-matrix may also be
brought into the form of Eq.~\eqref{eq:genform}.

Our perturbative computation was only attainable through the use
of a tailor made computer program which generated all \emph{planar}
graphs and performed the necessary term algebra in a massively
distributed fashion, which is reviewed in detail in App.~\ref{sec:gemstone}.
The program should also be applicable for similar problems in
planar perturbation theory at high orders, in particular for a
similar analysis in larger closed subsectors of the plane-wave
matrix model, such as $\su(2|3)$.

The motivation for this computation grew out of the close
relationship of the effective plane-wave matrix model Hamiltonian
to the dilatation operator of $\Ncal =4$ super Yang-Mills, which
is firm in the joint minimal $\su(2)$ subsector up to the three
loop order \cite{pwmt-integrability,beisert-su32}. This connection
stems from the consistent (classical) reduction of the full field
theory on a three-sphere \cite{pwmt-from-sym} and is seemingly
stable under quantum corrections (up to three-loops) resulting in
a ``renormalization'' of the relation between the matrix model
(dimensionless) mass parameter $M$ and the super Yang-Mills
coupling constant $g_{\rm YM}$
\be
  \Lambda = \Lambda_r + \frac{7}{2} \Lambda_r^2 - 11 \Lambda_r^3
  + \frac{1257}{16} \Lambda_r^4 + \order{\Lambda_r^5}
\ee
where $\Lambda=2\, N/M^3$ and $\Lambda_r=g_{\rm YM}^2\,
N/4\pi$. Here the highest order term was determined by requiring
``naive'' BMN scaling, i.e.~demanding that the four-loop
contribution to the effective Hamiltonian acts as a clean lattice
laplacian to the fourth power on well separated magnons. It
remains to be shown whether the established correspondence 
of plane-wave matrix theory to the
$\Ncal =4$ dilation operator of the full uncompactified field
theory extends beyond the three-loop level (in the planar sector).
For this the four-loop anomalous scaling dimensions of one or two
super Yang-Mills operators would need to be computed. Hence 
unfortunately our verdict on the perturbative BMN scaling of the
full $\Ncal =4$ gauge theory at four loops cannot be conclusive,
it, however,  does certainly question its existence beyond three-loops.
A better understanding of the inner workings of the planar plane-wave matrix
theory/$\Ncal =4$ super Yang-Mills correspondence on the basis of
an effective field theory approach in which one integrates out the 
higher Kaluza-Klein modes of the 4d gauge theory, would be of great
importance.

Our results also shed light on the structure of the ``wrapping''
interactions for long-range spin-chains, which are inaccessible
through the means of the asymptotic Bethe ansatz where the length
of the chain needs to be strictly larger than the spread of the
local Hamiltonian: In the considered $\su(2)$ subsector this
happens for the first time at four-loop order, where the
resulting spin-chain Hamiltonian involves five neighboring spin
interaction terms, whereas the shortest non-protected operator in
this subsector is a Konishi descendant of length four:
$\tr[Z^\dagger,W^\dagger]^2$. Its four-loop energy shift can
neither be accounted for by any ``natural'' prescription of how to
act with the five spin terms of the Hamiltonian on a length four
state nor by blindly applying the Bethe ansatz. One finds \be
  \delta\Delta_r = 12\Lambda_r - 48\Lambda_r^2 + 336\Lambda_r^3 -
  \begin{cases}
  2958\Lambda_r^4             & \mbox{(wrap around)} \\
  3026\Lambda_r^4             & \mbox{(discard long-range terms)} \\
  3054\Lambda_r^4             & \mbox{(Bethe ansatz)} \\
  \tfrac{12771}{4}\Lambda_r^4 & \mbox{(true value)}
  \end{cases}
\ee More work is needed to understand how this intrinsic
restriction of the asymptotic Bethe ansatz could be overcome.

Finally it would be very nice to establish a proof of the
integrability of the planar plane-wave matrix theory. Compared to
the situation for the $\Ncal =4$ super Yang-Mills field theory
this appears to be more tractable, as one has access to the full
non-perturbative Hamiltonian. However, efficient techniques to
isolate the planar large $N$ sector in a non-perturbative fashion
are lacking.


\paragraph{Acknowledgments}\hfill\bigbreak

\noindent We would like to thank Gleb Arutyunov, Niklas Beisert,
Virginia Dippel, Sergey Frolov and Matthias Staudacher for crucial discussions
and comments.

\noindent Moreover we are greatly indebted to the members of the
computational division of the GEO 600 gravitational wave detector
project, especially Maria Alessandra Papa and Steffen Grunewald,
for allowing us to use their Merlin cluster for part of our
calculations and to the experimental particle physics group at the
University of Wuppertal, especially Hendrik Hoeth, for computation
time on the presently largest German university computer -- the
ALiCEnext Linux cluster. Minor parts of the calculation have been
performed at the {\tt cip.physik.uni-muenchen.de} computer lab of
the University of Munich, so we want to thank both the manager as
well as the head administrator of that lab -- Sigmund Stintzing
and Susanna Maurer. Other parts have been done on workstation
machines of the AEI, hence we also want to thank the AEI's
administrative staff, especially the head Christa Hausmann-Jamin,
for their support. We also want to thank Peter Herzog for allowing
us to use the {\tt spotlight.de} webserver for our purposes which
played a key role in coordinating the distribution of parts of the
calculation to individual machines via HTTP. Furthermore, the
authors thank Klaus Aehlig for his permission to include his very
beautiful solution to a functional programming exercise problem
that plays a minor role in our approach into our source code.
Thomas Fischbacher acknowledges financial support by the German
Academic Exchange Service via the DAAD postdoc research program.
This work was partially supported by IISN - Belgium
(convention 4.4505.86), by the ``Interuniversity Attraction Poles
Programme -- Belgian Science Policy'' and by the European Commission
FP6 programme MRTN-CT-2004-005104, in which Thomas Fischbacher is
associated to V.U. Brussel.

\appendix

 \newpage
 

\section{Details on the Computation} \label{sec:gemstone}

\noindent
The problem solved by the \gemstone{} code 
(which was developed for this application) is the generation of planar
multiloop graphs from large sets of vertices with directed and colored
legs, which have a term algebra associated to them, with a highly
efficient algorithm that furthermore is implemented close to the
machine language level for maximum speed.

Considering the technical complexity of the calculation of the dilaton
operator at four loops, the authors have strong reason to believe that
this may well be the largest {\em symbolic} calculation performed so
far\footnote{There are far larger massively distributed
prime searches as well as encryption breaking attempts, but the
underlying questions should perhaps not be regarded as being of
symbolic nature.}.


The key to its successful completion was the development of
aggressively optimized symbolic algebra code. With this program, which
the authors named \gemstone{}\footnote{as compactness of code and
flawlessness are the most important virtues for such a task}, the
three-loop calculations that have been done
in~\cite{pwmt-integrability} (by using the already highly optimized
FORM~\cite{Vermaseren:2000nd} symbolic manipulation program and a more
conventional algorithm) could be checked in about $1/100$~of the
original calculation time.

As this code is at present the best tool available to systematically
do many more planar higher-loop matrix model calculations, and due to
its modularity should be easily adoptable to planar high-loop field
theory calculations as well, it might be quite interesting in itself
and has been included in the source archive of the {\tt arXiv.org}
preprint of this work at {\tt http://www.arxiv.org/e-print/hep-th/0412331}.

Furthermore, as the results presented in this work hinge on the
correctness of the algorithm as well as its implementation, the
scientific demand of verifiability gives another strong incentive to
make this code publicly available.

Finally, many of the approaches to the problem of making the
calculation fast are generic enough to be useful in many other
situations, even if not of direct relevance to the physics of this
problem. As this appendix may be interesting to computer scientists as
well, it is essentially self-contained.

\subsection{The Task}

\noindent
In a nutshell, the problem at hand consists of the determination of
perturbative quantum corrections to the effective Hamiltonian that
come from interaction diagrams which can be drawn without
self-intersections (i.e. are planar). Conceptually, it may be obtained
by a simple algebraic normal ordering procedure with subsequent
isolation of those contributions that can be associated to planar
graphs. While this approach is very simple, it is quite limited in
practice by excessive growth of the number of terms when increasing
the perturbative order. As both quantitatively and qualitatively new
phenomena appear when going to higher orders, doing such calculations
is not merely an academic exercise. Hence, one should make use of as
much of the specific structure of the problem as possible in order to
reduce the complexity of high-order calculations. In our case, this
especially means to make as much use of the planarity property as
possible right from the beginning, instead of generating all terms and
later throwing away contributions that can be identified as coming
from nonplanar graphs. While the number of nonplanar graphs also
increases dramatically with increasing number of vertices, there are
nevertheless far less planar than non-planar graphs in all problems of
interest.

We have to find all graphs with certain pre-defined properties,
associate a term to every graph, and sum all the contributions from
allowed graphs. Graphs are built from vertices, which may have up to
four legs, by successively adding vertices and fusing legs. Every
vertex leg has a direction (in-going or out-going, corresponding to
particle creation or annihilation) as well as a color (corresponding
to particle type). Only matching legs (same color, different
directions) may be fused. To every vertex, we associate an algebraic
factor depending on the type of the vertex. Leg fusion induces
algebraic transformations.

Vertices have three or four legs, except for some degenerate cases
where self-binding of a four-leg vertex gives a two-leg vertex. The
factors associated to three-leg vertices contain a single power of the
interaction strength $g$, while those associated to (possibly
degenerate) four-leg vertices contain a factor~$g^2$.  These two large
vertex classes are further subdivided by considering the energy shift
associated to particular vertices -- this is just the mass of the
particles created (represented by out-going legs) minus the mass of
the particles destroyed, and can be represented by a small
integer. With this additional distinction, there are~$15$ different
vertex classes, each containing between~$1$ and~$188$ different types
of vertices.

In a first step, we expand the Hamiltonian into a set of sequences of
vertex classes. As this is of minor relevance for the heart of the
calculation, this step is discussed separately in
section~\ref{sec:OpSeq}. For the fourth-order calculation, this yields
$338\,834$ individual sequences, each carrying an interaction strength
factor $g^8$.

Next, an initial closed chain of vertices with all legs on the outside
has to be chosen. Starting from this chain, we have to consider every
single sequence of vertex classes individually, finding all possible
realizations of that sequence as a sequence of vertex types. These
building blocks then have to be fused subsequently to the initial
chain in all possible ways, subject to some extra constraints. The
fundamental graph operations, such as fusing legs, induce
transformations on the associated terms that have to be kept track
of. Eventually, all terms from admissible final graphs have to be
summed.

\begin{figure}
\begin{center}
\includegraphics[height=11cm]{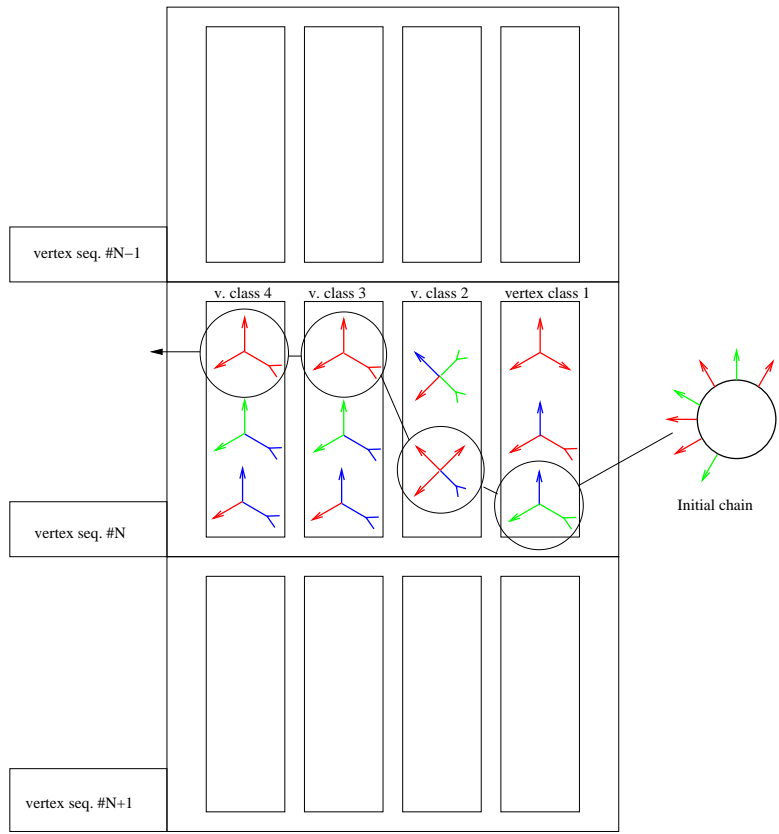}
\end{center}
\caption{\small Forming all Graphs: One particular choice where the last two vertices to be added are of the same type.}
\label{fig:GraphForming}
\end{figure}

As an example, the vertex class sequence consisting of four times the
largest class with $188$~four-leg vertex types in it alone gives rise
to $188^4=1\,249\, 198\,336$ individual sequences of vertex types. For
every single one of those, all possible ways to add these vertices in
sequence to the starting graph have to be taken into
consideration. Likewise for the other $338\,833$ sequences. This is
depicted schematically in figure~\ref{fig:GraphForming}.

For the determination of the four-loop dilatation operator, it
suffices to form all additions of all these sequences of vertices to a
suitably chosen initial chain of length eleven.

Vertices to be added may be turned, but not flipped over. Every final
graph must obey the following constraints:

\begin{enumerate}

\item It is planar.

\item All the open legs are on the outside.

\item Its construction by successive addition of vertices never involved
a step where an in-going leg remained open (as this would act on the
vacuum and give zero).

\end{enumerate}

As an extra difficulty, some legs are fermionic, which means that care
of extra factors~$-1$ has to be taken which occur in some (but not
all) situations when such legs are fused.

\subsection{Choice of the Platform}

\noindent As a basis for the implementation, the Objective
CaML~\cite{Leroy} system has been chosen (an early prototype was
written in LISP). This decision was based on the expectation that the
four-loop calculation might become almost prohibitively large, and
hence one should be able to make use of computer power donated by
volunteers to run the calculation in a highly parallelized scheme,
similar to efforts like {\tt SETI@Home}~\cite{setiathome}, {\tt
GIMPS}~\cite{GIMPS}, and many others. This requires the availability of
a high-quality native code compiler for a reasonable price (ideally
for free) for all intended target platforms, especially including
Linux/x86, Microsoft Windows, Mac OS X, and possibly even Linux on
non-PC hardware. This excludes LISP as an implementation language,
since there is no free LISP Compiler of sufficient strength available
for the Windows platform. Java as well as C (or C++) are not
considered viable options as well, as some of the algorithmic tricks
that had to be used are prohibitively clumsy to express in such
languages that do not provide sufficient support especially for lambda
abstraction; furthermore, code development and in particular debugging
is facilitated greatly by the availability of an interactive command
prompt where one can immediately test program components
individually. This is also not widely available for the well-known
mainstream languages.

The decision to use Objective CaML was further supported by Ocaml's
excellent record for being a versatile tool to quickly build solutions
for challenging problems~\cite{icfp}, availability of key algorithmic
libraries (hashes, rational numbers, etc.), as well as quite positive
experience with this system in an earlier research project from the
side of one of the authors.

\subsection{Planar Graph Generation} \label{sec:planar-graph-generation}

\noindent
As explained, we are interested in planar graphs only, as this
restriction will considerably constrain the number of graphs.

In terms of data structures, one attractive way to organize this
calculation is to represent every graph as a vector of connected
components, where every connected component is a vector of domains
(cells) that carry legs along their boundaries and are glued together
in such a way that all legs point to the inside. To every connected
component, we associate a particular term -- details of the term
algebra will be discussed in the next section.

The cells may themselves be represented as vectors of legs, and every
leg has to know about its type (direction and color) and furthermore
some kind of pointer (in a general, not implementation-specific sense)
into the term of the leg's connected component. In the implementation,
all this information is recorded in a single signed 30-bit integer per
leg. We make use of the observation that for a four-loop calculation,
we will introduce at most $8\cdot3=24$ legs from vertices in addition
to the~$11$ legs in the initial sequence. This means that we can
conveniently label legs by 6-bit numbers in the range~$0\ldots35$.
(Additional legs may be introduced by tracing fermionic loops of odd
length -- this gives an extra factor $\epsilon^{ijk}$ from the
projector. There may be at most two such factors in our calculations,
so we take labels for these indices from the upper end of the 6-bit
integer range.)

Fundamental operations on graphs are (1) the addition of a new vertex,
which initially is regarded as an individual connected component with
only one cell, (2) the composition of two connected components right
before legs between them are fused: this causes one entry to be
removed from the vector of all connected components and another
one to be enlarged, as well as corresponding transformation on the
vector of associated terms, where one entry is removed and multiplied
into another one, and (3) connection of cells by fusion of legs. This
reduces the number of cells in a connected component and as well
induces certain transformations on the associated term.

\begin{figure}
\begin{center}
\includegraphics[height=8cm]{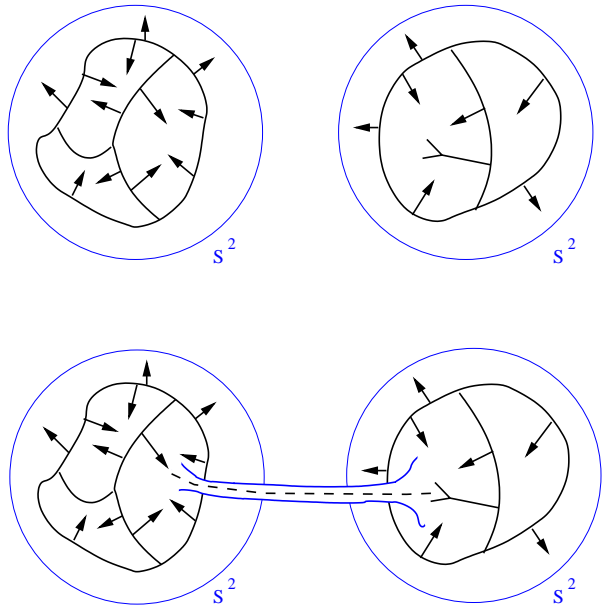}
\end{center}
\caption{\small Fusing legs}
\label{fig:Domains}
\end{figure}

Concerning the fusion of legs, it is helpful to think of each
individual connected components as living on a sphere. When legs
are fused, these spheres are punctured at the interior of the
partaking cells and then connected appropriately, see
figure~\ref{fig:Domains}.

Note that steps (1) and (3) introduce a choice among many different
possibilities. In (1) this comes from (in some cases, quite many)
different vertex types contributing to a vertex class, all of them
having individual leg structure, hence requiring to branch graph
generation, in (3) this comes from the possibility to connect the legs
from two domains in multiple different ways. In the implementation,
the technique of {\em continuation coding} turned out to be a very
convenient way to express the branchings in these steps of the
calculation. That is, one passes to the function $b$ that branches the
calculation a function $f$ which is to be called with an individual
branching and does all the remaining processing for this
branch. (``Passing around the entire rest of the calculation as a
function argument.'') This is then called multiple times, once for
every branch. The abstract idea of passing around code towards the
data that have to be processed -- like an inverse assembly line that
transports workers to partially completed pieces -- is a re-occurring
pattern in the design of the entire system. While it is nice and
highly expressive to write the code in such a way, dynamic memory
allocation (aka `consing') of functions with partially fixed
parameters (aka `closures') is a bit wasteful in terms of
efficiency. With a lot more work, some (limited) additional efficiency
improvement might be gained here.

A further subtlety comes from the requirement to properly treat signs
from fermionic loops. This is done by subsequently recording the types
of all vertex's legs (fermionic/non-fermionic) in a special bit-vector
in the graph (in the implementation, this is distributed out over two
30-bit integers) whenever a new vertex is added, and removing the
fermionic leg bits again whenever two such legs are being fused,
introducing an extra minus sign if there is an odd number of active
fermionic legs whose label lies between the labels of the fused
legs. (Care must be taken to ensure that one does not apply such
modifications to the graph data structure in such a way that they are
not being undone properly once we go back to do a different branch of
the calculation.)

A final trick is to make use of the property that in every vertex
addition step, the only legs that may be fused are those that were
introduced by the vertex. It may well be that this eventually leads to
the combination of (at worst) five different connected components 
(the vertex itself and four other components that bind to its four
in-going legs), but this can be done sequentially, and in every such
absorption step, we only have to remember the presently `hot'
connected component plus the cell where the vertex originally was
absorbed, and the positions of the vertex's original still unbound
legs in the new cell. As every leg can be labeled with a 6-bit
integer, and as we at most have four active (`hot') legs at any step
during binding, it becomes attractive to encode a list of four legs in
one 28-bit integer, seven bits per leg, with $127=1111111_2$ as a
special `end of list' stop code. Implementing primitive list
operations on such short lists of small integers brings the benefit
that no expensive memory management is required, plus a lot of work
can be done very close to the processor's register file. Furthermore,
should microprocessor design evolve into the highly desirable
direction to have certain parts of freely reconfigurable logic on the
CPU (cf. e.g.~\cite{fpgaCPU}) during the next decade, this specific
part of the calculation could easily be moved directly to the hardware
level, so that it could be done at virtually no cost at all.

The term algebra is fully abstract, which means that one may hook
arbitrary implementations that provide notions of fusing terms and
legs into the combinatorial engine. In particular, two term algebras
that are particularly nice for debugging are the trivial one that just
records all fusions in a list, and one that produces a {\tt
graphviz}~\cite{graphviz} input file for visual rendering of
individual graphs to check that the algorithm is working correctly.

\subsection{Operator Sequences}\label{sec:OpSeq}

\noindent The set of all vertex sequences that have to be applied
to the initial chain is determined by expanding the effective Hamiltonian
\eqref{eqn:eff-ham} in terms of vertices of definite energy
shift. This allows one to shift all projectors and propagators out of
the Hamilton operator.  The corresponding rules are ($V_1$ denoting
three-legged vertices and $V_2$ denoting four-legged vertices):

\begin{equation}
\begin{array}{lcl}
        V_1&\rightarrow&\sum_{k\in\{\pm1,\pm2,\pm3\}} V_{1,k}\\
        V_2&\rightarrow&\sum_{k\in\{-4\ldots 4\}} V_{2,k}\\
        P(E)V_{\{1,2\}m}&\rightarrow&V_{\{1,2\}m}P(E-m)\\
        \Delta(E)V_{\{1,2\}m}&\rightarrow&V_{\{1,2\}m}\Delta(E-m)\\
        P(E)|\psi\rangle&\rightarrow&\delta_{E,0}|\psi\rangle\\
        \Delta(0)|\psi\rangle&\rightarrow&0\\
        \Delta(E)|\psi\rangle&\rightarrow&1/E|\psi\rangle\quad(E\neq0)
\end{array}
\end{equation}
Thus, the one-loop operator (compare (B.5))
\begin{equation}
        T_2=P V_2+P V_1\Delta V_1
\end{equation}
is unfolded to a sum of seven terms:
\begin{equation}
\begin{array}{lcl}
T_2&=&-\frac13 V_{1,+3} V_{1,-3}\\
    &&-\frac12 V_{1,+2} V_{1,-2}\\
    &&-\phantom{\frac11} V_{1,+1} V_{1,-1}\\
    &&+\frac13 V_{1,-3} V_{1,+3}\\
    &&+\frac12 V_{1,-2} V_{1,+2}\\
    &&+\phantom{\frac11} V_{1,-1} V_{1,+1}\\
    &&+\phantom{\frac11} V_{2,0}.
\end{array}
\end{equation}
At four-loop order, the corresponding operator contains~$339$ summands,
which likewise unfold to~$338\,834$ contributions (compare (B.8)):
\begin{equation}
\begin{array}{lcl}
T_8&=&\phantom+\frac5{2187} V_{1,+3} V_{1,-3}  V_{1,+3} V_{1,-3}  V_{1,+3} V_{1,-3}  V_{1,+3} V_{1,-3}\\
    &&+\frac{29}{3888}  V_{1,+3} V_{1,-3}  V_{1,+3} V_{1,-3}  V_{1,+3} V_{1,-3}  V_{1,+2} V_{1,-2}\\
    &&+\frac{17}{243}  V_{1,+3} V_{1,-3}  V_{1,+3} V_{1,-3}  V_{1,+3} V_{1,-3}  V_{1,+1} V_{1,-1}\\
    &&+338\,830\;{\rm more}\\
    &&+\frac{1}{128} V_{2,-4} V_{2,-4} V_{2,+4} V_{2,+4}
\end{array}
\end{equation}
Every such factor~$V_{\cdot,\cdot}$ may again be a sum of up to~$188$
contributions, each one with its own leg structure.

While this is sufficiently easy to do with a small FORM program within
about 15 minutes of calculation time on 1.7 GHz Pentium-IV hardware,
we note that just as well here, it is possible with very little effort
to speed up this part of the calculation by a factor of about~300 by
using compiled (e.g. Lisp) code. As propagators and projectors are
shifted through the term, it makes sense to just record and operate on
the positions of all these factors that still are present instead of
generating new terms even for shifts that only introduce an extra
factor. One key observation is that there are only $15$~different
types of vertices, so one can encode a sequence of up to eight
vertices very conveniently in a 32-bit number. (As the authors
originally hoped to be able to go even beyond level four, and as the
Lisp system with which they performed this expansion does not support
64-bit numbers, they abused 52-bit floatingpoint mantissas for this
purpose.) These tags for vertex sequences are used as hash keys to
collect contributions to the various sequences.

\subsection{Term algebra}

\noindent Playing around with the graph engine shows that, even for small
problems, quite impressive amounts of data have to be processed by the
term algebra. As required by the trivial observation known as Amdahl's
law that it helps little for speeding up a program by orders of
magnitude to make just part of the code much faster that originally
consumed half of the calculation time, an aggressive implementation of
graph combinatorics must be matched by an equally aggressive term
algebra.

\subsubsection{Rules}

\noindent All the terms that occur in the calculation are sums of contributions
that consist of a rational coefficient, optionally with an extra
factor $i$ or $\sqrt2$, and a series of further factors from the
$\spin{(9)}$ algebra of types $\delta_{ij}$, $\delta_{ab}$,
$\epsilon_{ijk}$, $\epsilon_{m_1\ldots m_9}$,
$\gamma^i_{\alpha\beta}$, $\gamma^a_{\alpha\beta}$,
$\Pi^\pm_{\alpha\beta}$. Here, $i,j,\ldots$ denote $\SO(3)$ vector
indices, $a,b,\ldots$ denote $\SO(6)$ vector indices and $m,\alpha$ are
$\spin{(9)}$ vector and spinor indices for $\SO(3)\times \SO(6)\subset
\spin{(9)}$. A projection factor $\Pi^\pm$ =
$\frac{1}{2}\left(1\pm\frac{1}{6}\,i\gamma^{ijk}\epsilon_{ijk}\right)$
has to be inserted wherever two fermionic legs are fused -- cf. \eqref{eqn:commutation-relations}:
\[
c_\alpha c^\dag_\beta \rightarrow \frac{1}{2} \Pi^-_{\alpha\beta} .
\]
These projection matrices are shifted through a sequence of $\gamma$
matrices via the rule $\gamma^i\Pi^\pm=\Pi^\pm\gamma^i$,
$\gamma^a\Pi^\pm=\Pi^\mp\gamma^a$. Note that $(\Pi^\pm)^2 = \Pi^\pm$
and $\Pi^+ \Pi^- = 0$, as advertised for projectors.

As we work with a $\SU(2)$ subgroup sector of $\SU(4)=\Spin{(6)}$, we
further split the~$a$ indices via $a\rightarrow (W,\bar W, Y, \bar Y,
Z, \bar Z)$; we will then be only concerned with initial graphs formed
from out-going $W, Z$ legs.

Explicitly, our $\spin{(9)}$ algebra reduction rules are:

\begin{itemize}

\item (delta) $\delta_{?_1?_2} T_{\ldots ?_1\ldots} = T_{\ldots ?_2\ldots}$,
 $\delta_{aa}=6$, $\delta_{ii}=3$, $\delta_{\alpha\alpha}=16$,
 $\delta_{X\bar X}=\delta_{\bar X X}=\delta_{Y\bar Y}=\ldots=1$

\item (epsilon) $\epsilon_{ijk} \epsilon_{ilm} = \delta_{jl}\delta_{km}-\delta_{jm}\delta_{kl}$

\item (Clifford) $\gamma^i_{\alpha\beta}\gamma^j_{\beta\gamma}+\gamma^j_{\alpha\beta}\gamma^i_{\beta\gamma}=2\delta_{ij}\delta_{\alpha\gamma}$

\item (epsilon-9) $\delta_{\alpha_1\alpha_10}\prod_{k=1}^9\gamma^{i_k}_{\alpha_k\alpha_(k+1)}=16\,\epsilon^{(9)}_{i_1\ldots i_9}$,\qquad 
$\delta_{ij}\epsilon^{(9)}_{\ldots i \ldots j \ldots} = 0$

\end{itemize}

There would be further (obvious) reduction rules involving e.g.
$\epsilon^{(9)}\cdot\epsilon^{(9)}$, but we do not have to take care
of those, as they cannot occur up to four-loop order. Likewise, we do
not include rules of the type
$\epsilon^{(9)}\epsilon^{(3)}\rightarrow\epsilon^{(6)}$ -- the number
of such terms is expected to be quite limited, and treatment of such
special cases may always be deferred to more generic term manipulation
programs.

The Clifford and epsilon-9 rules are not used in the form given here,
but to determine beforehand a table of all the expansions of traces of
gamma matrices in terms of epsilons and deltas that can show up in the
calculation.

\subsubsection{Strategy}

\noindent
Aside from some special cases
($\epsilon\epsilon\rightarrow\sum\delta\delta$ and
$\tr\gamma\ldots\gamma\rightarrow\sum\ldots$), every graph contributes
to the coefficient of a single individual summand. Hence, we associate
to every connected component a tight encoding of a summand which
contains information about the coefficient, certain extra factors like
$i$ or $\sqrt{2}$, as well as $\epsilon_{ijk}$, $\epsilon_{m_1\ldots
m_9}$, $\delta_{mn}$ and $\gamma^m_{\alpha\beta}$ algebraic factors
and indices. Once a contribution has been generated, the new
coefficient is added to a hash table mapping normalized summands to
(references to) coefficients.

Using conventional symbolic algebra, fusing legs would be performed by
generating a new summand which is just the old one extended by an
extra $\delta_{mn}$ (or $\Pi^\pm_{\alpha\beta}$) factor and then
successively searching through all algebra reduction rules to find a
place in the new term where some such rule can be applied. One can do
much better by making use of the information that the original terms
have been totally reduced and therefore, if any new reduction became
possible, it has to include the newly added factor. Hence, we mark the
newly added factor as `hot' and successively chase it through all
possible reduction rules that such a hot factor may participate
in. Hotness is treated as contagious here: every factor that is
`activated' by a hot factor becomes hot itself.

In our particular case, such an approach stratifies the algebra in the
sense that the longest chain of reductions that a newly introduced
factor may generate is a hot gamma closing a trace which then is
expanded to a set of hot deltas, which connect different epsilons that
originally were connected to the gamma chain to produce another set of
hot deltas, which may rename open legs (on other $\delta$, $\epsilon$,
or $\gamma$ factors). No longer sequence of reductions than this
$\gamma\rightarrow\delta\rightarrow\epsilon\rightarrow\delta$ chain
may occur; in particular, there are no loops in our reduction rules
that may be performed an unspecified number of times. Thus, at the
heart of the term algebra lies a set of functions that represent this
chain of reductions which may be entered at an arbitrary place,
depending on the type of factor that has been added. As allocating
memory for the many intermediate terms that have to be generated while
a summand runs through this reduction chain would cause an excessive
number of expensive memory reclaiming steps (garbage collection), we
reduce dynamic memory allocation to a minimum by first copying
summands that have to be reduced to a statically allocated space of
sufficient size, which we call the `workbench'; term transformations
are directly applied to a term's copy on the workbench. The problem
that one reduction step may branch into many different individual
summands is then also easily dealt with: instead of a single workbench
area, we use multiple such areas, start by copying the summand to area
$\#0$, and whenever the calculation branches into multiple summands,
e.g. via
$\epsilon_{ijk}\epsilon_{imn}\rightarrow\delta_{jm}\delta_{kn}-\delta{jn}\delta_{km}$,
we copy the term that is currently processed in area $\#N$ to area
$\#(N+1)$ multiple times, once for every branch, add appropriate
modifications to it, and continue processing with the terms in place
$\#(N+1)$, one after the other. Eventually, when no more reductions
can be performed, we create a new summand out of the contents of the
workbench area we presently work at.

\begin{figure}
\begin{center}
\includegraphics[height=6cm]{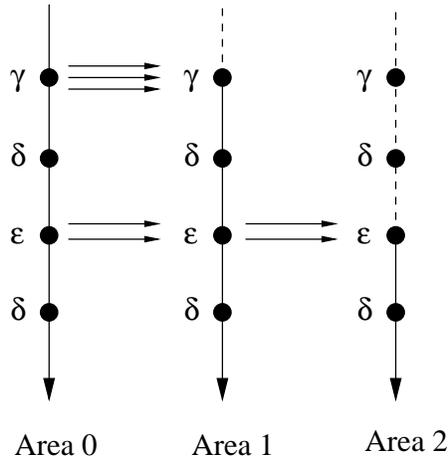}
\end{center}
\caption{\small Calculations on terms without dynamic memory management}
\label{fig:Workbench}
\end{figure}

This is depicted in diagram~\ref{fig:Workbench}. (Admittedly, this
description is slightly simplified.) The method employed here to chase
active parts through a term in order to avoid both unnecessary
checking whether some reduction rule can be applied as well as
dynamical memory management is quite general and can be applied in
every situation where reductions have to be performed on terms of
which one knows that they are derived from non-reducible terms by a
small number of local modifications. Indeed, it should even be
possible to implement a system that automatically compiles term
reduction rules to machine code that employs such a `waterfall'
scheme.

The drawback of the decision to branch graph generation to individual
summands obtained in $\epsilon\epsilon$ and $\tr\gamma$ reductions is
that some graph generation work is re-done spuriously that could be
avoided if one attributed sums, not products, to individual graphs. On
the other hand, this scheme helps cutting some branches in the graph
generation (to be discussed later). One should notice, however, that a
large number of vertices has to be used to construct a graph that
branches into many summands; as there are only at most eight vertices
in total, the amount of work that is unnecessarily multiplied is
limited to the tips of the graph of all possible choices. The most
important advantage if this scheme is, of course, that it is
comparatively easy to implement and debug.

As a further trick, one notes that besides the coefficients of
individual vertex sequences, and symmetrization factors $1/6$ that
come from $\Pi^\pm$, all the denominators that appear in the
calculation are small powers of two, with the numerators not becoming
excessively large. One can convince oneself that up to four loops,
floatingpoint calculations are exact on such numbers(!), so we can abuse
double precision floatingpoint numbers to represent coefficients for a
large part of the calculation and handle the other fractional factors
not of this form in a separate final step. Besides vastly simplified
memory management, this especially speeds up multiplication of
coefficients.

\subsubsection{On the Graph $\leftrightarrow$ Term interplay}

\noindent Planar graph generation drives the calculation.
Naively, one may consider to just memorize all the leg fusions during
graph generation and eventually do all of the calculation of the term
corresponding to a graph in the very last step. Alternatively, one may
do as much of the calculation as possible in every step throughout the
incremental generation of a graph. The advantage of the first approach
is that no term calculations at all have to be performed for such
intermediary graphs that cannot lead to a final graph of the desired
topology. The big disadvantage is that for different final graphs that
are derived from some common stem, the part of the calculation that
corresponds to this stem is done multiple times, once for each final
graph, although in principle it would have been sufficient to do it
only once.

The big advantage of the second approach is that we may be able to see
early that some stem cannot give any contribution anymore, e.g. if the
coefficient became zero due to occurrence of a factor $\epsilon_{iij}$
or a $\Pi^+\Pi^-$ projector combination, and we may save a lot of
expensive graph generation work for all graphs that are derived from
this stem. Unfortunately, we then have to do lots of spurious term
calculations for stems which will not produce a suitable final graph.

\begin{figure}
\begin{center}
\includegraphics[height=6cm]{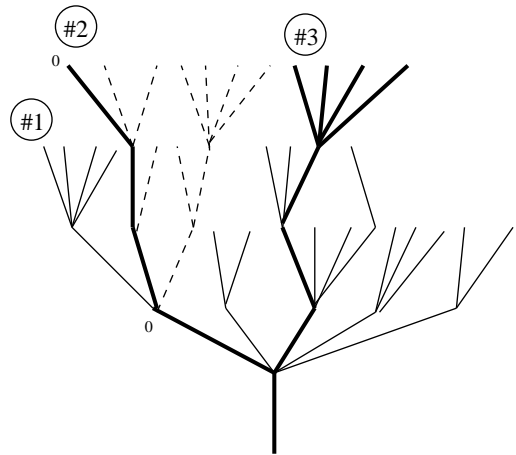}
\end{center}
\caption{\small On the value of lazy evaluation:
Every node corresponds to a partially constructed graph and its associated (lazy) term.
The root node is just the initial domain. Every link leads from a
graph to a graph which has been extended by one more vertex.
In (\#1), no algebraic calculations are performed, as graph generation
gets stuck with partial graphs that cannot be completed to final planar graphs.
In (\#2), a planar graph could be generated, and term algebra forces all the 
promises for calculations down to the root to be fulfilled (bold lines).
However, the result is zero, coming from a zero at level two that is discovered
in this step only. All the other graphs and partial graphs that would also derive from
this node are then not even generated, as the `known zero value' information is
used in graph generation. In (\#3), we get final nonzero contributions.
Note furthermore that intermediary results down the path to the root are only calculated once!}
\label{fig:Lazytree}
\end{figure}

Interestingly, it {\em is} possible to get the best of both approaches
by using so-called {\em lazy evaluation} (see
figure~\ref{fig:Lazytree}): instead of triggering calculations on
actual terms, we let the graph engine generate promises to do such
calculations if needed. Such a `promise' is basically a data structure
holding a flag whether the promise was fulfilled already, plus either
the value of this promise (if the flag says it indeed already was
fulfilled), or otherwise a function, plus arguments, that, when
evaluated, will produce that value. Once such a promise is requested
to be fulfilled (`forced'), the promise's flag will be set
accordingly, and the function and arguments will be replaced by the
promise's value, so that next time its value is needed, one can just
take it, without having to call the function again. Promises are
comparatively cheap to give -- this only requires a small amount of
consing. Conceptually, the idea behind such promises is quite similar
to the idea behind stock options. In practical applications, a
`promise' will typically promise to first require some other promises
it depends on to be fulfilled, and use their values to compute its own
value.

Here, we only carry promises for terms which depend on other promises
for terms through the depth-first graph generation until we actually
do generate a final graph. No expensive term calculations have been
done so far, especially not for futile partial graphs that cannot be
completed to a topologically interesting graph. Then, we take the
graph's promise for a term and force it. It may be that this is zero,
due to some promise in the stem producing a zero result, and we did
some small amount of unnecessary work in generating this
graph. However, when depth-first graph generation proceeds to the
other branches that share a stem whose term is zero with the present
graph's genealogy, we see that the promise for the term in the stem
already has been fulfilled and we can cut generating further graphs
from this particular stem.

One subtlety lies in the treatment of fermionic signs: the
determination of signs from fusion of fermionic legs has to be taken
out of the promise, as one has little control over the point in time
when a promise is being fulfilled, and the state of the flags denoting
which legs are unbound fermionic ones may be a very different one from
the state of those flags at the time when the promise was given (which
is the proper state to base sign decisions on).

\subsubsection{Implementation}

\noindent Due to memory access being a major bottleneck in today's CPUs,
and the hierarchical structure of caches in computer systems, one should take
some care not to unnecessarily waste space with data
representations. However, one can over-optimize here -- tight data
representations frequently come at the expense of more code. In the
present implementation, we use a single 8-bit byte (char) to represent
legs within the term algebra. The most significant bit encodes whether
a leg is of type $ijk$ or $abc$, the other bits encode the `name' of
the leg (a number). This works because fermionic $ABC$ legs can only
occur at the ends of chains of gamma matrices, which are stored in
special places. Leg type information is needed in two places: to
determine which extra number factor to introduce when reducing
$\delta^m_m$, and to determine whether a sign flip has to occur when
the rule $\gamma^a\Pi^\pm=\Pi^\mp\gamma^a$,
$\gamma^i\Pi^\pm=\Pi^\pm\gamma^i$ is used. Every chain of gamma
matrices in a term is encoded as a string (vector of bytes), where
byte~0 and~1 denote the names of the fermionic ends; the high-bits of
these two bytes encode whether the chain has an extra factor $\Pi^\pm$
at the left end as well as its particular type. All delta factors are
likewise encoded in a string holding pairs of bytes. Epsilon factors
$\epsilon_{ijk}$ combine three indices, hence we use one 30-bit signed
integer per factor, while we use a string to hold nine indices for
every $\epsilon_{m_1\ldots m_9}$ factor. As fermionic legs are just
those appearing inside gamma chains, hence easy to discern from
$\SO(3)$ $ijk$ and $\SO(6)$ $abc$ legs, we only have to discern the
latter two types. This we do by using the convention that all legs
given a number in the range $0,\ldots63$ (i.e. with non-set sixth bit)
are of type $a$, while those in the range $64\ldots126$ are of type
$i$.

On the workbench, we use extra-long arrays to hold these data and fill
pointers remembering how many entries are momentarily
valid. Furthermore, on the workbench, we have to hold additional
information about `hot' factors. A further subtlety concerns two
special types of indices: first, we get $ijk$ indices not introduced
by the legs of some vertex: we get these from the
$\gamma^{ijk}\epsilon_{ijk}$ contribution to $\Pi^\pm$ when closing a
gamma trace. Here, we take indices from the end of the numerical range
(with 127 being treated as special, as it also denotes `no index' in
graph generation) 126,125,\ldots, remembering how many such factors we
already introduced for the present graph. Second, as we want to split
the $a$ index to $Z,\bar Z, W,\bar W,Y,\bar Y$, we let indices in the
range $58\ldots63$ denote these special indices. As these split index
ranges are one-dimensional, they do not provide special handles
associating them with parts of the term; we do not need them here
anymore.

\subsection{On the Calculation}

\noindent
The size of this calculation is larger by orders of magnitude than any
other symbolic computation the authors are aware of. Such jumps in
complexity are quite generically accompanied by the discovery of new
effects that did not occur on smaller scales. In this particular case,
removing flaws from the process of running such large calculations
eventually turned out to be more demanding than removing bugs in the
code. The main reasons were:

\begin{enumerate}

\item Without a dedicated supercomputer time budget for this project,
 computation time had to be borrowed from a variety of different
 sources.

\item The job queuing systems used at some sites turned out
 to be well adapted to the needs of numerical simulations,
 but not symbolic computations.

\item Some especially hard parts of the calculation
 had to be split multiple times.

\item Perturbation theory calculations are notoriously hard to debug if
 the only sign of an error is that the overall sum of data from
 $O(10\,000)$ individual files, which take many CPU-years to produce,
 is nonsense.

\end{enumerate}

All in all, performing the actual calculation took about half a year.

\subsubsection{Performance}

\noindent Profiling a three-loop sample calculation shows that
with all of these tricks incorporated, the code spends most of the
time in graph generation, but there is no single hot spot where
further optimization may be useful. One might wonder whether it would
be possible to further speed up graph generation by adding more
heuristic checks that allow to determine early -- given a set of
remaining vertices that have to be added to a partially finished graph
-- whether there will be any way to eventually produce a graph with
only one connected component carrying exactly one cell with open
legs, the idea being that as soon as situations occur like legs of
type $A$ and $B$ that will not be bound anywhere anymore going to
different cells, one may stop graph generation early. Experiments have
shown that this is an important optimization if the term algebra is so
expensive that it slows down partial graph generation. As long as
there is only very little dynamic memory management going on inside
the graph engine, the extra effort to do such early checks is large
enough to destroy any advantage of this approach. Again, this
situation would change should reconfigurable logic become available on
mainstream CPUs. Then, one might be able to perform such checks at
very little extra cost.

In table~$\ref{tab:Sizes}$, we display the sizes of some calculations,
done in FORM and with \gemstone{} -- note that one should not try to
naively deduce a statement about relative quality of these symbolic
manipulation systems from such timings, as very different algorithms
are used for the different systems. Despite the not unimpressive size
of the number of non-zero contributions (roughly, the number of
diagrams), one should not be misled by them: the hard part of the work
lies in generating graphs, the vast majority of which turn out to give
no contribution. This can be glimpsed from the fact that average
\gemstone{} term generation speed drops from about $640/s$ for $ZZ$ at
three loops to about $20/s$ for the length-eleven chain at four loops,
but can be as high as $1050/s$ at three loops or even $5100/s$ at four
loops for individual sequences. If we assume that these `fluctuations'
in term generation speed are roughly in the ballpark of the square
root of the number of total (mostly non-permissible) graphs per
non-vanishing term, we may wildly guess that the total length-eleven
chain calculation at four-loop order would have had somewhat like half
a quadrillion possible Feynman graphs. At least, it is a funny
coincidence to note that by doing the estimate this way from a
`measurement', one arrives at the same number ($4.3\cdot 10^{14}$)
which is obtained by just multiplying the naive number of possible
combinations for the four-times-four-legs sequence for the largest
vertex class ($188^4$) with the number of different vertex sequences
($338\,834$). As this second estimate does not take such factors as
the initial chain length, multiple ways to add one vertex to a graph,
and variations in complexity between vertex sequences into account,
this is indeed only a coincidence.

{\small
\begin{table}
\ \\
\begin{tabular}{|llrlll|}
\hline
Initial & Loop  & Non-Zero           & Approx. Calc Time        & System&Determines\\
Chain   & Order & Terms\hfill        & (equiv. CPU-hours,       &       &Dilatation\\
        &       &                    & 2 GHz AMD Athlon)        &       &Operator\\
\hline
\hline
{\tiny ZZ}      & 2     & $1\,150$ & \hfill $1.3\cdot10^{-4}$  & \gemstone{}&No\\
{\tiny WWWWZZZ} & 2     & $4\,729$ & \hfill $6.0\cdot10^{-4}$  & \gemstone{}&Yes\\
{\tiny N.A.}    & 2     & N.A.     & \hfill $1.6\cdot10^{-2}$  & \mbox{\tt FORM}&Yes\\
\hline
\hline
{\tiny ZZ}      & 3     & $575\,916$ & \hfill $0.24$  & \gemstone{}&No\\
{\tiny WWZWZWZ} & 3     & $2\,764\,630$ & \hfill $1.6$  & \gemstone{}&Yes\\
{\tiny N.A.}       & 3  & N.A.  & \hfill $177$  & \mbox{\tt FORM}&Yes\\
\hline
\hline
{\tiny ZZ}      & 4     & $736\,439\,974$ & \hfill $1\,051$  & \gemstone{}&No\\
{\tiny WWZZ}    & 4     & $1\,601\,839\,076$         & \hfill $4\,170$  & \gemstone{}&Wrapping 1\\
{\tiny WZWZ}    & 4     & $1\,669\,740\,684$         & \hfill $4\,161$  & \gemstone{}&Wrapping 2\\
{\tiny WWZWZZ}    & 4     & $2\,777\,838\,350$       & \hfill $11\,560$  & \gemstone{}&No\\
{\tiny WWWZZZZZWWZ} & 4 & $6\,638\,616\,118$         & \hfill $88\,812$ & \gemstone{}&Upto Wrapping\\
\hline
\end{tabular}
\\
\ \\
\caption{Sizes of various calculations}
\label{tab:Sizes}
\end{table}
}

\subsubsection{Distributed Computing}

\noindent Evidently, such four-loop calculations
are far too demanding to be performed on a single
machine. Fortunately, they naturally decompose into many individual
independent parts -- one for each sequence of vertices -- and hence
are trivially parallelizable. (One may wonder how much time is wasted
by doing graph generation over and over again for vertex sequences
that share a long common beginning. With reasonable assumptions, one
can estimate this spurious effort to contribute only at most a few
per-cent to total calculation time.)

\begin{figure}
\begin{center}
\includegraphics[height=7cm]{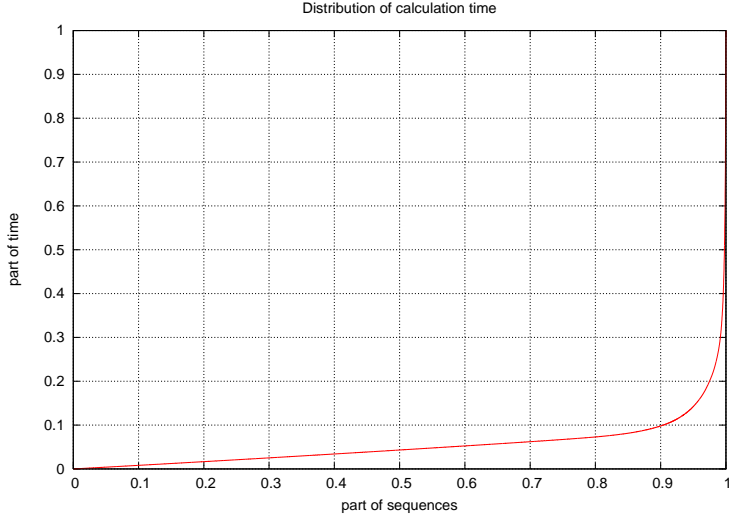}
\end{center}
\caption{\small Accumulated calculation time vs. part of sequences
 for $ZZ$ at three-loop order }
\label{GemstoneTiming}
\end{figure}

Profiling a three-loop calculation shows that while some vertex
sequences can be calculated much faster than others, there is no clear
jump in the distribution of calculation times of individual
pieces. Nevertheless, about 50\% of the work is done for 0.5\% of all
vertex sequences, while 90\% of all sequences take only 10\% of the
time. This is shown in figure~\ref{GemstoneTiming}.

It is tempting to try to provide additional mechanisms to further
parallelize work done on individual chains, and the nature of the
calculation with its many branchings suggests quite directly how one
may achieve this. As we do not know a priori where the hot spots are,
we want to be able to signal a running calculation to break itself up.
So far, this is only supported very rudimentarily: if a running
calculation process receives an alarm signal (SIGALRM, either sent
externally or set as an alarm timeout via a command-line option), the
signal handler sets a flag which tells the graph engine to record all
work that has to be done from that instant on in a list of deferred
calculations, which is then serialized, gzip-compressed,
base64-encoded, and written to disk.\footnote{There are indications of
problems with some versions of the Ocaml serializer, so this feature
should be used with care.} (One may regard this as an ad-hoc approach
to serializing a continuation.) The \gemstone{} program is able to
split a serialized calculation into multiple pieces of (hopefully)
similar size.

As serialization of a running computation is a comparatively hairy
action, we implemented as well a much simpler fallback method which
involves splitting vertex classes into two parts. After identification
of those pieces of the calculation that are hard to do, a complete
expansion of all such $N$ vertex sequences of length $L$ into
$N\cdot2^L$ parts is indeed sufficient to bring the calculation time
of every individual piece down to less than one week.

\subsubsection{Security}

\noindent
So far, the \gemstone{} system has only been used in controlled
environments. Technically, it would be easy to extend the system in
such a way that donated computer time from individual volunteers can
be employed; the source already contains some first support for
exchanging data via HTTP requests directly from the client. This
raises some concerns in conjunction with both credibility of the
overall result and security of participants' and organizers'
computers. First of all, when even only one single machine produces an
erroneous partial result -- be it out of mischief or hardware
unreliability, e.g. due to bad RAM or CPU overclocking, this will not
be noticed until the final result is available. Hence, if such a
scheme is feasible at all, then at least, every part of the
calculation would have to be done twice by independent
volunteers. Second, the problem of handling unexpectedly large pieces
of the calculation has to be addressed. While it would be possible to
let a client machine produce an image of the current state of its
calculation and send it back so that it can be broken up and sent to
further volunteers, this must not be permitted. Besides the
occasionally prohibitive size of serialized calculations, a major
concern is that the organizer of such a distributed calculation has
little or no control over internal safety checks performed by the
serializer. Thus, a mischievous participant may send in a specially
manipulated data file which, upon re-serialization, may corrupt the
Ocaml heap and in the worst case lead to arbitrary code execution for
a different volunteer. The perhaps safest way to address this problem
is to accept some redundant work and recursively decompose any large
problem for which a client sends the information that it cannot be
done within a given time-frame into smaller sub-problems by splitting
vertex classes. This also has the big advantage that one may exert
full control over the locations of these splits -- hardly possible if
they are mostly controlled by a timeout.

\subsubsection{On the Code}

\noindent
While the main program is written in Objective Caml for portability,
some of the Ocaml source files are generated by Perl and Lisp,
especially if they had to be translated from Mathematica. (For some
pieces for which speed is not essential, the machine generated code is
quite ad-hoc and certainly not as efficient as it might be.) Thus,
building the entire system will unfortunately require quite a large
software development infrastructure.

As this system is work in progress, there are quite some places where
the code is just as clean as it has to be in order to be maintainable;
depending on further applications of this codebase, it may or may not
become desirable to do some extensive re-structurings especially of
the documentation in order to enhance readability.

\subsection{Five Loops}

\noindent Can this head-on approach be used to do the five loop calculation?
One does not have to consider initial chains of length greater
than~$13$ in order to fix the five-loop dilatation operator. As such a
calculation may introduce at most nine extra indices from three
$\epsilon^{ijk}$ via fermion loops of odd length, we would have to
deal with at most $3\cdot10+9+13=42$ indices, well below the built-in
maximum of $63$ in \gemstone. Furthermore, if one is interested only
in the result and does not care about final states with fermions,
which will cancel in the total sum anyway, small changes to
\gemstone{} that drop these terms would make it run with a very small
memory footprint, hence maybe even suited for networked video game
consoles.

While \gemstone{} not only is considerable faster than any other
method used so far, but also scales better, the relative computational
effort for three loops is $~3\,000$ times larger than for two loops,
and the effort for four loops is again larger by a factor $~60\,000$.
So it is perhaps not unreasonable to estimate that the size of the
five-loop calculation should be at least one million times larger than
the largest ones presented here. Hence, one would probably need
something in the range of 100-500 million fast machines to do it. This
is roughly in the ballpark of the total technical capability of
mankind today.

\subsection{Results}

\noindent For the sake of reproducibility of our methods and findings,
we include raw results of all the large four-loop computer
calculations, as well as for the calculations determining the
dilatation operator at two and three loops.

\bigbreak

\noindent
$WWWWZZZ$ at two loops:
\begin{equation}
\begin{array}{lr}
WWWWZZZ& -44 \\
WWWZWZZ& 26 \\
WWWZZWZ& 26 \\
WWZWWZZ& -8 \\
\end{array}
\end{equation}

\noindent
$WWZWZWZ$ at three loops:
\begin{equation}
\begin{array}{lr}
WWWWZZZ&16\\
WWWZWZZ&-612\\
WWWZZWZ&-612\\
WWZWWZZ&-1464\\
WWZWZWZ&2672
\end{array}
\end{equation}

\noindent
$ZZ$ at four loops:
\begin{equation}
0
\end{equation}

\noindent
$WZWZ$ at four loops:
\begin{equation} \label{eqn:gemstoneZWZW}
\begin{array}{lr}
WWZZ& 88224 \\
WZWZ& -88224
\end{array}
\end{equation}

\noindent
$WWZZ$ at four loops:
\begin{equation} \label{eqn:gemstoneZZWW}
\begin{array}{lr}
WWZZ& -44112 \\
WZWZ& 44112
\end{array} 
\end{equation}

\noindent
$WWZWZZ$ at four loops:
\begin{equation} \label{eqn:gemstoneWWZWZZ}
\begin{array}{lr}
WZWZWZ& 22291\\
WWWZZZ& 12803\\
WWZWZZ& -73356\\ 
WWZZWZ& 38262
\end{array}
\end{equation}

$WWWZZZZZWWZ$ at four loops:
\begin{equation} \label{eqn:gemstone11}
\begin{array}{lr}
WWWWWZZZZZZ& -3448 \\
WWWWZWZZZZZ& -5640 \\
WWWWZZWZZZZ& 52 \\
WWWWZZZWZZZ& 52 \\
WWWWZZZZWZZ& -1000 \\
WWWWZZZZZWZ& 22399 \\
WWWZWWZZZZZ& 22615 \\
WWWZWZWZZZZ& 24 \\
WWWZWZZZZWZ& -80 \\
WWWZZWWZZZZ& 76 \\
WWWZZWZZWZZ& -32 \\
WWWZZWZZZWZ& 928 \\
WWWZZZWWZZZ& -28 \\
WWWZZZWZWZZ& 376 \\
WWWZZZWZZWZ& -5012 \\
WWWZZZZWWZZ& -2968 \\
WWWZZZZWZWZ& 21819 \\
WWWZZZZZWWZ& -63260 \\
WWZWWZWZZZZ& 22195 \\
WWZWWZZWZZZ& -5064 \\
WWZWWZZZWZZ& 928 \\
WWZWWZZZZWZ& -104 \\
WWZWZWWZZZZ& -5168 \\
WWZWZWZWZZZ& 432 \\
WWZWZWZZWZZ& -32 \\
WWZWZZWWZZZ& -28 \\
WWZWZZZWWZZ& -32
\end{array}
\end{equation}

This last result receives contributions from individual sequences such
as e.g.
\[
\begin{array}{lr}
WWWZZZZWZWZ&    3747744\\
WWWWZZZZZWZ&    3747744\\
WWWZWWZZZZZ&    3747744\\
WWZWWZWZZZZ&    3747744\\
WWWZZZZZWWZ& -292115824
\end{array}
\]
(from the sequence
$-\frac{7}{4}V_{1,+1}V_{1,-1}V_{1,+1}V_{2,+0}V_{1,+1}V_{1,-1}V_{1,-1}$)
or
\[
\begin{array}{lr}
WWWZZZZZWWZ&2673/560
\end{array}
\]
(from the sequence
$-\frac{1}{6720}V_{1,+3}V_{1,+1}V_{1,+3}V_{2,+1}V_{1,-3}V_{1,-3}V_{1,-2}$).
Hence the fact that the result contains only small, allowed, integer
contributions should be regarded as a highly nontrivial first
indication of its correctness.


\section{Perturbation expansion of effective Hamiltonian} \label{sec:perturbation-theory}

\noindent
The effective Hamiltonian acting on the $\su(2)$ subsector is given by a perturbative expansion
\be \label{eqn:eff-ham}
  T = \sum_{k=0}^\infty T_{2k} \qquad \mbox{with } T_{2k} \propto \frac{1}{M^{3k-1}} \; ,
\ee
where $T_{2k}$ is called the $k$-loop contribution. As building blocks we need an operator $P_E$ which projects onto the subspace of states with free energy $E$, i.e.
\be
  P_E \ket{F} = \delta_{EF} \ket{F} \quad \mbox{for eigenstate $\ket{F}$ of $H_0$ with eigenvalue $F$}  
\ee
and a ``propagator''
\be
  \Delta_E = \sum_{F\not=E} \frac{P_F}{E - F} \; .
\ee
With these definitions, the terms in \eqref{eqn:eff-ham} read as follows:
\begin{align}
T_0 = &\ H_0 \\
T_2 = &\ \sum_E P_E \bigl[ V_1 \Delta_E V_1 + V_2 \bigr] P_E \\
T_4 = &\ \sum_E P_E \bigl[ V_1 \Delta_E V_1 \Delta_E V_1 \Delta_E V_1 \nonumber \\[-4mm]
      & \qquad           + V_1 \Delta_E V_1 \Delta_E V_2
                         + V_1 \Delta_E V_2 \Delta_E V_1
                         + V_2 \Delta_E V_1 \Delta_E V_1
                         + V_2 \Delta_E V_2 \nonumber \\
      & \qquad           - V_1 \Delta_E^2 V_1 P_E T_2 \bigr] P_E \\ 
T_6 = &\ \sum_E P_E \bigl[ V_1 \Delta_E V_1 \Delta_E V_1 \Delta_E V_1 \Delta_E V_1 \Delta_E V_1 \nonumber \\[-4mm]
      & \qquad           + V_1 \Delta_E V_1 \Delta_E V_1 \Delta_E V_1 \Delta_E V_2
                         + V_1 \Delta_E V_1 \Delta_E V_1 \Delta_E V_2 \Delta_E V_1 \nonumber \\
      & \qquad           + V_1 \Delta_E V_1 \Delta_E V_2 \Delta_E V_1 \Delta_E V_1
                         + V_1 \Delta_E V_2 \Delta_E V_1 \Delta_E V_1 \Delta_E V_1 \nonumber \\
      & \qquad           + V_2 \Delta_E V_1 \Delta_E V_1 \Delta_E V_1 \Delta_E V_1 \nonumber \\
      & \qquad           + V_1 \Delta_E V_1 \Delta_E V_2 \Delta_E V_2
                         + V_1 \Delta_E V_2 \Delta_E V_1 \Delta_E V_2
                         + V_1 \Delta_E V_2 \Delta_E V_2 \Delta_E V_1 \nonumber \\
      & \qquad           + V_2 \Delta_E V_1 \Delta_E V_1 \Delta_E V_2
                         + V_2 \Delta_E V_1 \Delta_E V_2 \Delta_E V_1
                         + V_2 \Delta_E V_2 \Delta_E V_1 \Delta_E V_1 \nonumber \\
      & \qquad           + V_2 \Delta_E V_2 \Delta_E V_2 \nonumber \\
      & \qquad           - \bigl( V_1 \Delta_E^2 V_1 \Delta_E V_1 \Delta_E V_1
                                + V_1 \Delta_E V_1 \Delta_E^2 V_1 \Delta_E V_1
                                + V_1 \Delta_E V_1 \Delta_E V_1 \Delta_E^2 V_1 \nonumber \\
      & \qquad\quad\;           + V_1 \Delta_E^2 V_1 \Delta_E V_2
                                + V_1 \Delta_E^2 V_2 \Delta_E V_1
                                + V_2 \Delta_E^2 V_1 \Delta_E V_1 \nonumber \\
      & \qquad\quad\;           + V_1 \Delta_E V_1 \Delta_E^2 V_2
                                + V_1 \Delta_E V_2 \Delta_E^2 V_1
                                + V_2 \Delta_E V_1 \Delta_E^2 V_1 \nonumber \\
      & \qquad\quad\;           + V_2 \Delta_E^2 V_2 
                           \bigr) P_E T_2 \nonumber \\
      & \qquad           + V_1 \Delta_E^3 V_1 P_E T_2 P_E T_2 - V_1 \Delta_E^2 V_1 P_E T_4 \bigr] P_E
\end{align}

\be \label{eqn:eff-ham-t8}
{\scriptsize
\begin{split}
T_8 = \sum_E P \Bigl[ \; & 
    V_1 \Delta V_1 \Delta V_1 \Delta V_1 \Delta V_1 \Delta V_1 \Delta V_1 \Delta V_1 \\[-3.5mm]
& + V_1 \Delta V_1 \Delta V_1 \Delta V_1 \Delta V_1 \Delta V_1 \Delta V_2
  + V_1 \Delta V_1 \Delta V_1 \Delta V_1 \Delta V_1 \Delta V_2 \Delta V_1 \\
& + V_1 \Delta V_1 \Delta V_1 \Delta V_1 \Delta V_2 \Delta V_1 \Delta V_1
  + V_1 \Delta V_1 \Delta V_1 \Delta V_2 \Delta V_1 \Delta V_1 \Delta V_1 \\
& + V_1 \Delta V_1 \Delta V_2 \Delta V_1 \Delta V_1 \Delta V_1 \Delta V_1
  + V_1 \Delta V_2 \Delta V_1 \Delta V_1 \Delta V_1 \Delta V_1 \Delta V_1 \\
& + V_2 \Delta V_1 \Delta V_1 \Delta V_1 \Delta V_1 \Delta V_1 \Delta V_1 \\
& + V_1 \Delta V_1 \Delta V_1 \Delta V_1 \Delta V_2 \Delta V_2
  + V_1 \Delta V_1 \Delta V_1 \Delta V_2 \Delta V_1 \Delta V_2
  + V_1 \Delta V_1 \Delta V_1 \Delta V_2 \Delta V_2 \Delta V_1 \\
& + V_1 \Delta V_1 \Delta V_2 \Delta V_1 \Delta V_1 \Delta V_2
  + V_1 \Delta V_1 \Delta V_2 \Delta V_1 \Delta V_2 \Delta V_1
  + V_1 \Delta V_1 \Delta V_2 \Delta V_2 \Delta V_1 \Delta V_1 \\
& + V_1 \Delta V_2 \Delta V_1 \Delta V_1 \Delta V_1 \Delta V_2
  + V_1 \Delta V_2 \Delta V_1 \Delta V_1 \Delta V_2 \Delta V_1
  + V_1 \Delta V_2 \Delta V_1 \Delta V_2 \Delta V_1 \Delta V_1 \\
& + V_1 \Delta V_2 \Delta V_2 \Delta V_1 \Delta V_1 \Delta V_1
  + V_2 \Delta V_1 \Delta V_1 \Delta V_1 \Delta V_1 \Delta V_2
  + V_2 \Delta V_1 \Delta V_1 \Delta V_1 \Delta V_2 \Delta V_1 \\
& + V_2 \Delta V_1 \Delta V_1 \Delta V_2 \Delta V_1 \Delta V_1
  + V_2 \Delta V_1 \Delta V_2 \Delta V_1 \Delta V_1 \Delta V_1
  + V_2 \Delta V_2 \Delta V_1 \Delta V_1 \Delta V_1 \Delta V_1 \\
& + V_1 \Delta V_1 \Delta V_2 \Delta V_2 \Delta V_2
  + V_1 \Delta V_2 \Delta V_1 \Delta V_2 \Delta V_2
  + V_1 \Delta V_2 \Delta V_2 \Delta V_1 \Delta V_2 \\
& + V_1 \Delta V_2 \Delta V_2 \Delta V_2 \Delta V_1
  + V_2 \Delta V_1 \Delta V_1 \Delta V_2 \Delta V_2
  + V_2 \Delta V_1 \Delta V_2 \Delta V_1 \Delta V_2 \\
& + V_2 \Delta V_1 \Delta V_2 \Delta V_2 \Delta V_1
  + V_2 \Delta V_2 \Delta V_1 \Delta V_1 \Delta V_2
  + V_2 \Delta V_2 \Delta V_1 \Delta V_2 \Delta V_1 \\
& + V_2 \Delta V_2 \Delta V_2 \Delta V_1 \Delta V_1 \\
& + V_2 \Delta V_2 \Delta V_2 \Delta V_2 \\
& - \bigl(
           V_1 \Delta V_1 \Delta V_1 \Delta V_1 \Delta V_1 \Delta^2 V_1
         + V_1 \Delta V_1 \Delta V_1 \Delta V_1 \Delta^2 V_1 \Delta V_1
         + V_1 \Delta V_1 \Delta V_1 \Delta^2 V_1 \Delta V_1 \Delta V_1 \\
&\quad\; + V_1 \Delta V_1 \Delta^2 V_1 \Delta V_1 \Delta V_1 \Delta V_1
         + V_1 \Delta^2 V_1 \Delta V_1 \Delta V_1 \Delta V_1 \Delta V_1 \\
&\quad\; + V_1 \Delta V_1 \Delta V_1 \Delta V_1 \Delta^2 V_2
         + V_1 \Delta V_1 \Delta V_1 \Delta V_2 \Delta^2 V_1
         + V_1 \Delta V_1 \Delta V_1 \Delta^2 V_1 \Delta V_2 \\
&\quad\; + V_1 \Delta V_1 \Delta V_1 \Delta^2 V_2 \Delta V_1
         + V_1 \Delta V_1 \Delta V_2 \Delta V_1 \Delta^2 V_1
         + V_1 \Delta V_1 \Delta V_2 \Delta^2 V_1 \Delta V_1 \\
&\quad\; + V_1 \Delta V_1 \Delta^2 V_1 \Delta V_1 \Delta V_2
         + V_1 \Delta V_1 \Delta^2 V_1 \Delta V_2 \Delta V_1
         + V_1 \Delta V_1 \Delta^2 V_2 \Delta V_1 \Delta V_1 \\
&\quad\; + V_1 \Delta V_2 \Delta V_1 \Delta V_1 \Delta^2 V_1
         + V_1 \Delta V_2 \Delta V_1 \Delta^2 V_1 \Delta V_1
         + V_1 \Delta V_2 \Delta^2 V_1 \Delta V_1 \Delta V_1 \\
&\quad\; + V_1 \Delta^2 V_1 \Delta V_1 \Delta V_1 \Delta V_2
         + V_1 \Delta^2 V_1 \Delta V_1 \Delta V_2 \Delta V_1
         + V_1 \Delta^2 V_1 \Delta V_2 \Delta V_1 \Delta V_1 \\
&\quad\; + V_1 \Delta^2 V_2 \Delta V_1 \Delta V_1 \Delta V_1
         + V_2 \Delta V_1 \Delta V_1 \Delta V_1 \Delta^2 V_1
         + V_2 \Delta V_1 \Delta V_1 \Delta^2 V_1 \Delta V_1 \\
&\quad\; + V_2 \Delta V_1 \Delta^2 V_1 \Delta V_1 \Delta V_1
         + V_2 \Delta^2 V_1 \Delta V_1 \Delta V_1 \Delta V_1 \\
&\quad\; + V_1 \Delta V_1 \Delta V_2 \Delta^2 V_2
         + V_1 \Delta V_1 \Delta^2 V_2 \Delta V_2
         + V_1 \Delta V_2 \Delta V_1 \Delta^2 V_2
         + V_1 \Delta V_2 \Delta V_2 \Delta^2 V_1 \\
&\quad\; + V_1 \Delta V_2 \Delta^2 V_1 \Delta V_2
         + V_1 \Delta V_2 \Delta^2 V_2 \Delta V_1
         + V_1 \Delta^2 V_1 \Delta V_2 \Delta V_2
         + V_1 \Delta^2 V_2 \Delta V_1 \Delta V_2 \\
&\quad\; + V_1 \Delta^2 V_2 \Delta V_2 \Delta V_1
         + V_2 \Delta V_1 \Delta V_1 \Delta^2 V_2
         + V_2 \Delta V_1 \Delta V_2 \Delta^2 V_1
         + V_2 \Delta V_1 \Delta^2 V_1 \Delta V_2 \\
&\quad\; + V_2 \Delta V_1 \Delta^2 V_2 \Delta V_1
         + V_2 \Delta V_2 \Delta V_1 \Delta^2 V_1
         + V_2 \Delta V_2 \Delta^2 V_1 \Delta V_1
         + V_2 \Delta^2 V_1 \Delta V_1 \Delta V_2 \\
&\quad\; + V_2 \Delta^2 V_1 \Delta V_2 \Delta V_1
         + V_2 \Delta^2 V_2 \Delta V_1 \Delta V_1 
         + V_2 \Delta V_2 \Delta^2 V_2
         + V_2 \Delta^2 V_2 \Delta V_2
    \bigr) P T_2 \\
& + \bigl(
           V_1 \Delta V_1 \Delta V_1 \Delta^3 V_1
         + V_1 \Delta V_1 \Delta^2 V_1 \Delta^2 V_1
         + V_1 \Delta V_1 \Delta^3 V_1 \Delta V_1
         + V_1 \Delta^2 V_1 \Delta V_1 \Delta^2 V_1 \\
&\quad\; + V_1 \Delta^2 V_1 \Delta^2 V_1 \Delta V_1
         + V_1 \Delta^3 V_1 \Delta V_1 \Delta V_1 \\
&\quad\; + V_1 \Delta V_1 \Delta^3 V_2
         + V_1 \Delta V_2 \Delta^3 V_1
         + V_1 \Delta^2 V_1 \Delta^2 V_2
         + V_1 \Delta^2 V_2 \Delta^2 V_1 \\
&\quad\; + V_1 \Delta^3 V_1 \Delta V_2
         + V_1 \Delta^3 V_2 \Delta V_1
         + V_2 \Delta V_1 \Delta^3 V_1
         + V_2 \Delta^2 V_1 \Delta^2 V_1 \\
&\quad\; + V_2 \Delta^3 V_1 \Delta V_1
         + V_2 \Delta^3 V_2
    \bigr) P T_2 P T_2 \\
& - \bigl( V_1 \Delta V_1 \Delta V_1 \Delta^2 V_1
         + V_1 \Delta V_1 \Delta^2 V_1 \Delta V_1
         + V_1 \Delta^2 V_1 \Delta V_1 \Delta V_1 \\
&\quad\; + V_1 \Delta V_1 \Delta^2 V_2
         + V_1 \Delta V_2 \Delta^2 V_1
         + V_1 \Delta^2 V_1 \Delta V_2
         + V_1 \Delta^2 V_2 \Delta V_1
         + V_2 \Delta V_1 \Delta^2 V_1
         + V_2 \Delta^2 V_1 \Delta V_1 \\
&\quad\; + V_2 \Delta^2 V_2 \bigr) P T_4 \\
& - V_1 \Delta^4 V_1 P T_2 P T_2 P T_2 \\
& + V_1 \Delta^3 V_1 P \bigl(T_4 P T_2 + T_2 P T_4 \bigr) \\[-1mm]
& - V_1 \Delta^2 V_1 P T_6 \; \Bigr] P
\end{split}
}
\ee
\newpage

\noindent For the actual computation it is convenient to split up the vertices $V_1$ and $V_2$ into parts of definite energy shift
\be
  \comm{H_0}{V_{i,\delta E}} = \delta E \, V_{i,\delta E} \qquad i=1,2 \; . 
\ee
Since this implies
\begin{align}
  \Delta_E V_{i,\delta E} & = V_{i,\delta E} \Delta_{E-\delta E} \\
  P_E V_{i,\delta E}      & = V_{i,\delta E} P_{E - \delta E}
\end{align}
we are able to remove all projectors and propagators from the formulas above by moving them to the right end and using $P_E P_F = \delta_{EF} P_E$.

The PWMT vertices \eqref{eqn:matrix-theory} have components
\be
  V_1 = \sum_{\textstyle \atopfrac{k=-3}{k\not=0}}^3 V_{1,k M} \quad \mbox{and} \quad V_2 = \sum_{k=-4}^4 V_{2,k M}
\ee
where pieces with positive and negative energy shift are related by $\left(V_{i,\delta E}\right)^\dag = V_{i,-\delta E}$. We now list the vertices as they read after 
\emph{planar normal ordering}, i.e. all planar self-contractions of the vertices are already taken into account:
\begin{align}
 V_{1,M}  = & - \tfrac{3}{\sqrt{8M}} \eps_{ijk} \tr b_i^\dag b_j^\dag b_k
              + \tfrac{2i}{\sqrt{M}} (\gamma_a)_{\alpha\beta} \tr \theta_\alpha^\dag \theta_\beta^\dag a_a
              + i\sqrt{\tfrac{2}{M}} (\gamma_i)_{\alpha\beta} \tr \theta_\alpha^\dag \comm{b_i^\dag}{\theta_\beta} \\[4mm]
 V_{1,2M} = & - \tfrac{2i}{\sqrt{M}} (\gamma_a)_{\alpha\beta} \tr \theta_\alpha^\dag \theta_\beta^\dag a_a^\dag \\[4mm]
 V_{1,3M} = & + \tfrac{1}{\sqrt{8M}} \eps_{ijk} \tr b_i^\dag b_j^\dag b_k^\dag \\[4mm]
 V_{2,0}  = &   \tfrac{99N^2}{4} + 13N \tr a_a^\dag a_a + 7N \tr b_i^\dag b_i \nonumber \\
            & - \tfrac{1}{2M^2} \tr \comm{b_i^\dag}{b_j} \comm{b_i^\dag}{b_j}
              + \tfrac{1}{8M^2} \tr \comm{b_i^\dag}{b_i} \comm{b_j^\dag}{b_j}
              - \tfrac{1}{4M^2} \tr \comm{b_i^\dag}{b_j^\dag} \comm{b_i}{b_j} \nonumber \\
            & - \tfrac{2}{M^2}  \tr \comm{a_a^\dag}{a_b} \comm{a_a^\dag}{a_b}
              + \tfrac{1}{2M^2} \tr \comm{a_a^\dag}{a_a} \comm{a_b^\dag}{a_b}
              - \tfrac{1}{M^2}  \tr \comm{a_a^\dag}{a_b^\dag} \comm{a_a}{a_b} \nonumber \\
            & - \tfrac{1}{2M^2} \tr \comm{a_a^\dag}{b_i^\dag} \comm{a_a}{b_i}
              - \tfrac{1}{2M^2} \tr \comm{a_a^\dag}{b_i} \comm{a_a}{b_i^\dag} \\[4mm]
 V_{2,M}  = & - \tfrac{13N}{2}  \tr a_a^\dag a_a^\dag
              + \tfrac{1}{M^2}  \tr \comm{a_a^\dag}{a_b^\dag} \comm{a_a^\dag}{a_b} \nonumber \\
            & - \tfrac{1}{4M^2} \tr \comm{a_a}{b_i^\dag} \comm{a_a}{b_i^\dag}
              + \tfrac{1}{2M^2} \tr \comm{a_a^\dag}{b_i^\dag} \comm{a_a^\dag}{b_i} \\[4mm]
 V_{2,2M} = & - \tfrac{7N}{2}    \tr b_i^\dag b_i^\dag
              + \tfrac{1}{4M^2} \tr \comm{b_i^\dag}{b_j^\dag} \comm{b_i^\dag}{b_j} \nonumber \\
            & - \tfrac{1}{4M^2} \tr \comm{a_a^\dag}{a_b^\dag} \comm{a_a^\dag}{a_b^\dag}
              + \tfrac{1}{2M^2} \tr \comm{a_a^\dag}{b_i^\dag} \comm{a_a}{b_i^\dag} \\[4mm]
 V_{2,3M} = & - \tfrac{1}{4M^2} \tr \comm{a_a^\dag}{b_i^\dag} \comm{a_a^\dag}{b_i^\dag} \\[4mm]
 V_{2,4M} = & - \tfrac{1}{4M^2} \tr \comm{b_i^\dag}{b_j^\dag} \comm{b_i^\dag}{b_j^\dag}
\end{align} 

These are the building blocks for planar graphs (see Fig.~\ref{fig:GraphForming}), e.g.
\psfrag{alpha}{$\alpha$}
\psfrag{beta}{$\beta$}
\psfrag{iii}{$i$}
\be
  \tr \theta_\alpha^\dag b_i^\dag \theta_\beta = \raisebox{-6mm}{\includegraphics*{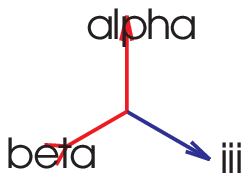}}
\ee
The arrows represent clockwise the oscillators in a trace, creation operators are out-going arrows and annihilation operators are in-going ones, the type of field 
determines the coloring. In the actual computer implementation we have split the $\so(6)$ index $a$ according to \eqref{eqn:so6-redefined} into six different colors.


\section{Higher charges} \label{sec:higher-charges}

\noindent
We now give the next two charges \eqref{eqn:higher-charges} beyond the spin-chain Hamiltonian $Q_2$ \eqref{eqn:spin-chain-hamiltonian},\eqref{eqn:dilop-renormalized}. 
They are not uniquely determined from demanding that they commute with each other and with the Hamiltonian. The remaining freedom has been fixed in order to match the 
eigenvalue formulas \eqref{eqn:eigenvalue-formulas}. 
\begin{align}
Q_{3,2} =\ 
& i( \{0,1\} - \{1,0\} ) \\[3mm]
Q_{3,4} =\ 
& -8i( \{0,1\} - \{1,0\} )
  +2i( \{0,1,2\} - \{2,1,0\} ) \\[3mm]
Q_{3,6} =\ 
& 73i( \{0,1\} - \{1,0\} )
  -28i( \{0,1,2\} - \{2,1,0\} ) \nonumber \\
& -2i( \{0,1,3\} + \{0,2,3\} - \{0,3,2\} - \{1,0,3\} ) \nonumber \\
& +5i( \{0,1,2,3\} - \{3,2,1,0\} )
  -i( \{0,2,1,3\} - \{1,0,3,2\} )  \nonumber \\
& -i( \{0,1,3,2\} - \{0,3,2,1\} + \{1,0,2,3\} - \{2,1,0,3\} ) \\[3mm]
Q_{3,8} =\ 
& -(751+\alpha)i( \{0,1\} - \{1,0\} )
  +\tfrac{1}{2}i(701+4\alpha)( \{0,1,2\} - \{2,1,0\} )  \nonumber \\
& +\tfrac{199}{4}i( \{0,1,3\} + \{0,2,3\} - \{0,3,2\} - \{1,0,3\} )  \nonumber \\
& -i(106+\alpha)( \{0,1,2,3\} - \{3,2,1,0\} )
  +19i( \{0,2,1,3\} - \{1,0,3,2\} )  \nonumber \\
& +\tfrac{1}{2}i(41-\alpha)( \{0,1,3,2\} - \{0,3,2,1\} + \{1,0,2,3\} - \{2,1,0,3\} )  \nonumber \\
& +2i( \{0,1,4\} + \{0,3,4\} - \{0,4,3\} - \{1,0,4\} )
  -4i( \{0,1,3,4\} - \{1,0,4,3\} )  \nonumber \\
& -4i( \{0,1,2,4\} + \{0,2,3,4\} - \{0,4,3,2\} - \{2,1,0,4\} )  \nonumber \\
& +14i( \{0,1,2,3,4\} - \{4,3,2,1,0\} )
  +i\alpha( \{0,3,2,1,4\} - \{1,0,2,4,3\} )  \nonumber \\
& +\tfrac{1}{4}i(1+2\alpha)( \{0,2,1,3,2\} + \{1,0,2,1,3\} - \{1,0,3,2,1\} - \{2,1,0,3,2\} )  \nonumber \\
& -\tfrac{1}{2}i(6+\alpha)( \{0,1,3,2,4\} + \{0,2,1,3,4\} - \{1,0,4,3,2\} - \{2,1,0,4,3\} )  \nonumber \\
& -\tfrac{1}{2}i(6-\alpha)( \{0,1,2,4,3\} - \{0,4,3,2,1\} + \{1,0,2,3,4\} - \{3,2,1,0,4\} )
\end{align}
\begin{align}
Q_{4,2} =\ 
& -\tfrac{4}{3} \{\}
  +\tfrac{8}{3} \{0\}
  -\tfrac{2}{3}( \{0,1\} + \{1,0\} ) \nonumber \\
& -\tfrac{2}{3}( \{0,2,1\} + \{1,0,2\} )
  +\tfrac{2}{3}( \{0,1,2\} + \{2,1,0\} ) \\[3mm]
Q_{4,4} =\ 
& 20 \{\}
  -\tfrac{124}{3} \{0\}
  +\tfrac{34}{3}( \{0,1\} + \{1,0\} )
  +\tfrac{8}{3} \{0,2\} \nonumber \\
& -\tfrac{26}{3}( \{0,1,2\} + \{2,1,0\} )
  +\tfrac{22}{3}( \{0,2,1\} + \{1,0,2\} )
  -\tfrac{4}{3} \{1,0,2,1\} \nonumber \\
& +2( \{0,1,2,3\} + \{3,2,1,0\} )
  -\tfrac{2}{3}( \{0,2,1,3\} + \{1,0,3,2\} ) \nonumber \\
& -\tfrac{2}{3}( \{0,1,3,2\} + \{0,3,2,1\} + \{1,0,2,3\} + \{2,1,0,3\} ) \\[3mm]
Q_{4,6} =\ 
& -252 \{\}
  +\tfrac{1604}{3} \{0\}
  -154( \{0,1\} + \{1,0\} )
  -\tfrac{160}{3} \{0,2\} \nonumber \\
& +\tfrac{332}{3}( \{0,1,2\} + \{2,1,0\} )
  -\tfrac{244}{3}( \{0,2,1\} + \{1,0,2\} )
  +\tfrac{64}{3} \{1,0,2,1\} \nonumber \\
& -\tfrac{20}{3} \{0,3\}
  +4( \{0,1,3\} + \{0,2,3\} + \{0,3,2\} + \{1,0,3\} ) \nonumber \\
& -38( \{0,1,2,3\} + \{3,2,1,0\} )
  +\tfrac{38}{3}( \{0,2,1,3\} + \{1,0,3,2\} ) \nonumber \\
& +\tfrac{34}{3}( \{0,1,3,2\} + \{0,3,2,1\} + \{1,0,2,3\} + \{2,1,0,3\} ) \nonumber \\
& +\tfrac{4}{3}( \{0,1,4,3\} + \{1,0,3,4\} )
  -\tfrac{4}{3}( \{0,1,3,4\} + \{1,0,4,3\} ) \nonumber \\
& +\tfrac{4}{3}( \{0,2,1,4\} + \{0,2,4,3\} + \{0,3,2,4\} + \{1,0,2,4\} ) \nonumber \\
& -\tfrac{4}{3}( \{0,1,2,4\} + \{0,2,3,4\} + \{0,4,3,2\} + \{2,1,0,4\} ) \nonumber \\
& +6( \{0,1,2,3,4\} + \{4,3,2,1,0\} )
  +\tfrac{2}{3}( \{0,3,2,1,4\} + \{1,0,2,4,3\} ) \nonumber \\
& +\tfrac{2}{3}( \{0,2,1,4,3\} + \{1,0,3,2,4\} )
  +\tfrac{2}{3}( \{0,1,4,3,2\} + \{2,1,0,3,4\} ) \nonumber \\
& -\tfrac{4}{3}( \{0,2,1,3,2\} + \{1,0,2,1,3\} + \{1,0,3,2,1\} + \{2,1,0,3,2\} ) \nonumber \\
& -2( \{0,1,3,2,4\} + \{0,2,1,3,4\} + \{1,0,4,3,2\} + \{2,1,0,4,3\} ) \nonumber \\
& -2( \{0,1,2,4,3\} + \{0,4,3,2,1\} + \{1,0,2,3,4\} + \{3,2,1,0,4\} ) \\[3mm]
Q_{4,8} =\ 
& +\tfrac{2}{3}(4763+2\alpha) \{\}
  -\tfrac{2}{3}(10349+16\alpha) \{0\}
  +\tfrac{1}{3}(6139+22\alpha)( \{0,1\} + \{1,0\} ) \nonumber \\
& +\tfrac{5}{3}(529+4\alpha) \{0,2\}
  -\tfrac{835}{3} \{1,0,2,1\}  \nonumber \\
& -\tfrac{1}{3}(4391+12\alpha)( \{0,1,2\} + \{2,1,0\} )
  +(961-2\alpha)( \{0,2,1\} + \{1,0,2\} ) \nonumber \\
& +\tfrac{2}{3}(271+2\alpha) \{0,3\}
  -\tfrac{1}{3}(323+4\alpha)( \{0,1,3\} + \{0,2,3\} + \{0,3,2\} + \{1,0,3\} ) \nonumber \\
& +\tfrac{1}{3}(1793+6\alpha)( \{0,1,2,3\} + \{3,2,1,0\} )
  -\tfrac{1}{6}(1163+16\alpha)( \{0,2,1,3\} + \{1,0,3,2\} ) \nonumber \\
& +\tfrac{1}{2}(59+2\alpha)( \{0,2,1,3,2\} + \{1,0,2,1,3\} + \{1,0,3,2,1\} + \{2,1,0,3,2\} ) \nonumber \\
& +12 \{0,4\}
  -\tfrac{40}{3} \{0,2,4\}
  -8( \{0,1,4\} + \{0,3,4\} + \{0,4,3\} + \{1,0,4\} ) \nonumber \\
& -\tfrac{1}{6}(227+2\alpha)( \{0,2,1,4\} + \{0,2,4,3\} + \{0,3,2,4\} + \{1,0,2,4\} ) \nonumber \\
& -\tfrac{1}{6}(955-4\alpha)( \{0,1,3,2\} + \{0,3,2,1\} + \{1,0,2,3\} + \{2,1,0,3\} ) \nonumber \\
& +\tfrac{1}{6}(275+2\alpha)( \{0,1,2,4\} + \{0,2,3,4\} + \{0,4,3,2\} + \{2,1,0,4\} ) \nonumber \\
& -\tfrac{1}{6}(207-4\alpha)( \{0,1,4,3\} + \{1,0,3,4\} )
  +\tfrac{1}{6}(271-4\alpha)( \{0,1,3,4\} + \{1,0,4,3\} ) + \nonumber
\end{align}
\begin{align}
& +\tfrac{1}{3}(145-\alpha)( \{0,1,2,4,3\} + \{0,4,3,2,1\} + \{1,0,2,3,4\} + \{3,2,1,0,4\} ) \nonumber \\
& -\tfrac{2}{3}(234+\alpha)( \{0,1,2,3,4\} + \{4,3,2,1,0\} ) \nonumber \\
& +\tfrac{8}{3}( \{0,3,2,4,3\} + \{1,0,2,1,4\} )
  -\tfrac{1}{3}(41-2\alpha)( \{0,2,1,4,3\} + \{1,0,3,2,4\} ) \nonumber \\
& -\tfrac{46}{3}( \{0,3,2,1,4\} + \{1,0,2,4,3\} )
  -\tfrac{47}{3}( \{0,1,4,3,2\} + \{2,1,0,3,4\} ) \nonumber \\
& +\tfrac{1}{3}(148+\alpha)( \{0,1,3,2,4\} + \{0,2,1,3,4\} + \{1,0,4,3,2\} + \{2,1,0,4,3\} ) \nonumber \\
& -\tfrac{1}{6}(11-4\alpha)( \{0,2,1,3,2,4\} + \{2,1,0,4,3,2\} ) \nonumber \\
& -\tfrac{19}{6}( \{0,1,3,2,4,3\} + \{1,0,2,1,3,4\} + \{1,0,4,3,2,1\} + \{3,2,1,0,4,3\} ) \nonumber \\
& +\tfrac{1}{2}( \{0,3,2,1,4,3\} + \{1,0,2,1,4,3\} + \{1,0,3,2,1,4\} + \{1,0,3,2,4,3\} ) \nonumber \\
& +\tfrac{4}{3}( \{0,1,4,5\} + \{1,0,5,4\} )
  -\tfrac{4}{3}( \{0,1,5,4\} + \{1,0,4,5\} ) \nonumber \\
& -\tfrac{4}{3}( \{0,2,1,5\} + \{0,3,5,4\} + \{0,4,3,5\} + \{1,0,2,5\} ) \nonumber \\
& +\tfrac{4}{3}( \{0,1,2,5\} + \{0,3,4,5\} + \{0,5,4,3\} + \{2,1,0,5\} ) \nonumber \\
& +\tfrac{4}{3}( \{0,2,3,5,4\} + \{0,3,2,1,5\} + \{0,4,3,2,5\} + \{1,0,2,3,5\} ) \nonumber \\
& +\tfrac{4}{3}( \{0,2,1,3,5\} + \{0,2,4,3,5\} + \{0,3,2,5,4\} + \{1,0,3,2,5\} ) \nonumber \\
& +\tfrac{4}{3}( \{0,1,4,3,5\} + \{0,2,1,4,5\} + \{1,0,2,5,4\} + \{1,0,3,5,4\} ) \nonumber \\
& +\tfrac{4}{3}( \{0,1,3,5,4\} + \{0,2,1,5,4\} + \{1,0,2,4,5\} + \{1,0,4,3,5\} ) \nonumber \\
& +\tfrac{4}{3}( \{0,1,3,2,5\} + \{0,2,5,4,3\} + \{0,3,2,4,5\} + \{2,1,0,3,5\} ) \nonumber \\
& +\tfrac{4}{3}( \{0,1,2,5,4\} + \{0,1,5,4,3\} + \{1,0,3,4,5\} + \{2,1,0,4,5\} ) \nonumber \\
& -4( \{0,1,2,4,5\} + \{0,1,3,4,5\} + \{1,0,5,4,3\} + \{2,1,0,5,4\} ) \nonumber \\
& -4( \{0,1,2,3,5\} + \{0,2,3,4,5\} + \{0,5,4,3,2\} + \{3,2,1,0,5\} ) \nonumber \\
& +\tfrac{2}{3}(2-\alpha)( \{0,4,3,2,1,5\} + \{1,0,2,3,5,4\} )
  +\tfrac{2}{3}(2+\alpha)( \{0,2,1,4,3,5\} + \{1,0,3,2,5,4\} ) \nonumber \\
& +\tfrac{4}{3}( \{0,2,1,3,5,4\} + \{0,3,2,1,5,4\} + \{1,0,2,4,3,5\} + \{1,0,4,3,2,5\} ) \nonumber \\
& -\tfrac{1}{6}(13+4\alpha)( \{1,0,2,1,3,2\} + \{2,1,0,3,2,1\} )
  -\tfrac{1}{6}(5+4\alpha)( \{0,2,1,4,3,2\} + \{2,1,0,3,2,4\} ) \nonumber \\
& +\tfrac{1}{3}(2+\alpha)( \{0,1,4,3,2,5\} + \{0,3,2,1,4,5\} + \{1,0,2,5,4,3\} + \{2,1,0,3,5,4\} ) \nonumber \\
& +\tfrac{1}{3}(2-\alpha)( \{0,1,3,2,5,4\} + \{0,2,1,5,4,3\} + \{1,0,3,2,4,5\} + \{2,1,0,4,3,5\} ) \nonumber \\
& +\tfrac{56}{3}( \{0,1,2,3,4,5\} + \{5,4,3,2,1,0\} )
  -\tfrac{16}{3}( \{0,1,3,2,4,5\} + \{2,1,0,5,4,3\} ) \nonumber \\
& +\tfrac{4}{3}( \{0,1,2,5,4,3\} + \{0,1,5,4,3,2\} + \{2,1,0,3,4,5\} + \{3,2,1,0,4,5\} ) \nonumber \\
& -\tfrac{1}{3}(18+\alpha)( \{0,1,2,4,3,5\} + \{0,2,1,3,4,5\} + \{1,0,5,4,3,2\} + \{3,2,1,0,5,4\} ) \nonumber \\
& -\tfrac{1}{3}(18-\alpha)( \{0,1,2,3,5,4\} + \{0,5,4,3,2,1\} + \{1,0,2,3,4,5\} + \{4,3,2,1,0,5\} )
\end{align}


\section{Spin chain spectrum} \label{sec:spectrum}

\noindent
In this appendix we list the eigenvalues $q_{2,3,4}$ of the charges $Q_{2,3,4}$ for spin-chains of length $L$ and magnon number $M$. The eigenvalues have been 
computed once by directly applying the corresponding operators to explicit states, and a second time by solving the Bethe equations for the momenta\footnote{We are 
grateful to V.~Dippel for providing us some of these solutions from her unpublished work.} and using the following eigenvalue formulas
\be
  q_i = \sum_{j=1}^M q_i(p_j)
\ee
with \\[-5mm]
\begin{subequations} \label{eqn:eigenvalue-formulas}
\begin{align}
  q_2(p) & = 8                  \sin^2(\tfrac{p}{2})
             - 32   \Lambda_r   \sin^4(\tfrac{p}{2}) \nonumber \\
     & \quad + 256  \Lambda_r^2 \sin^6(\tfrac{p}{2})
             - 2560 \Lambda_r^3 \sin^8(\tfrac{p}{2}) \; , \label{eqn:q2} \\[2mm]
  q_3(p) & = 8                  \sin(p) \sin^2(\tfrac{p}{2})
             - 64   \Lambda_r   \sin(p) \sin^4(\tfrac{p}{2}) \nonumber \\
     & \quad + 640  \Lambda_r^2 \sin(p) \sin^6(\tfrac{p}{2})
             - 7168 \Lambda_r^3 \sin(p) \sin^8(\tfrac{p}{2}) \label{eqn:q3} \; , \\[2mm]
  q_4(p) & = \tfrac{32}{3}                 \sin(\tfrac{3p}{2}) \sin^3(\tfrac{p}{2})
             - 128             \Lambda_r   \sin(\tfrac{3p}{2}) \sin^5(\tfrac{p}{2}) \nonumber \\
     & \quad + 1536            \Lambda_r^2 \sin(\tfrac{3p}{2}) \sin^7(\tfrac{p}{2})
             -\tfrac{57344}{3} \Lambda_r^3 \sin(\tfrac{3p}{2}) \sin^9(\tfrac{p}{2}) \; . \label{eqn:q4}
\end{align}
\end{subequations}
These formulas are in accordance with the general all-loop expressions for $q_i$ given in \cite{bds-long-range,frolov-gleb-matthias,gtcharges}.

The stars ``$***$'' in the tables below indicate that the higher loop eigenvalues could not be determined due to wrapping issues. The value $q_{2,8} = 
-\tfrac{12771}{4}\Lambda_r^3$ for the state $(L,M)=(4,2)$ is also subject to wrapping, but was determined in a separate calculation \eqref{eqn:dilop-onto-Konishi}. If 
we used the eigenvalue formula \eqref{eqn:q2} one would get the wrong value $q_{2,8} = -3054\Lambda_r^3$.

``Singular'' means that the Bethe equations become singular for the lowest order solution of the Bethe momenta $p_i$. In these cases one should use the inhomogeneous 
Bethe ansatz as in \cite{bds-long-range}.

Unpaired states are annihilated by $Q_3$, hence $q_3 = 0$. In these cases we give the parity of the corresponding multiplet in the column for $q_3$. If $Q_3$ does not 
annihilate a state, then it maps the state to the partner state of opposite parity. As the generic $Q_3$ and the parity operator do not commute (they anti-commute), 
we cannot diagonalize both operators simultaneously. In these cases we print the pair of eigenvalues of $Q_3$ which are always the negative of each other. All Bethe 
momenta are also opposite to each other for a degenerate pair of states.

\newcommand{\La}{}
\newcommand{\Lb}{\Lambda_r}
\newcommand{\Lc}{\Lambda_r^2}
\newcommand{\Ld}{\Lambda_r^3}
\begin{table}[h]
\begin{center}
\begin{scriptsize}
\begin{tabular}{|c|c|p{35mm}|p{25mm}|p{30mm}|p{25mm}|} \hline
$L $ & $M$ & $q_2$ & $q_3$ & $q_4$ & Bethe momenta \\ \hline
$4$  & $2$ & ${12}\La$                                & $+$ & $0\La$                         & $p_1 = \tfrac{2\pi}{3}$ \\
     &     & ${-48}\Lb$                               &     & $***$                          & $p_2 = -p_1$ \\
     &     & ${+336}\Lc$                              &     &                                & \\
     &     & ${-\tfrac{12771}{4}}\Ld$                 &     &                                & \\ \hline
$5$  & $2$ & ${8}\La$                                 & $-$ & $\tfrac{16}{3}\La$             & $p_1 = \tfrac{\pi}{2}$ \\
     &     & ${-24}\Lb$                               &     & $-32\Lb$                       & $p_2 = -p_1$ \\
     &     & ${+136}\Lc$                              &     & $***$                          & \\
     &     & ${-1024}\Ld$                             &     &                                & \\ \hline
\end{tabular}
\end{scriptsize}
\caption{Eigenvalues of $Q_2$,$Q_3$,$Q_4$}
\label{tab:quantumnumbers1}
\end{center}
\end{table}

\begin{table}
\begin{center}
\begin{scriptsize}
\begin{tabular}{|c|c|p{35mm}|p{25mm}|p{30mm}|p{25mm}|} \hline
$L $ & $M$ & $q_2$ & $q_3$ & $q_4$ & Bethe momenta \\ \hline
$6$  & $2$ & ${2(5-\sqrt{5})}\La$                     & $+$ & $-\tfrac{10}{3}(1-\sqrt{5})\La$   & $p_1 = \tfrac{2\pi}{5}$ \\
     &     & ${-(34-10\sqrt{5})}\Lb$                  &     & $+12(5+3\sqrt{5})\Lb$             & $p_2 = -p_1$ \\
     &     & ${+\tfrac{1}{5}(1170-414\sqrt{5})}\Lc$   &     & $-6(127+65\sqrt{5})\Lc$           & \\
     &     & ${-\tfrac{1}{5}(10695-4134\sqrt{5})}\Ld$ &     & $***$                             & \\ \cline{3-6}
     &     & ${2(5+\sqrt{5})}\La$                     & $+$ & $-\tfrac{10}{3}(1+\sqrt{5})\La$   & $p_1 = \tfrac{4\pi}{5}$ \\
     &     & ${-(34+10\sqrt{5})}\Lb$                  &     & $+12(5-3\sqrt{5})\Lb$             & $p_2 = -p_1$ \\
     &     & ${+\tfrac{1}{5}(1170+414\sqrt{5})}\Lc$   &     & $-6(127-65\sqrt{5})\Lc$           & \\
     &     & ${-\tfrac{1}{5}(10695+4134\sqrt{5})}\Ld$ &     & $***$                             & \\ \cline{2-6}
     & $3$ & ${12}\La$                                & $-$ & $-12\La$                          & ``singular'' \\
     &     & ${-36}\Lb$                               &     & $+144\Lb$                         & \\
     &     & ${+252}\Lc$                              &     & $-1620\Lc$                        & \\
     &     & ${-2484}\Ld$                             &     & $***$                             & \\ \hline
$7$  & $2$ & ${4}\La$                                 & $-$                  & $\tfrac{8}{3}\La$      & $p_1 = \tfrac{\pi}{3}$ \\
     &     & ${-6}\Lb$                                &                      & $-10\Lb$               & $p_2 = -p_1$ \\
     &     & ${+\tfrac{37}{2}}\Lc$                    &                      & $+\tfrac{81}{2}\Lc$    & \\
     &     & ${-\tfrac{335}{4}}\Ld$                   &                      & $-\tfrac{759}{4}\Ld$   & \\ \cline{3-6}
     &     & ${12}\La$                                & $-$                  & $0\La$                 & $p_1 = \tfrac{2\pi}{3}$ \\
     &     & ${-42}\Lb$                               &                      & $-18\Lb$               & $p_2 = -p_1$ \\
     &     & ${+\tfrac{555}{2}}\Lc$                   &                      & $-\tfrac{513}{2}\Lc$   & \\
     &     & ${-\tfrac{9465}{4}}\Ld$                  &                      & $+\tfrac{13311}{4}\Ld$ & \\ \cline{2-6}
     & $3$ & ${10}\La$                                & $\mp\sqrt{15}\La$    & $-\tfrac{10}{3}\La$    & $p_1 = \pm1.16\pm0.93i$ \\
     &     & ${-30}\Lb$                               & $\pm8\sqrt{15}\Lb$   & $+50\Lb$               & $p_2 = \pm1.16\mp0.93i$ \\
     &     & ${+200}\Lc$                              & $\mp69\sqrt{15}\Lc$  & $-570\Lc$              & $p_3 = -p_1-p_2$ \\
     &     & ${-\tfrac{3565}{2}}\Ld$                  & $\pm668\sqrt{15}\Ld$ & $+\tfrac{13245}{2}\Ld$ & \\ \hline
$8$  & $2$ & ${3.01}\La$      & $+$          & $1.70\La$           & $p_1 = \tfrac{2\pi}{7}$ \\
     &     & ${-3.32}\Lb$     &              & $-4.83\Lb$          & $p_2 = -p_1$ \\
     &     & ${+7.66}\Lc$     &              & $+14.81\Lc$         & \\
     &     & ${-27.30}\Ld$    &              & $-53.90\Ld$         & \\ \cline{3-6}
     &     & ${9.78}\La$      & $+$          & $4.42\La$           & $p_1 = \tfrac{4\pi}{7}$ \\
     &     & ${-29.22}\Lb$    &              & $-26.67\Lb$         & $p_2 = -p_1$ \\
     &     & ${+167.64}\Lc$   &              & $+176.12\Lc$        & \\
     &     & ${-1256.15}\Ld$  &              & $-1204.33\Ld$       & \\ \cline{3-6}
     &     & ${15.21}\La$     & $+$          & $-15.46\La$         & $p_1 = \tfrac{6\pi}{7}$ \\
     &     & ${-59.45}\Lb$    &              & $+187.49\Lb$        & $p_2 = -p_1$ \\
     &     & ${+456.70}\Lc$   &              & $-2190.93\Lc$       & \\
     &     & ${-4430.55}\Ld$  &              & $+26640.90\Ld$      & \\ \cline{2-6}
     & $3$ & ${8}\La$         & $\mp4\La$    & $\tfrac{4}{3}\La$   & $p_1 = \pm0.96\pm0.59i$ \\
     &     & ${-20}\Lb$       & $\pm28\Lb$   & $-4\Lb$             & $p_2 = \pm0.96\mp0.59i$ \\
     &     & ${+112}\Lc$      & $\mp210\Lc$  & $+16\Lc$            & $p_3 = -p_1 -p_2$ \\
     &     & ${-842}\Ld$      & $\pm1743\Ld$ & $-\tfrac{52}{3}\Ld$ & \\ \cline{3-6}
     &     & ${12}\La$        & $-$          & $-12\La$            & ``singular'' \\
     &     & ${-36}\Lb$       &              & $+132\Lb$           & \\
     &     & ${+264}\Lc$      &              & $-1536\Lc$          & \\
     &     & ${-2592}\Ld$     &              & $+19056\Ld$         & \\ \cline{2-6}
     & $4$ & ${6.49}\La$      & $+$          & $-0.93\La$          & $p_1 = 1.13+0.87i$ \\
     &     & ${-7.56}\Lb$     &              & $+10.15\Lb$         & $p_2 = 1.13-0.87i$ \\
     &     & ${+10.22}\Lc$    &              & $-55.32\Lc$         & $p_3 = -p_1$ \\
     &     & ${+10.25}\Ld$    &              & $+285.14\Ld$        & $p_4 = -p_2$ \\ \cline{3-6}
     &     & ${10.90}\La$     & $+$          & $-11.67\La$         & $p_1 = 2.98$ \\
     &     & ${-31.76}\Lb$    &              & $+129.44\Lb$        & $p_2 = -1.07i$ \\
     &     & ${+249.76}\Lc$   &              & $-1528.91\Lc$       & $p_3 = -p_1$ \\
     &     & ${-2538.27}\Ld$  &              & $+19065.27\Ld$      & $p_4 = -p_2$ \\ \cline{3-6}
     &     & ${22.60}\La$     & $+$          & $-8.73\La$          & $p_1 = 2.63$ \\
     &     & ${-88.68}\Lb$    &              & $+148.40\Lb$        & $p_2 = 1.52$ \\
     &     & ${+636.03}\Lc$   &              & $-1951.77\Lc$       & $p_3 = -p_1$ \\
     &     & ${-5933.98}\Ld$  &              & $+25244.26\Ld$      & $p_4 = -p_2$ \\ \hline
\end{tabular}
\end{scriptsize}
\let\thetable=3
\caption{Eigenvalues of $Q_2$,$Q_3$,$Q_4$ (cont.)}
\label{tab:quantumnumbers2}
\end{center}
\end{table}

\begin{table}
\begin{center}
\begin{scriptsize}
\begin{tabular}{|c|c|p{35mm}|p{21mm}|p{37mm}|p{25mm}|} \hline
$L $ & $M$ & $q_2$ & $q_3$ & $q_4$ & Bethe momenta \\ \hline
$9$  & $2$ & ${4(2-\sqrt{2})}\La$                       & $-$               & $-\tfrac{8}{3}(1-\sqrt{2})\La$             & $p_1 = \tfrac{\pi}{4}$ \\
     &     & ${-(26-17\sqrt{2})}\Lb$                    &                   & $+10(4-3\sqrt{2})\Lb$                      & $p_2 = -p_1$ \\
     &     & ${+\tfrac{1}{8}(179-993\sqrt{2})}\Lc$      &                   & $-\tfrac{21}{4}(88-63\sqrt{2})\Lc$         & \\
     &     & ${-\tfrac{1}{32}(51376-36103\sqrt{2})}\Ld$ &                   & $+\tfrac{1}{48}(260000-184411\sqrt{2})\Ld$ & \\ \cline{3-6}
     &     & ${8}\La$                                   & $-$               & $\tfrac{16}{3}\La$                         & $p_1 = \tfrac{\pi}{2}$ \\
     &     & ${-20}\Lb$                                 &                   & $-32\Lb$                                   & $p_2 = -p_1$ \\
     &     & ${+98}\Lc$                                 &                   & $+204\Lc$                                  & \\
     &     & ${-639}\Ld$                                &                   & $-\tfrac{4192}{3}\Ld$                      & \\ \cline{3-6}
     &     & ${4(2+\sqrt{2})}\La$                       & $-$               & $-\tfrac{8}{3}(1+\sqrt{2})\La$             & $p_1 = \tfrac{3\pi}{4}$ \\
     &     & ${-(26+17\sqrt{2})}\Lb$                    &                   & $+10(4+3\sqrt{2})\Lb$                      & $p_2 = -p_1$ \\
     &     & ${+\tfrac{1}{8}(179+993\sqrt{2})}\Lc$      &                   & $-\tfrac{21}{4}(88+63\sqrt{2})\Lc$         & \\
     &     & ${-\tfrac{1}{32}(51376+36103\sqrt{2})}\Ld$ &                   & $+\tfrac{1}{48}(260000-184411\sqrt{2})\Ld$ & \\ \cline{2-6}
     & $3$ & ${6.45}\La$                                & $\pm3.44\La$      & $2.67\La$                                  & $p_1 = \mp0.83\mp0.43i$ \\
     &     & ${-13.18}\Lb$                              & $\mp20.01\Lb$     & $-15.38\Lb$                                & $p_2 = \mp0.83\pm0.43i$ \\
     &     & ${+60.73}\Lc$                              & $\pm124.66\Lc$    & $+103.80\Lc$                               & $p_3 = -p_1-p_2$ \\
     &     & ${-378.34}\Ld$                             & $\mp859.22\Ld$    & $-748.03\Ld$                               & \\ \cline{3-6}
     &     & ${11.04}\La$                               & $\pm2.99\La$      & $-6.89\La$                                 & $p_1 = \mp1.28\mp1.26i$ \\
     &     & ${-33.50}\Lb$                              & $\mp23.92\Lb$     & $+83.02\Lb$                                & $p_2 = \mp1.28\pm1.26i$ \\
     &     & ${+231.93}\Lc$                             & $\pm223.38\Lc$    & $-952.18\Lc$                               & $p_3 = -p_1-p_2$ \\
     &     & ${-2149.91}\Ld$                            & $\mp2345.23\Ld$   & $+11361.10\Ld$                             & \\ \cline{3-6}
     &     & ${16.51}\La$                               & $\pm8.61\La$      & $-9.12\La$                                 & $p_1 = \pm2.98$ \\
     &     & ${-55.32}\Lb$                              & $\mp53.35\Lb$     & $+120.35\Lb$                               & $p_2 = \pm1.15$ \\
     &     & ${+383.33}\Lc$                             & $\pm414.82\Lc$    & $-1471.62\Lc$                              & $p_3 = -p_1-p_2$ \\
     &     & ${-3494.75}\Ld$                            & $\mp3879.59\Ld$   & $+18242.92\Ld$                             & \\ \cline{2-6}
     & $4$ & ${10}\La$                                  & $\mp\sqrt{11}\La$ & $-\tfrac{22}{3}\La$                        & $p_1 = \pm2.63$ \\
     &     & ${-30}\Lb$                                 & $\pm\tfrac{84}{11}\sqrt{11}\Lb$       & $+86\Lb$               & $p_2 = \mp0.53\pm0.88i$ \\
     &     & ${+220}\Lc$                                & $\mp\tfrac{8495}{121}\sqrt{11}\Lc$    & $-982\Lc$              & $p_3 = \mp1.77$ \\
     &     & ${-\tfrac{4185}{2}}\Ld$                    & $\pm\tfrac{978121}{1331}\sqrt{11}\Ld$ & $+\tfrac{23371}{2}\Ld$ & $p_4 = -p_1-p_2-p_3$ \\ \cline{3-6}
     &     & ${4(3-\sqrt{3})}\La$                       & $-$               & $0\La$                                     & $p_1 = 0.94+0.55i$ \\
     &     & ${-18(2-\sqrt{3})}\Lb$                     &                   & $+(24-12\sqrt{3})\Lb$                      & $p_2 = 0.94-0.55i$ \\
     &     & ${+\tfrac{1}{2}(456-255\sqrt{3})}\Lc$      &                   & $-(360-198\sqrt{3})\Lc$                    & $p_3 = -p_1$ \\
     &     & ${-\tfrac{1}{4}(7908-4541\sqrt{3})}\Ld$    &                   & $+\tfrac{1}{2}(9540-5417\sqrt{3})\Ld$      & $p_4 = -p_2$ \\ \cline{3-6}
     &     & ${4(3+\sqrt{3})}\La$                       & $-$               & $0\La$                                     & $p_1 = 2.28$ \\
     &     & ${-18(2+\sqrt{3})}\Lb$                     &                   & $+(24-12\sqrt{3})\Lb$                      & $p_2 = 1.28$ \\
     &     & ${+\tfrac{1}{2}(456+255\sqrt{3})}\Lc$      &                   & $-(360-198\sqrt{3})\Lc$                    & $p_3 = -p_1$ \\
     &     & ${-\tfrac{1}{4}(7908+4541\sqrt{3})}\Ld$    &                   & $+\tfrac{1}{2}(9540+5417\sqrt{3})\Ld$      & $p_4 = -p_2$ \\ \hline
\end{tabular}
\end{scriptsize}
\let\thetable=3
\caption{Eigenvalues of $Q_2$,$Q_3$,$Q_4$ (cont.) }
\label{tab:quantumnumbers3}
\end{center}
\end{table}


\end{document}